\documentclass[twocolumn,pra,letterpaper,showpacs,citesort]{revtex4}
\usepackage{hyperref}
\usepackage{latexsym,amsmath,amssymb,amsfonts,graphicx,color,amsthm}
\usepackage{multirow}
\usepackage{graphicx}
\input{Qcircuit}

\def\be{\begin{eqnarray}}
\def\ee{\end{eqnarray}}
\DeclareMathOperator{\erf}{erf}

\newcommand{\eps}{\mathcal{\varepsilon}}

\newcommand{\const}{\mathrm{const}}

\newcommand{\eff}{\text{eff}}
\newcommand{\conv}{\text{conv}}

\begin{document}
\title{Protected gates for superconducting qubits}
\author{Peter Brooks, Alexei Kitaev, and John Preskill}
\affiliation{Institute for Quantum Information and Matter, California Institute of Technology, Pasadena, CA 91125, USA}
\pacs{03.67.Pp, 03.67.Lx, 85.25.Hv, 85.25.Cp}

\begin{abstract}
We analyze the accuracy of quantum phase gates acting on ``0-$\pi$ qubits'' in superconducting circuits, where the gates are protected against thermal and Hamiltonian noise by continuous-variable quantum error-correcting codes. The gates are executed by turning on and off a tunable Josephson coupling between an $LC$ oscillator and a qubit or pair of quits; assuming perfect qubits, we show that the gate errors are exponentially small when the oscillator's impedance $\sqrt{L/C}$ is large compared to $\hbar/4e^2 \approx 1~k\Omega$. The protected gates are not computationally universal by themselves, but a scheme for universal fault-tolerant quantum computation can be constructed by combining them with unprotected noisy operations. We validate our analytic arguments with numerical simulations. 
\end{abstract}

\maketitle

\section{Introduction}\label{sec:intro}
Building a scalable quantum computer is a formidable challenge because quantum systems decohere readily and because their interactions are hard to control accurately, yet we hope to succeed someday by prudently applying the principles of quantum error correction and fault-tolerant quantum computing. In the standard ``software'' approach to quantum fault-tolerance \cite{Shor1996}, the deficiencies of noisy quantum hardware (if not {\sl too} noisy) are overcome through clever circuit design, while in the alternative ``topological'' approach \cite{Kitaev1997}, the hardware itself is intrinsically resistant to decoherence. Both approaches exploit the idea that logical qubits can be stored and processed reliably when suitably encoded in a quantum system with many degrees of freedom; perhaps both approaches will be employed together in future quantum computing systems. 

The best known version of the topological approach is based on nonabelian anyons, with quantum information stored in the fusion spaces of the anyons and processed by braiding the anyons, but it is important to search for other ways to realize quantum hardware such that intrinsic robustness results from how the information is physically encoded. One intriguing possibility is to use superconducting circuits for this purpose. Specifically, several authors \cite{DoucotVidal2002,IoffeFeigelman2002,Kitaev2006} have proposed designs for a superconducting ``0-$\pi$ qubit,'' a circuit containing Josephson junctions. The circuit's energy is a function of the superconducting phase difference $\theta$ between the two leads of the circuit, and there are two nearly degenerate ground states, localized near $\theta =0$ and $\theta = \pi$ respectively. The splitting of this degeneracy is exponentially small as a function of extensive system parameters, and stable with respect to weak local perturbations. Thus the 0-$\pi$ qubit should be highly resistant to decoherence arising from local noise.

For reliable quantum computing we need not just very stable qubits, but also the ability to apply very accurate nontrivial quantum gates to the qubits. A method for achieving protected single-qubit and two-qubit phase gates acting on 0-$\pi$ qubits, exploiting the error-correcting properties of a continuous-variable quantum code \cite{GottesmanKitaevPreskill2001}, was suggested in \cite{Kitaev2006}, and it was claimed that the gate errors can be exponentially small as a function of extensive system parameters. In this paper we further develop and explore the ideas behind this protected gate. 

Protected phase gates are executed by turning on and off a tunable Josephson coupling between an $LC$ oscillator and a qubit or pair of qubits. Assuming the qubits are perfect, we show, using analytic arguments validated by numerical simulations, that the gate errors are exponentially small when the oscillator's impedance $\sqrt{L/C}$ is large compared to $\hbar/4e^2 \approx 1~k\Omega$, where $L$ is the inductance and $C$ is the capacitance of the oscillator. The gates are robust against small deformations of the device Hamiltonian and against small thermal fluctuations of the oscillator. The very large inductance in the superconducting oscillator, which is crucial for the high gate accuracy, may be quite challenging to achieve in practice, but the potential rewards are correspondingly substantial.

The internal structure of the 0-$\pi$ qubit is not relevant to our analysis, but for completeness we nevertheless explain in Sec.~\ref{sec:qubit} the idea behind the qubit design proposed in \cite{Kitaev2006}, which also requires a large inductance in a superconducting circuit. We describe how a protected phase gate is executed in Sec.~\ref{sec:gate}, and in Sec.~\ref{sec:sketch} we outline our strategy for estimating the gate accuracy. We review the properties of continuous-variable quantum error-correcting codes in Sec.~\ref{sec:code}, and explain in Sec.~\ref{sec:overrotation} how the code provides protection against imperfect timing in the pulse that executes the gate. We analyze contributions to the gate error due to diabatic transitions and squeezing in Sec.~\ref{sec:diabatic} and Sec.~\ref{sec:squeezing}, then compare our predictions with numerical simulations in Sec.~\ref{sec:numerics}. We discuss robustness with respect to thermal effects in Sec.~\ref{sec:temperature} and with respect to Hamiltonian perturbations in Sec.~\ref{sec:anharmonic}. In Sec.~\ref{sec:universal} we explain how to obtain a complete scheme for universal fault-tolerant quantum computation by augmenting the protected phase gates with measurements and unprotected noisy phase gates. Sec.~\ref{sec:conclusions} contains our conclusions, and some further details are contained in Appendices.

\section{The 0-$\pi$ qubit}\label{sec:qubit}
For most of this paper we may treat the 0-$\pi$ qubit as a black box, disregarding its internal structure. But here we will briefly explain the concept underlying the proposal in \cite{Kitaev2006}. 

\begin{figure}[!ht]
\begin{center}
\includegraphics[width=0.4\textwidth]{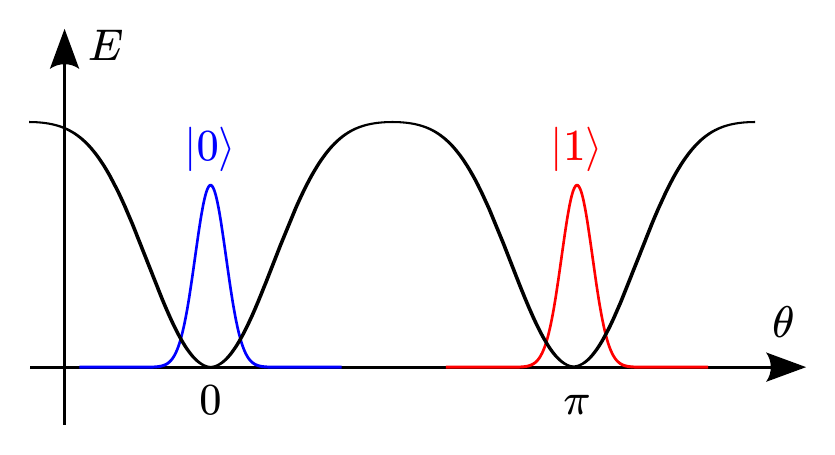}
\end{center}
\caption{\label{fig:zero-pi} (Color online) The 0-$\pi$ qubit. The energy $E(\theta)$ of a superconducting circuit is a periodic function with period $\pi$ of the phase difference $\theta$ between its two leads, aside from exponentially small corrections. The two basis states $\{|0\rangle,|1\rangle\}$ of the qubit, localized near the minima of the energy at $\theta = 0$ and $\theta = \pi$ respectively, are nearly degenerate.}
\end{figure}

We consider a two-lead superconducting circuit, whose energy $E(\theta)$ is a function of the phase difference $\theta$ between the two leads. Here $\theta$ is a periodic variable with period $2\pi$, but for a suitably constructed circuit, $E(\theta)$ will actually be very nearly a function with period $\pi$, apart from exponentially small corrections. This function has minima at $\theta=0$ and $\theta = \pi$, separated by a high barrier, so that there are two well localized states centered at $\theta = 0$ and $\theta = \pi$ respectively, which we may take to be the basis states $|0\rangle$ and $|1\rangle$ of an encoded qubit, as shown in Fig.~\ref{fig:zero-pi}. The high tunneling barrier suppresses bit flip errors, but the key feature of the qubit is that the $\pi$ periodicity, and hence the degeneracy of the two states, is robust against generic local perturbations, so that dephasing of coherent superpositions of $|0\rangle$ and $|1\rangle$ is also highly suppressed.

\begin{figure}[!ht]
\begin{center}
\includegraphics[width=0.5\textwidth]{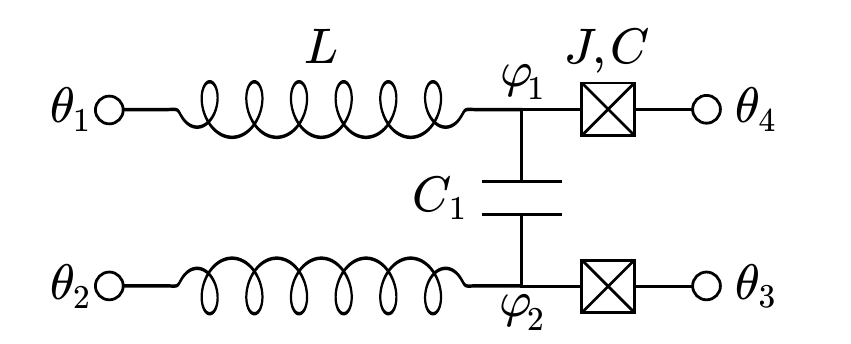}
\end{center}
\caption{\label{fig:four-leads} Two-rung superconducting circuit underlying the 0-$\pi$ qubit. If $\sqrt{L/C}$ is large, $C_1$ is large compared to $C$, and $JC$ is not too large, then the circuit's energy is a function of  the combination of phases $\left(\theta_2 + \theta_4\right) - \left(\theta_1+\theta_3\right)$, aside from corrections that are exponentially small in $\sqrt{L/C}$.}
\end{figure} 

To understand the qubit's properties, first consider the four-lead circuit shown in Fig.~\ref{fig:four-leads}. This circuit has two identical rungs, connected by a large capacitance $C_1$. Each rung consists of a Josephson junction, with Josephson energy $J$ and intrinsic capacitance $C$, connected in series with a inductance $L$, chosen such that $\sqrt{L/C}$ is large compared to the natural unit of impedance $\hbar / (2e)^2 \approx 1.03 ~k\Omega$, and hence much larger than its ``geometric'' value $4\pi/c\approx 377 ~\Omega$ (where $c$ is the speed of light), the impedance of free space. Achieving such a ``superinductance'' may be an engineering challenge, but we take it for granted here that it is possible. The properties of a single rung, which can operate as an adiabatic switch when $J$ varies, is discussed in more detail in Appendix \ref{app:switch}.

We denote the value of the superconducting phase on the circuit's four leads as $\theta_1$, $\theta_2$, $\theta_3$, $\theta_4$ as shown, and the phase on either side of the capacitor connecting the rungs by $\varphi_1$, $\varphi_2$. Then the phase $\varphi_+ = \left(\varphi_1 + \varphi_2\right)/2$ is insensitive to the value of the capacitance $C_1$, which we assume is much larger than $C$. Therefore the sum $\varphi_+$ is a ``light'' variable with large fluctuations (assuming $JC$ is not too large), while in contrast the difference  $\varphi_- = \varphi_1 - \varphi_2$, which does feel the effect of the large capacitance $C_1$, is a well localized ``heavy'' variable. We assume that phase slips through the inductors are suppressed, so that we may regard $\varphi_{\pm}$ as real variables rather than periodic phase variables with period $2\pi$.

A circuit with capacitance $C_{\rm conv}$ and inductance $L_{\rm conv}$ has Hamiltonian 
\begin{equation}
H = \frac{q^2}{2C_{\rm conv}} + \frac{\Phi^2}{2L_{\rm conv}},
\end{equation}
where $q$ is the charge on the capacitor and $\Phi$ is the magnetic flux linking the circuit. We use the subscript ``conv'' to indicate that capacitance and inductance are expressed here in conventional units, while we will find it more convenient to use rationalized units such that
\begin{equation}
C = C_{\rm conv}/(2e)^2,\quad L = L_{\rm conv} /\left(\hbar/2e\right)^2, 
\end{equation}
so that
\begin{equation}\label{eq:hamiltonian-rationalized}
H = \frac{Q^2}{2C} + \frac{\varphi^2}{2L},
\end{equation}
where $Q = q/2e$ is charge expressed in units of the Cooper pair charge $2e$, and $\varphi =  (2e/\hbar)\Phi$ is the superconducting phase, such that $\varphi= 2\pi$ corresponds to the quantum $h/2e$ of magnetic flux. Then $[\varphi,Q]= i$, and 
\begin{eqnarray}
\sqrt{L/C}& = &\sqrt{L_{\rm conv}/C_{\rm conv}}~/ ~\left(\hbar/4 e^2\right)\nonumber\\
&\approx& \sqrt{L_{\rm conv}/C_{\rm conv}}~/~(1.03~ k\Omega)
\end{eqnarray}
is dimensionless. The ground state of the Hamiltonian Eq.~(\ref{eq:hamiltonian-rationalized}), with energy $E_0= 1/2\sqrt{LC}$, has Gaussian wave function $\psi(\varphi)$ such that
\begin{eqnarray}
|\psi(\varphi)|^2 \propto e^{-\varphi^2/2\langle \varphi^2\rangle},
\end{eqnarray}
where 
\begin{eqnarray}
\langle \varphi^2\rangle = \frac{1}{2}\sqrt{\frac{L}{C}}, 
\end{eqnarray}
and hence
\begin{eqnarray}
\langle \cos\varphi\rangle = e^{-\langle \varphi^2\rangle/2} = \exp\left(-\frac{1}{4}\sqrt{\frac{L}{C}}\right).
\end{eqnarray}
For $\sqrt{L/C} \gg 1$, the ground state wave function is very broad and the wiggles of the cosine nearly average out aside from an exponentially small correction.

The effective capacitance controlling the phase $\varphi_+$ is $C_{\rm eff} = 2C$ and the effective inductance is $L_{\rm eff} = L/2$. Therefore, in the circuit's ground state we have 
\begin{eqnarray}
\langle \varphi_+^2\rangle =  \frac{1}{2}\sqrt{\frac{L_{\rm eff}}{C_{\rm eff}}} =  \frac{1}{4}\sqrt{\frac{L}{C}}.
\end{eqnarray}
The dependence of the Josephson energy on the strongly fluctuating light variable $\varphi_+$ is proportional to  
\begin{eqnarray}
\langle \cos \varphi_+ \rangle = \exp\left(-\frac{1}{8}\sqrt{\frac{L}{C}}\right),
\end{eqnarray}
which is negligible when $\sqrt{L/C}$ is large. We therefore need only consider the dynamics of the well localized heavy variable $\phi_-$, which locks to the value 
\begin{eqnarray}
\phi_-=\left(\theta_4 - \theta_1\right) - \left(\theta_3-\theta_2\right)
= \left(\theta_2 + \theta_4\right) - \left(\theta_1+\theta_3\right)
\end{eqnarray}
determined by the phases on the leads, so that the energy stored in the circuit is
\begin{eqnarray}
E = f\left(\theta_2 + \theta_4 - \theta_1 - \theta_3\right) +O\left(\exp\left(-\frac{1}{8}\sqrt{\frac{L}{C}}\right)\right),
\end{eqnarray}
where $f(\theta)$ is a periodic function with period $2\pi$.

\begin{figure}[!ht]
\begin{center}
\includegraphics[width=0.45\textwidth]{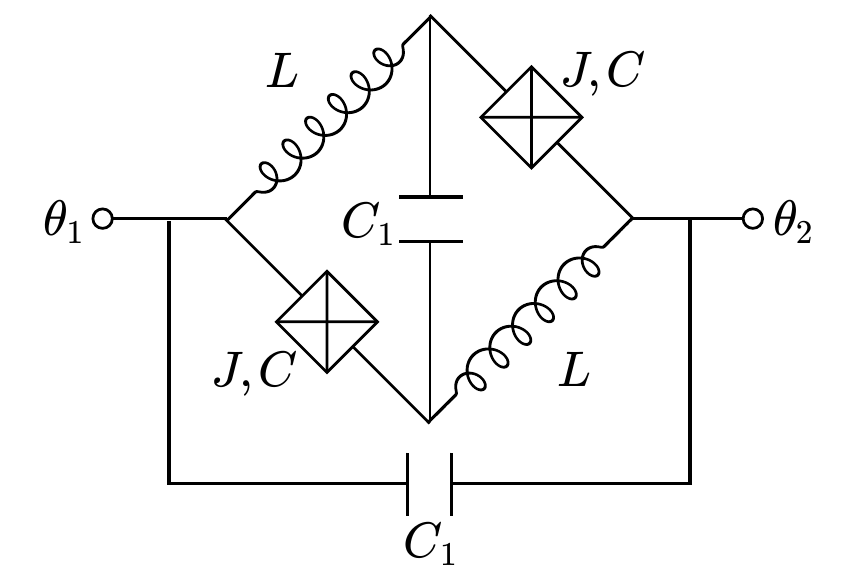}
\end{center}
\caption{\label{fig:qubit-circuit} The circuit for the 0-$\pi$ qubit is obtained from the circuit in Fig.~\ref{fig:four-leads} by twisting one of the rungs and connecting the leads, thus identifying $\theta_2$ with $\theta_4$ and $\theta_1$ with $\theta_3$. In addition, another large capacitance is added to further suppress tunneling events that change $\theta_2-\theta_1$ by $\pi$.}
\end{figure} 

Now, to devise a qubit, we twist the upper rung relative to the lower one and connect the leads as shown in Fig.~\ref{fig:qubit-circuit}, thus identifying $\theta_2$ with $\theta_4$ and $\theta_1$ with $\theta_3$. In addition, we add another large capacitance to ensure that tunneling events changing $\theta_2-\theta_1$ by $\pi$ are heavily suppressed. The energy of the resulting circuit is 
\begin{eqnarray}\label{eq:qubit-periodicity}
E = f(2(\theta_2 - \theta_1))+ \cdots
\end{eqnarray}
where the ellipsis represents exponentially small corrections. Thus the energy is very nearly a periodic function with period $\pi$ of the phase difference $\theta_2-\theta_1$, with two nearly degenerate minima as in Fig.~\ref{fig:zero-pi}.

This robust degeneracy derives from the ``superinducting'' properties of each rung, {\em i.e.}, the large value of $\sqrt{L/C}$. One way to achieve a superinductor, suggested in \cite{Kitaev2006}, is to construct a long chain of $N$ Josephson junctions, each with Josephson coupling $\bar J$ and capacitance $\bar C$. Then the inductance of the chain is linear in $N$, and the capacitance is proportional to $1/N$, so $\sqrt{L/C} \propto N$, and the breaking of the degeneracy is exponentially small in the chain length. This suppression arises because the correction terms in Eq.~(\ref{eq:qubit-periodicity}) that break the $\pi$-periodicity are associated with quantum tunneling from one end to the other in the two-rung ladder. We also require $\bar J \bar C$ to be large, to suppress phase slips due to tunneling across the chain, thus ensuring that $\varphi_+$ can be regarded as a real variable rather than a periodic variable with period $2\pi$. 

An impedance $\sqrt{L/C}\approx 20$ has been achieved using long chains of devices \cite{Devoret2009,Masluk2012,Gershenson2012}. Another possibility for achieving large $\sqrt{L/C}$ is to use a long wire, thick enough to suppress phase slips, built from an amorphous superconductor with a large kinetic inductance. Whatever method is used, reaching, say, $\sqrt{L/C}$ of order 100 may be quite challenging, but in this paper we take it for granted that a robust 0-$\pi$ qubit can be realized. In fact, our scheme for implementing accurate quantum gates will also be based on superinducting circuits.

We will need to be able to measure the qubit, in either the standard $\{|0\rangle,|1\rangle\}$ basis (measurement of the Pauli operator $Z$) or in the dual basis $\{|0\rangle \pm |1\rangle\}$ (measurement of the Pauli operator $X$). In principle the $Z$ measurement could be performed by connecting the two leads of the qubit with a Josephson junction, while inserting $1/4$ of a flux quantum through the loop; then the current through the junction is proportional to $\sin\left(\theta_2-\theta_1 - \pi/2\right)$, with sign dependent on whether $\theta_2-\theta_1$ is $0$ or $\pi$. 

For measuring $X$, we envision ``breaking'' the connection between $\theta_1$ and $\theta_3$ and then measuring the charge conjugate to the phase difference $\theta_1 - \theta_3$. The energy of the circuit is $f(\theta_1+\theta_3 - 2\theta_2)$, so that if $\theta_1$ advances adiabatically by $2\pi$ with $\theta_3$ fixed, then $\theta_2$ advances by $\pi$; if $X=1$ the wave function is invariant and if $X=-1$ the wave function changes sign. Correspondingly, the dual charge is either an even or odd multiple of $1/2$. In practice, the $X$ and $Z$ measurements are bound to be noisy, but the limitations on measurement accuracy can be overcome by repeating the measurements or by using appropriate coding schemes, as we describe in Sec.~\ref{sec:universal}.

\section{Phase gate}\label{sec:gate}

Following \cite{Kitaev2006}, we will explain how to execute with high fidelity the single-qubit phase gate $\exp\left(i \frac{\pi}{4}Z\right)$ and the two-qubit phase gate $\exp\left(i  \frac{\pi}{4}Z\otimes Z\right)$. These gates are not sufficient by themselves for universal quantum computation, but we will discuss in Sec.~\ref{sec:universal} how they can be used as part of a universal fault-tolerant scheme.

\begin{figure}[!ht]
\begin{center}
\includegraphics[width=0.35\textwidth]{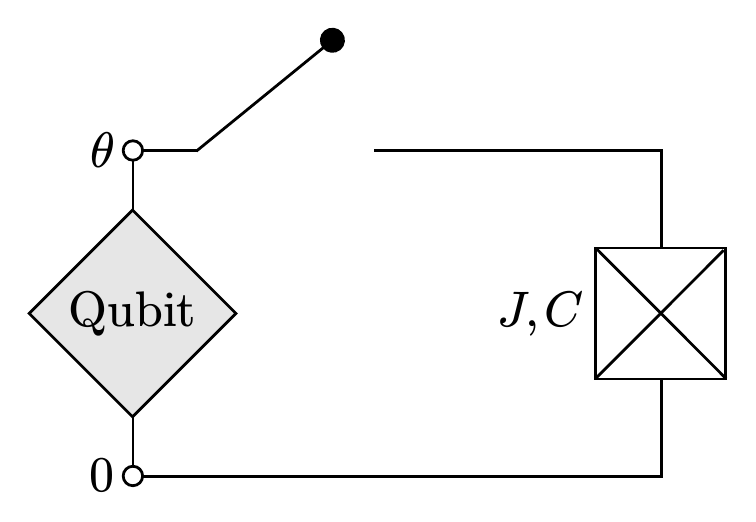}
\end{center}
\caption{\label{fig:unprotected} A phase gate can be applied to a qubit by coupling it to a Josephson junction, but the gate is not protected against pulse errors and other noise sources.}
\end{figure}

First, for contrast, consider an example of an unprotected single-qubit gate implementation. As shown in Fig.~\ref{fig:unprotected} we could close a switch that couples the qubit for time $t$ to a Josephson junction with Josephson coupling $J$, in effect turning on a term $J\cos\theta = JZ$ in the Hamiltonian, where $\theta\in \{0,\pi\}$ is the phase difference across the qubit. After time $t$ the unitary transformation $\exp(-itJZ)$ has been applied. Thus by choosing the time $t$ appropriately we can rotate the qubit about the $z$ axis by any desired angle. However, this gate is sensitive to errors in the pulse that closes and opens the switch, and to other fluctuations in the circuit parameters. 

\begin{figure}[!ht]
\begin{center}
\includegraphics[width=0.35\textwidth]{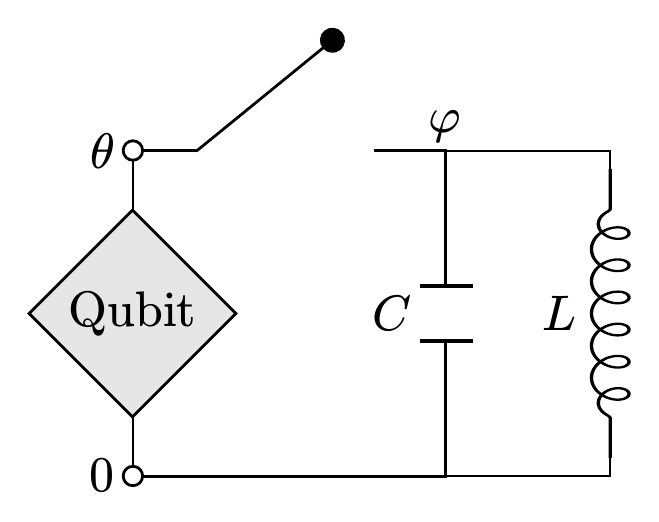}
\end{center}
\caption{\label{fig:protected} A protected phase gate is executed by coupling a qubit (or a pair of qubits connected in series) to a ``super-quantum'' $LC$ circuit with $\sqrt{L/C}\gg 1$. }
\end{figure}

\begin{figure}[!ht]
\begin{center}
\includegraphics[width=0.35\textwidth]{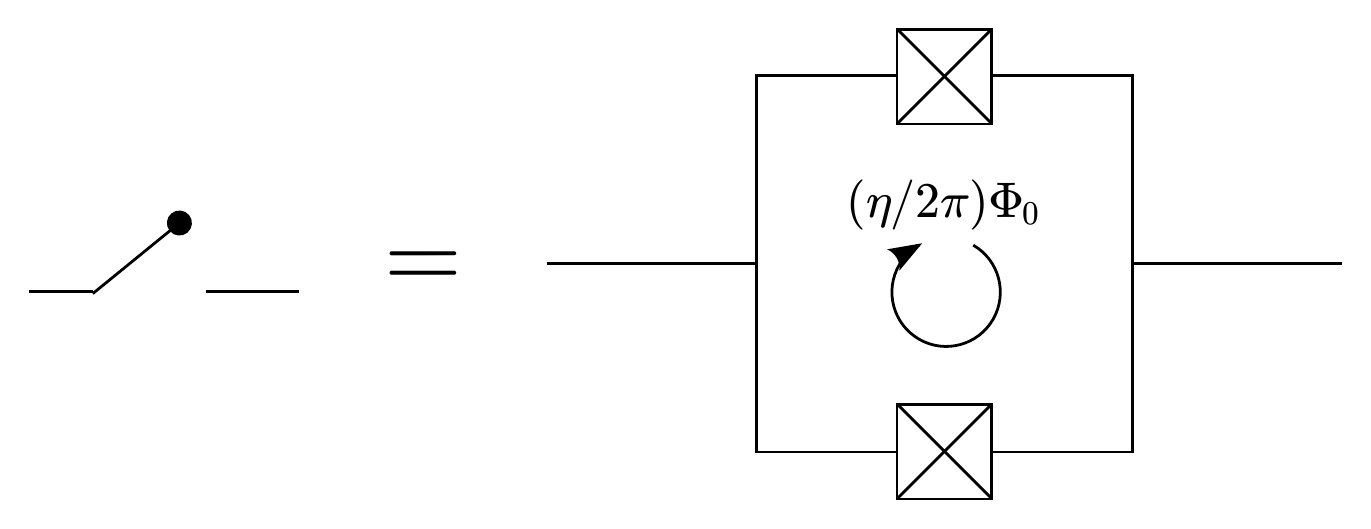}
\end{center}
\caption{\label{fig:switch} An effective Josephson junction can be tuned by adjusting the flux $(\eta/2\pi)\Phi_0$ inserted in a circuit containing two identical junctions.}
\end{figure} 

The protected single-qubit phase gate is executed as shown in Fig.~\ref{fig:protected} by coupling the qubit to a ``superinductive'' $LC$ circuit via a switch that pulses on and off. The switch is actually a tunable Josephson junction, which can be realized, as in Fig.~\ref{fig:switch}, by a loop containing two identical junctions, each with Josephson coupling $J$, linked by the magnetic flux $(\eta/2\pi)\Phi_0$, where $\Phi_0=h/2e$ is the flux quantum. The Josephson energy of this tunable junction is
\be
E(\theta,\eta) &=& -J\cos(\theta - \eta/2) -J\cos(\theta + \eta/2) \nonumber\\
&=& -2 J\cos(\eta/2)\cos\theta = -J_{\rm eff}(\eta)\cos\theta
\ee
where $\theta$ is the phase difference between the two leads on the loop. Thus the switch is ``on'' for $\eta = 0$ and ``off'' for $\eta = \pi$. The ``off'' setting can be fairly soft --- it is good enough for $J_{\rm eff}$ to be comparable to $1/L$ rather than strictly zero --- while in the ``on'' position we require $J_{\rm eff} C$ to be large. The inductance $L$ and capacitance $C$ of the circuit are unrelated to the inductance and capacitance for the 0-$\pi$ qubit discussed in Sec.~\ref{sec:qubit}, though we will again demand that $\sqrt{L/C}\gg 1$. From now on we will assume the 0-$\pi$ qubit is perfect, and will focus on realizing the robust phase gate under this assumption. 

Using the same normalization conventions as in Sec.~\ref{sec:qubit}, the Hamiltonian for the circuit can be expressed as
\be
H(t) = \frac{Q^2}{2C} + \frac{\varphi^2}{2L} - J(t)\cos(\varphi - \theta),
\ee
where now $J(t)$ is the time-dependent effective Josephson coupling of the tunable junction, $\theta$ is the phase difference across the qubit, and $\varphi$ is the phase difference across the inductor. We assume that phase slips through the circuit are strongly suppressed, so that $\varphi$ can be regarded as a real variable rather than a periodic phase variable --- when $\varphi$ winds by $2\pi$ the flux linking the $LC$ circuit increases by one flux quantum. Depending on whether the state of the qubit is $|0\rangle$ or $|1\rangle$, the phase $\theta$ is either 0 or $\pi$; hence, the Hamiltonian can be expressed as
\be\label{eq:hamiltonian}
H_{0,1}(t) = \frac{Q^2}{2C} + \frac{\varphi^2}{2L} \mp J(t)\cos\varphi,
\ee
with the $\mp$ sign conditioned on the qubit's state. 

Suppose for now that the initial state $|\psi^{\rm in}\rangle$ of the oscillator is its ground state, a Gaussian wave function with $\langle \varphi^2\rangle = \frac{1}{2}\sqrt{L/C}$ and $\langle Q^2 \rangle = \frac{1}{2}\sqrt{C/L}$. (Other harmonic oscillator energy eigenstates will be considered in Sec.~\ref{sec:temperature}.) Because $\sqrt{L/C}\gg 1$, the wave function is broad in $\varphi$ space and narrow in $Q$ space. Hence when the switch pulses on, the contribution to the expectation of the energy arising from the cosine potential is highly suppressed by the factor
\be\label{eq:cosine-expect}
\langle \cos\varphi\rangle = e^{-\langle \varphi^2\rangle /2} = \exp\left(-\frac{1}{4}\sqrt{\frac{L}{C}}\right).
\ee
Correspondingly, the energy is very insensitive to the state of the qubit, which determines the sign of the cosine potential. This suppression factor determines the characteristic scale of the error in the phase gate. 

\begin{figure}[!ht]
\begin{center}
\includegraphics[width=0.4\textwidth]{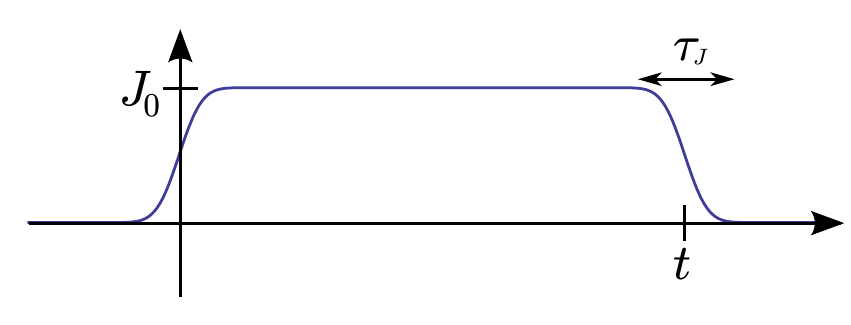}
\end{center}
\caption{\label{fig:pulse} (Color online) The profile of the tunable Josephson coupling $J(t)$ in the execution of the protected phase gate.} 
\end{figure} 

\begin{figure}[!ht]
\begin{center}
\includegraphics[width=0.45\textwidth]{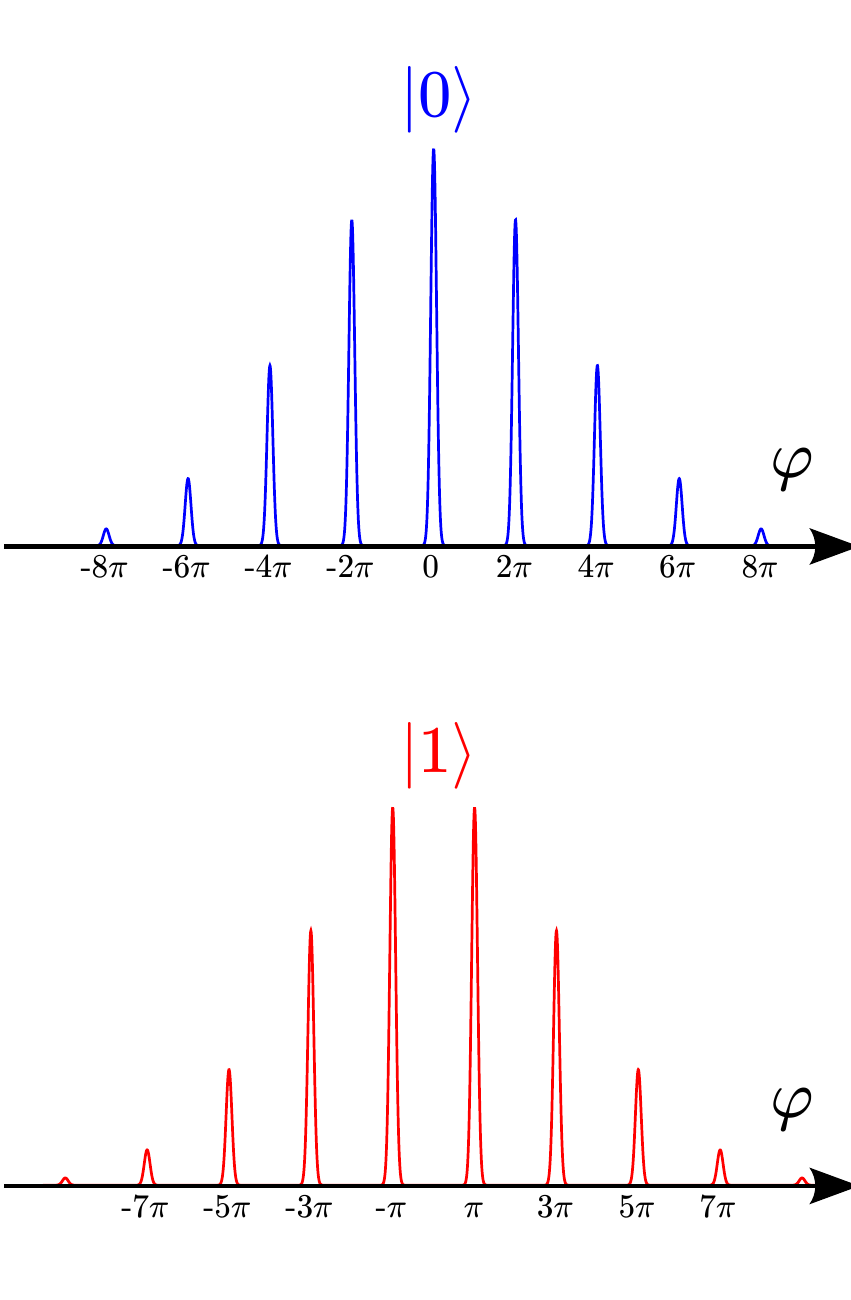}
\end{center}
\caption{\label{fig:grids} (Color online) Coupling the qubit to the oscillator prepares a grid state in $\varphi$ space, a superposition of narrowly peaked functions governed by a broad envelope function. The peaks occur where $\varphi$ is an even multiple of $\pi$ if the qubit's state is $|0\rangle$, and where $\varphi$ is an odd multiple of $\pi$ if the qubit's state is $|1\rangle$.} 
\end{figure}

Schematically, the tunable Josephson coupling $J(t)$ has the form shown in Fig.~\ref{fig:pulse} --- it starts at zero, ramps on smoothly, maintains the value $J_0$ for a time $t\approx L/\pi$, and then ramps off smoothly. The characteristic time $\tau_J$ for the coupling to ramp on and off is subject to some constraints which we will specify shortly. With $J_0$ at its steady state value, phase slips (tunneling events between successive minima of the cosine potential) are suppressed by the WKB factor
\be
\exp\left( -\int_0^{2\pi} d\varphi \sqrt{2J_0C(1-\cos\varphi)}\right)=\exp(-8\sqrt{J_0C}).\nonumber\\
\ee
We assume that $\sqrt{J_0C}$ is large enough so that phase slips can be safely neglected. In addition, we assume that $J(t)$ ramps up slowly enough to prepare adiabatically the ground state in each local minimum of the cosine potential, yet quickly enough to prevent the state from collapsing to just a few local minima with the smallest values of $\varphi^2/2L$. Thus, as $J(t)$ turns on, the initial state of the oscillator evolves to become a ``grid state'' as shown in Fig.~\ref{fig:grids}, a superposition of narrowly peaked functions governed by a broad envelope function. The width of the broad envelope is $\langle \varphi^2\rangle \approx \frac{1}{2}\sqrt{L/C}\gg 1$ as for the oscillator's initial state, while the width of each narrow peak is $\langle (\varphi-\varphi_0)^2\rangle\approx\frac{1}{2}\sqrt{1/J_0C}\ll 1$, the width of the ground state supported near the local minimum of the cosine potential. 

If the state of the 0-$\pi$ qubit is $|0\rangle$ and the coefficient of the cosine in Eq.~(\ref{eq:hamiltonian}) is negative, then the narrow peaks occur where $\varphi$ is an even multiple of $\pi$. We denote this grid state of the oscillator as $|0_C\rangle$; the subscript stands for ``code,'' since as we will explain latter this state can be regarded as a basis state for a quantum error-correcting code. If the state of the qubit is $|1\rangle$ and the coefficient of the cosine is positive, then the narrow peaks occur where $\varphi$ is an odd multiple of $\pi$; in that case we denote the grid state as $|1_C\rangle$. Thus, if the initial state of the 0-$\pi$ qubit is $a|0\rangle+b|1\rangle$, then when $J(t)$ turns on the joint state of the qubit and oscillator evolves according to
\be
\left(a|0\rangle + b|1\rangle\right) |\psi^{\rm in}\rangle \rightarrow a |0\rangle\otimes |0_C\rangle + b|1\rangle\otimes |1_C\rangle.
\ee

A diabatic transition that excites the oscillator in the cosine well is most likely to occur while $J(t)C$ is approximately one and the frequency of oscillations in the well is approximately $1/C$. Landau-Zener theory therefore indicates that the probability $P_{\rm diab}$ of such a transition scales like
\be
P_{\rm diab}(\tau_J) \sim \exp\left(-({\rm constant})\frac{\tau_J}{C}\right). 
\ee
where $\tau_J$ is the characteristic time for $J(t)$ to ramp on. (We will discuss this error in more detail in Sec.~\ref{sec:diabatic}.) Since diabatic effects also contribute to the error in the phase gate, we require $\tau_J \gg C$. Indeed, the diabatic error is comparable to the intrinsic error in Eq.~(\ref{eq:cosine-expect}) for
\be
\tau_J \sim \sqrt{LC};
\ee
that is, when the ramping time is of order the period of the $LC$ oscillator. During this ramping time, the envelope function of the Gaussian grid state is squeezed somewhat in $\varphi$ space (and correspondingly spreads somewhat in $Q$ space), but stays broad enough for the intrinsic error to remain heavily suppressed. In Sec.~\ref{sec:squeezing} we argue that the error arising from squeezing scales like
\be\label{eq:squeeze-scaling}
P_{\rm sq}(\tau_J) \sim \exp\left(-({\rm constant})\frac{L}{\tau_J}\right);
\ee
hence it too is comparable to the intrinsic error for $\tau_J\sim \sqrt{LC}$.

After the Gaussian grid state has been prepared, the Josephson coupling $J(t)$ maintains its steady-state value $J_0$ for a time $t \equiv L\tilde t/\pi$, where $\tilde t$ is a rescaled time variable. While the coupling is on, each narrowly peaked function is stabilized by the strongly confining cosine potential, but the state is subjected to the Gaussian operation $e^{-it \varphi^2/2L} = e^{-i\tilde t\varphi^2/2\pi}$ due to the harmonic potential $\varphi^2/2L$, which alters the relative phases of the peaks. As $\tilde t$ increases the oscillator states
$|0_C\rangle$ and $|1_C\rangle$ evolve, but when $\tilde t$ reaches 1, each returns to its initial value, apart from a state-dependent geometric phase. For the grid state $|0_C\rangle$, the peaks in $\varphi$ space occur at $\varphi = 2\pi n$ where $n$ is an integer, and the Gaussian operation
\be
|\varphi=2\pi n\rangle \rightarrow e^{-2\pi i \tilde t n^2}|\varphi=2\pi n\rangle
\ee
acts trivially. But for the grid state $|1_C\rangle$, the peaks occur at $\varphi = 2\pi(n+\tfrac{1}{2})$, and the operation
\be
|\varphi=2\pi n\rangle \rightarrow e^{-2\pi i \tilde t \left(n+\tfrac{1}{2}\right)^2}|\varphi=2\pi(n+\tfrac{1}{2})\rangle
\ee 
therefore modifies the phase of the state by the factor $-i$. Hence the joint state of the qubit and oscillator becomes
\be
 a |0\rangle\otimes |0_C\rangle +b|1\rangle\otimes |1_C\rangle\rightarrow a|0\rangle\otimes |0_C\rangle -ib|1\rangle\otimes |1_C\rangle.\nonumber\\
\ee

To complete the execution of the phase gate, the tunable coupling $J(t)$ ramps down from $J_0$ to zero, again with a characteristic time scale $\tau_J$ subject to the constraints specified above. As the coupling turns off, the state $|0_C\rangle$ of the oscillator evolves to $|\psi_0^{\rm fin}\rangle$ and the state $|1_C\rangle$ evolves to $|\psi_1^{\rm fin}\rangle$; the final joint state of the qubit and oscillator is 
\be\label{eq:gate-psi-final}
&& a |0\rangle\otimes |0_C\rangle -ib|1\rangle\otimes |1_C\rangle\nonumber\\ 
&& \rightarrow a|0\rangle\otimes |\psi_0^{\rm fin}\rangle -ib|1\rangle\otimes |\psi_1^{\rm fin}\rangle.
\ee
Thus, a perfect phase gate $\exp\left(i\tfrac{\pi}{4}Z\right)$ has been applied to the qubit if $|\psi_0^{\rm fin}\rangle = |\psi_1^{\rm fin}\rangle$. If on the other hand $|\langle \psi_1^{\rm fin}|\psi_0^{\rm fin}\rangle| < 1$, then the qubit and oscillator are entangled in the final state, compromising the gate fidelity. Even if $|\langle \psi_1^{\rm fin}|\psi_0^{\rm fin}\rangle| =1 $ so there is no entanglement, the gate may be imperfect because the phase of $\langle \psi_1^{\rm fin}|\psi_0^{\rm fin}\rangle$ deviates from zero.

We will argue that under appropriate conditions $\langle \psi_1^{\rm fin}|\psi_0^{\rm fin}\rangle\approx 1$ to extremely high accuracy so that the phase gate is nearly perfect. Note that we need not require the final state of the oscillator to match the initial state $|\psi^{\rm in}\rangle$; noise terms in the Hamiltonian may excite the oscillator, but the phase gate is still highly reliable as long as the oscillator's final state depends only very weakly on the state of the 0-$\pi$ qubit, {\em i.e.}, on whether the sign of $J(t)$ is positive or negative. Indeed, the oscillator serves as a reservoir that absorbs the entropy introduced by noise. If not too badly damaged, the oscillator can be reused a few times for the execution of additional protected gates. Eventually, though, it will become too highly excited, and will need to be cooled before being employed again.

A gate error may arise if the coupling between qubit and oscillator remains on for too long or too short a time, {\em i.e.}, if $\tilde t=1+ \varepsilon$ rather than $\tilde t=1$. But we will see that such timing errors do not much compromise the performance of the gate when $\varepsilon$ is small; specifically, the gate error is $\exp\left(-\tfrac{1}{4}\sqrt{L/C}\right)\times O(1)$ provided $|\varepsilon| < 2\pi \left(L/C\right)^{-3/4}$. Slightly overrotating or underrotating contributes to the damage suffered by the oscillator, but without much enhancing the sensitivity of the oscillator's final state to the qubit's state, and hence without much reducing the fidelity of the gate. We study the consequences of overrotation/underrotation in Sec.~\ref{sec:overrotation}, and confirm our findings using numerical simulations in Sec.~\ref{sec:numerics}. We also argue, in Sec.~\ref{sec:temperature} and Sec.~\ref{sec:anharmonic}, that the phase gate is robust against a sufficiently small nonzero temperature and against small perturbations in the Hamiltonian.

The two-qubit phase gate $\exp(i\frac{\pi}{4}Z\otimes Z)$ is executed using a similar procedure, but where now two qubits connected in series are coupled to the $LC$ oscillator. The total phase difference across the pair of qubits is either $0$ for the states $|0\rangle\otimes|0\rangle$ and $|1\rangle\otimes |1\rangle $, in which case the oscillator evolves to the final $|\psi_0^{\rm fin}\rangle$, or $\pi$ for the states $|0\rangle\otimes|1\rangle$ and $|1\rangle\otimes |0\rangle $, in which case the oscillator evolves to the final state $|\psi_1^{\rm fin}\rangle$. Again, the gate is executed perfectly if $|\psi_0^{\rm fin}\rangle = |\psi_1^{\rm fin}\rangle$. 

Let us summarize the sufficient conditions for the phase gate to be well protected. Just as for the realization of the 0-$\pi$ qubit itself, the execution of the protected phase gate relies on the construction of a ``superinducting'' circuit with $\sqrt{L/C} \gg 1$. This is a daunting engineering challenge as we have already noted at the end of Sec.~\ref{sec:qubit}. To ensure high gate accuracy, we also assume that the steady state value $J_0$ of the Josephson coupling between the 0-$\pi$ qubit and the oscillator satisfies $\sqrt{J_0 C} \gg 1$ and that the characteristic time scale $\tau_J$ for the coupling to ramp on and off is $O(\sqrt{LC})$; 
thus $\tau_J$ is also small compared to the time $L/\pi$ needed to execute the gate. Under these conditions, the error in the phase gate scales as $\exp\left(-O\left(\sqrt{L/C}~\right)\right)$, and is stable with respect to small fluctuations in the implementation of the gate.

\section{Sketch of the error estimate}
\label{sec:sketch}

A noisy quantum gate realizes a quantum operation ${\cal N}_{\rm actual}$, and a useful way to quantify the error in the gate is to specify the deviation $\| {\cal N}_{\rm actual} - {\cal N}_{\rm ideal}\|_{\diamond}$ from the ideal gate ${\cal N}_{\rm ideal}$ in the ``diamond norm''  \cite{KitaevDiamond}. As explained in Appendix \ref{app:error}, for the protected phase gate this diamond norm distance (assuming there are no bit flips) is 
\be
\| {\cal N}_{\rm actual} - {\cal N}_{\rm ideal}\|_{\diamond} = |1 - \langle \psi_1^{\rm fin}|\psi_0^{\rm fin}\rangle|
\ee
where $|\psi_{0,1}^{\rm fin}\rangle$ denotes the final state of the oscillator when $|0\rangle,|1\rangle$ is the state of the 0-$\pi$ qubit, as in Eq.~(\ref{eq:gate-psi-final}). Thus we assess the gate accuracy by estimating the deviation of $\langle \psi_0^{\rm fin}|\psi_1^{\rm fin}\rangle$ from 1.

To perform this estimate we track how the oscillator states $|\psi_0(t)\rangle$ and $|\psi_1(t)\rangle$ are related through three stages of evolution:
\be
&&|\psi^{\rm in}\rangle \stackrel{{J(t)} \atop {\rm turns ~on}}{\longrightarrow} |\psi_{0}^{\rm begin}\rangle \stackrel{J(t) = J_0}{\longrightarrow}|\psi_{0}^{\rm end}\rangle\stackrel{{J(t)}\atop  {\rm turns~off }}{\longrightarrow} |\psi_{0}^{\rm fin}\rangle,\nonumber\\
&&|\psi^{\rm in}\rangle \stackrel{{J(t)} \atop {\rm turns ~on}}{\longrightarrow} |\psi_{1}^{\rm begin}\rangle \stackrel{J(t) = J_0}{\longrightarrow}-i|\psi_{1}^{\rm end}\rangle\stackrel{{J(t)}\atop  {\rm turns~off }}{\longrightarrow} -i|\psi_{1}^{\rm fin}\rangle\nonumber\\
\ee
In the first stage $J(t)$ ramps on and the grid states are prepared -- the initial state $|\psi^{\rm in}\rangle$ evolves to $|\psi_0^{\rm begin}\rangle$ if the 0-$\pi$ qubit's state is $|0\rangle$ and to $|\psi_1^{\rm begin}\rangle$ if the qubit's state is $|1\rangle$. In the second stage $J(t)=J_0$ and the grid state $|\psi^{\rm begin}_{0}\rangle$ evolves to $|\psi^{\rm end}_{0}\rangle$ while the grid state $|\psi^{\rm begin}_{1}\rangle$ evolves to $-i|\psi^{\rm end}_{1}\rangle$, where ideally $|\psi_{0,1}^{\rm end}\rangle = |\psi_{0,1}^{\rm begin}\rangle$. In the third stage $J(t)$ ramps off and the grid states $|\psi_{0,1}^{\rm end}\rangle$ evolve to the final oscillator states $|\psi_{0,1}^{\rm fin}\rangle$, where ideally $|\psi_1^{\rm fin}\rangle = |\psi_0^{\rm fin}\rangle$. 

Consider the first (or third) stage of the evolution, where the coupling $J(t)$ ramps on (or off) in a time of order $\tau_J$. If $\tau_J$ is not too large compared to the period $2\pi\sqrt{LC}$ of the oscillator, then the harmonic potential term $\varphi^2/2L$ may be treated perturbatively during this evolution stage. Hence, in first approximation the Hamiltonian is one of 
\begin{eqnarray}\label{eq:truncated-H}
H_0 = \frac{Q^2}{2C} - J(t)\cos\varphi,\quad H_1 = \frac{Q^2}{2C} + J(t)\cos\varphi,
\end{eqnarray}
depending on whether the state of the 0-$\pi$ qubit is $|0\rangle$ or $|1\rangle$. This Hamiltonian commutes with the operator $e^{-2\pi i Q}$, which translates $\varphi$ by $2\pi$; therefore $e^{-2\pi i Q}$ and the Hamiltonian can be simultaneously diagonalized. We may express the eigenvalue of this translation operator as $e^{-2\pi i q}$, where $q = Q-[Q] \in [-\frac{1}{2},\frac{1}{2}]$ is the conserved Bloch momentum, and $[Q]$ denotes the nearest integer to $Q$; thus $[Q]$ labels the distinct bands in the Hamiltonian's spectrum. 

A diabatic transition between bands may be excited while $J(t)$ varies, changing the value of $[Q]$ by an integer, most likely $\pm 1$. If such transitions occur with nonnegligible probability, the final state of the oscillator will contain, in addition to a primary peak supported near $Q=0$, also secondary peaks supported near $Q=\pm 1$; the phases of the secondary peaks depend on whether the Hamiltonian is $H_0$ or $H_1$, and therefore diabatic transitions contribute to the gate error. The probability of a diabatic transition cannot be computed precisely, but as we will explain in Sec.~\ref{sec:diabatic} it can be analyzed semi-quantitatively, and is very small if $\tau_J$ is sufficiently large. 

For the purpose of discussing this diabatic error and other contributions to the deviation of $\langle \psi_1^{\rm fin}|\psi_0^{\rm fin}\rangle$ from 1, we will find it useful to consider the operator 
\be\label{barX-define}
\bar X &\equiv& (-1)^{[Q]} = \Pi^Q_{\rm even} - \Pi^Q_{\rm odd}\nonumber\\
&\equiv& 2\Pi^Q_{\rm even} - I\equiv I - 2\Pi^Q_{\rm odd}.
\ee
Here $\Pi^Q_{\rm even}$ projects onto values of $Q$ such that the nearest integer value $[Q]$ is even and $\Pi^Q_{\rm odd}$ projects onto values of $Q$ such that $[Q]$ is odd. We denote this operator as $\bar X$ because it can be regarded as the error-corrected Pauli operator $\sigma^X$ acting on a qubit encoded in the Hilbert space of the oscillator, as we explain in Sec.~\ref{sec:code}. (Note that $\bar X^2 = I$.) Another (related) important property is that, since $e^{\mp i \varphi}$ translates $Q$ by $\pm 1$, $\bar X$ anticommutes with $\cos\varphi$:
\be
\bar X \cos\varphi \bar X = - \cos\varphi.
\ee
Our argument showing that $|\psi_1^{\rm fin}\rangle \approx |\psi_0^{\rm fin}\rangle$ has two main elements. On the one hand we use approximate symmetries and properties of grid states to see that $|\psi_1(t)\rangle \approx \bar X |\psi_0(t)\rangle$ at each stage of the oscillator's evolution, so that in particular $|\psi_1^{\rm fin}\rangle \approx \bar X |\psi_0^{\rm fin}\rangle$. On the other hand we argue that if the time scale $\tau_J$ for $J(t)$ to turn on and off is suitably chosen, then the oscillator's final state is mostly supported near $Q=0$, so that in particular $\bar X|\psi_0^{\rm fin}\rangle \approx |\psi_0^{\rm fin}\rangle$. 

We note that the approximate Hamiltonians $H_0$ and $H_1$ in Eq.~(\ref{eq:truncated-H}) are related by
\begin{equation}
H_1 = \bar X H_0 \bar X.
\end{equation}
By integrating the Schr\"odinger equation using the Hamiltonian $H_0$ or $H_1$ during the first stage of evolution while $J(t)$ ramps on, we obtain the unitary time evolution operators $U_{0}$, $U_{1}$, which are related by
\begin{equation}
U_{1} = \bar X U_{0} \bar X.
\end{equation}
Thus the initial oscillator state $|\psi^{\rm in}\rangle$ evolves to one of the states
\begin{eqnarray}
&&|\psi_0^{\rm begin}\rangle = U_0 |\psi^{\rm in}\rangle \nonumber\\
&&| \psi_1^{\rm begin} \rangle = U_1 |\psi^{\rm in}\rangle = \bar X U_0\bar X|\psi^{\rm in}\rangle,
\end{eqnarray}
and therefore 
\begin{eqnarray}
\label{eq:almost-X-conjugate}
\langle \psi_1^{\rm begin}|\bar X |\psi_0^{\rm begin}\rangle &=& \langle \psi^{\rm in} |\bar X|\psi^{\rm in}\rangle 
= \langle \psi^{\rm in}|I-2\Pi^Q_{\rm odd}|\psi^{\rm in}\rangle \nonumber\\
&=& 1 - 2\langle \psi^{\rm in} |\Pi^Q_{\rm odd}|\psi^{\rm in}\rangle.
\end{eqnarray}
We conclude that if the initial state is almost fully supported on even values of $[Q]$ (for example, the oscillator ground state, a Gaussian in $Q$-space with width much less than $1/2$), then $| \psi_1^{\rm begin}\rangle$ is very close to $\bar X|\psi_0^{\rm begin}\rangle$. 

So far we have ignored the effects of the quadratic term $\varphi^2/2L$ in the potential. This term can cause the wave function to broaden in $Q$-space and be squeezed in $\varphi$ space, but we argue in Sec.~\ref{sec:squeezing} that this squeezing is a relatively small effect, so that the conclusion $| \psi_1^{\rm begin}\rangle\approx \bar X|\psi_0^{\rm begin}\rangle$ still holds accurately. Specifically, the contribution to the gate error due to squeezing scales as in Eq.~(\ref{eq:squeeze-scaling}), and hence becomes comparable to the other sources of error when we choose $\tau_J\sim\sqrt{LC}$.


During the second stage of the evolution, while $J(t) = J_0$ is held constant, distinct peaks in the grid state acquire relative phases, and the condition $|\psi_1(t)\rangle = \bar X |\psi_0(t)\rangle$ becomes badly violated. However, after a time $t\approx L/\pi$, the initial states $|\psi_0^{\rm begin}\rangle$ and $|\psi_1^{\rm begin}\rangle$ are restored, aside from the state dependent phase $-i$, and hence $|\psi_1^{\rm end}\rangle = \bar X |\psi_0^{\rm end}\rangle$ apart from a small error. Equivalently, the beginning states
\be
|\psi_{\pm}^{\rm begin}\rangle  = \frac{1}{\sqrt{2}}\left (|\psi_0^{\rm begin}\rangle \pm |\psi_1^{\rm begin}\rangle \right)
\ee
are very nearly $\bar X$ eigenstates with eigenvalues $\pm 1$, and this property is preserved by the ending states
\be
|\psi_{\pm}^{\rm end}\rangle  = \frac{1}{\sqrt{2}}\left (|\psi_0^{\rm end}\rangle \pm |\psi_1^{\rm end}\rangle \right).
\ee
The $\bar X$ eigenvalues of these states are highly stable with respect to timing errors in the gate, in which the coupling is left on for too short or too long a time, because these states are approximate codewords of a quantum code, well protected against logical phase errors. We study the errors resulting from imperfect timing in detail in Sec.~\ref{sec:overrotation}, because they can be calculated explicitly and are the dominant errors in some parameter regimes. 

In the third stage of the evolution, as in the first stage, it is a good first approximation to ignore the harmonic $\varphi^2/2L$ term in the potential as the coupling $J(t)$ ramps off. Using this approximation, the time evolution operators $V_{0,1}$ obtained by integrating the Schr\"odinger equation during the third stage when the state of the 0-$\pi$ qubit is $|0\rangle,|1\rangle$, are related by
\be
V_1 = \bar X V_0\bar X; 
\ee
hence the final oscillator states are 
\be
|\psi_0^{\rm fin}\rangle = V_0 |\psi_0^{\rm end}\rangle,\quad 
|\psi_1^{\rm fin}\rangle = \bar X V_0 \bar X|\psi_1^{\rm end}\rangle
\ee
and we conclude that 
\be
\langle \psi_1^{\rm fin}|\bar X |\psi_0^{\rm fin}\rangle = \langle \psi_1^{\rm end} |\bar X|\psi_0^{\rm end}\rangle.
\ee
Again, this conclusion is not modified much when the $\varphi^2/2L$ term is properly taken into account, so we may infer that the condition $|\psi_0(t)\rangle \approx \bar X |\psi_1(t)\rangle$ is well preserved during the final stage of evolution.

We have now seen that $|\psi_1^{\rm fin}\rangle \approx \bar X|\psi_0^{\rm fin}\rangle$, and it remains to show that $\bar X|\psi_0^{\rm fin}\rangle \approx |\psi_0^{\rm fin}\rangle$. This condition will be well satisfied provided that the final state $|\psi_0^{\rm fin}\rangle$ of the oscillator, like the initial state $|\psi^{\rm in}\rangle$, is almost fully supported in the interval $Q\in [-\frac{1}{2},\frac{1}{2}]$. Logical errors may occur because of diabatic transitions between bands, which may change $Q$ by an odd integer, or because of spreading in $Q$ space, which may enhance the tails of the wave function outside $[-\frac{1}{2},\frac{1}{2}]$. However, if diabatic transitions are rare and spreading is modest, as we expect if $\tau_J$ lies in the appropriate range, then the gate will be highly accurate. 

That our criterion for achieving $|\psi_1^{\rm fin}\rangle \approx |\psi_0^{\rm fin}\rangle$ involves the operator $\bar X$, which has a sharp discontinuity at $Q=\frac{1}{2} +{\rm integer}$, is really an artifact of an insufficiently careful treatment of diabatic transitions. The transitions occur with enhanced probability for $Q$ close to $\frac{1}{2} +{\rm integer}$, replacing the sharp edge in $Q$ space by a rounded step with width of order $C/\tau_J$, as we will explain in Sec.~\ref{sec:diabatic}. 

\section{Encoding a qubit in an oscillator}
\label{sec:code}

A continuous-variable quantum error-correcting code \cite{GottesmanKitaevPreskill2001} underlies the robustness of the protected phase gate. The theory of quantum codes is not really essential for understanding our estimate of the gate accuracy, but this theory provides motivation for our construction of the protected gate, as well a convenient language for explaining how it works. Therefore, we will now review some of the relevant features of a code first described in \cite{GottesmanKitaevPreskill2001}.

In the version of the code we will use, a two-dimensional encoded qubit is embedded in the infinite-dimensional Hilbert space of a harmonic oscillator with position $\varphi$ and conjugate momentum $Q$ satisfying $[\varphi,Q] = i$. The code space can be specified as the simultaneous eigenspace with eigenvalue 1 of the two commuting operators
\be
M_Z = e^{2i\varphi},\quad M_X = e^{-2\pi i Q};
\ee
we say that $M_Z$ and $M_X$ are the generators of the code's ``stabilizer group.'' Using the identify $e^Ae^B = e^{[A,B]}e^B e^A$ (where $A$ and $B$ commute with their commutator), we can easily verify that $M_X$ and $M_Z$ commute. The logical Pauli operators acting on the encoded qubit are
\be
\bar Z = e^{i\varphi}, \quad \bar X = e^{-i\pi Q}.
\ee
One sees that $\bar X$ and $\bar Z$ commute with  the stabilizer generators $M_X$ and $M_Z$, and hence preserve the code space; furthermore they anticommute with one another, as the logical Pauli operators should. 

\begin{figure}[!ht]
\begin{center}
\includegraphics[width=0.45\textwidth]{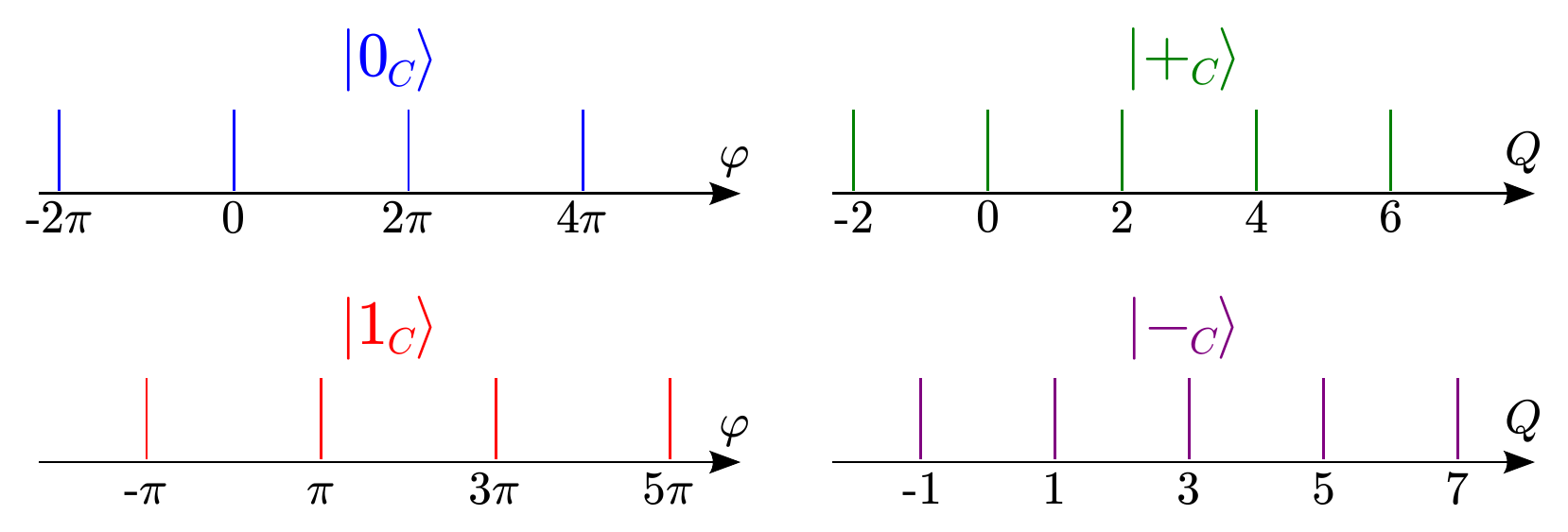}
\end{center}
\caption{\label{fig:ideal-codewords} (Color online) Ideal codewords of the continuous variable code. The $\bar Z=\pm 1$ eigenstates $|0_C\rangle,|1_C\rangle$, expressed in $\varphi$ space, are uniform superpositions of position eigenstates with $\varphi$ an even or odd multiple of $\pi$, respectively.  The $\bar X=\pm 1$ eigenstates $|+_C\rangle,|-_C\rangle$, expressed in $Q$ space, are uniform superpositions of momentum eigenstates with $Q$ an even or odd integer, respectively. }
\end{figure}

The (unnormalizable) state $|0_C\rangle^{\rm ideal}$ is the unique $\bar Z$ eigenstate with eigenvalue 1 in the code space. The condition $\bar Z=1$ requires the variable $\varphi$ to be an integer multiple of $2\pi$, and the condition $M_X=1$ requires the codeword to be invariant under translation of $\varphi$ by $2\pi$. Hence $|0_C\rangle^{\rm ideal}$ is represented in $\varphi$ space as the uniform superposition of delta functions
\be
|0_C\rangle^{\rm ideal} = \sum_{n=-\infty}^\infty |\varphi=2\pi n\rangle;
\ee
the $\bar Z=-1$ eigenstate $|1_C\rangle^{\rm ideal} = \bar X |0_C\rangle^{\rm ideal}$, obtained from $|0_C\rangle^{\rm ideal}$ by displacing $\varphi$ by $\pi$, is
\be
|1_C\rangle^{\rm ideal} = \sum_{n=-\infty}^\infty |\varphi=2\pi (n+\tfrac{1}{2})\rangle.
\ee
Similarly, the $X=\pm 1$ eigenstates $|\pm_C\rangle^{\rm ideal}$, invariant under translation of $Q$ by 2, are represented in $Q$ space as
\be
|+_C\rangle^{\rm ideal} &=& \sum_{n=-\infty}^\infty |Q=2n\rangle,\nonumber\\
|-_C\rangle^{\rm ideal} &=& \sum_{n=-\infty}^\infty |Q=2(n+\tfrac{1}{2})\rangle.\nonumber\\
\ee
See Fig.~\ref{fig:ideal-codewords}.

Weak noise may displace $\varphi$ slightly, but the codewords $|0_C\rangle^{\rm ideal}$ and $|1_C\rangle^{\rm ideal}$ remain perfectly distinguishable, and the error is correctable, as long as the value of $\varphi$ shifts by less than $\pi/2$ in either direction. Similarly, a shift in $Q$ by less than 1/2 is correctable. In principle we could diagnose the error by measuring $M_Z$ to determine the value of $\varphi$ modulo $\pi$, and $M_X$ to determine the value of $Q$ modulo 1, then perform active error correction by applying the minimal shifts in $\varphi$ and $Q$ that return the damaged code state to the code space. (In our protected phase gate, however, the error correction will be passive rather than active.)

\begin{figure}[!ht]
\begin{center}
\includegraphics[width=0.5\textwidth]{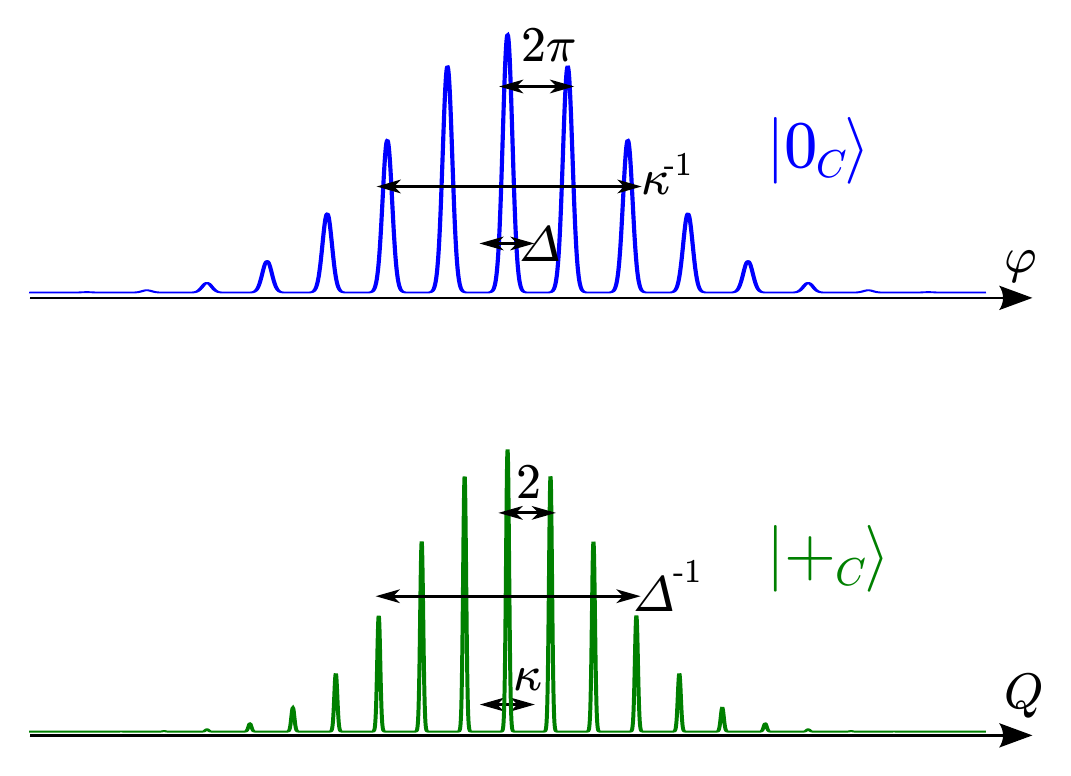}
\end{center}
\caption{\label{fig:approx-codewords} (Color online) Approximate codewords of the continuous variable code. The codeword $|0_C\rangle$, expressed in $\varphi$ space, is a superposition of Gaussian peaks, each of width $\Delta$, governed by a broad Gaussian envelope with width $\kappa^{-1}$. The codeword $|+_C\rangle$, expressed in $Q$ space, is a superposition of Gaussian peaks, each of width $\kappa$, governed by a broad Gaussian envelope with width $\Delta^{-1}$. } 
\end{figure}

The unnormalizable ideal codewords, with infinite $\langle \varphi\rangle$ and $\langle Q\rangle$, are unphysical. But if we coherently apply Gaussian distributed small shifts in $\varphi$ and $Q$ to the ideal codewords, we obtain the normalizable approximate codewords shown in Fig.~\ref{fig:approx-codewords}. The wave function in $\varphi$ space is a superposition of narrow Gaussians, each of width $\Delta\ll \pi/2$ ({\em i.e.},  $\langle(\delta\varphi)^2\rangle = \frac{1}{2}\Delta^2$, where $\delta\varphi$ denotes the deviation from the center of the narrow Gaussian), governed by a broad Gaussian envelope with width $\kappa^{-1} \gg 2$ ({\em i.e.}, $\langle \varphi^2\rangle = \frac{1}{2}\kappa^{-2}$). The Fourier dual wave function in $Q$ space is a superposition of narrow Gaussians each of width $\kappa$ ({\em i.e.}, $\langle (\delta Q)^2 \rangle = \frac{1}{2}\kappa^2$, where $\delta Q$ denotes the deviation from the center of the narrow Gaussian), governed by a broad Gaussian envelope with width $\Delta^{-1}$ ({\em i.e.}, $\langle Q^2 \rangle = \frac{1}{2}\Delta^{-2}$). If $\Delta$ and $\kappa$ are sufficiently small, these approximate codewords retain good error correction properties. However, there is now an intrinsic error arising from the tails of the narrow Gaussians, with the probability of a logical $\bar Z$ error (a shift in $\varphi$ by more than $\pi/2$) suppressed by $e^{-\pi^2/4\Delta^2}$, and the probability of a logical $\bar X$ error (a shift in $Q$ by more than $1/2$) suppressed by $e^{-1/4\kappa^2}$.

Note that in Sec.~\ref{sec:sketch} we used the notation $\bar X = (-1)^{[Q]}$ for the logical $X$ operator, where $[Q]$ denotes the nearest integer to the real variable $Q$. The operator $(-1)^{[Q]}$ acts in the same way as $e^{-i\pi Q}$ on ideal codewords for which $Q$ is an integer. By expressing the logical operator as $\bar X = (-1)^{[Q]}$ we are implicitly correcting phase errors that displace  $Q$. That is, a $Q$ eigenstate is decoded by shifting $Q$ to the nearest integer value, and the eigenvalue of $\bar X$ is determined by this ideal shifted value of $Q$, rather than the actual value of $Q$ prior to the shift. 

The first step in the execution of the protected phase gate described in Sec.~\ref{sec:gate} is the preparation of just such approximate codewords. If the state of the 0-$\pi$ qubit is $|0\rangle$, then the approximate $\bar Z = 1$ eigenstate $|0_C\rangle$ is prepared, and if the state of the 0-$\pi$ qubit is $|1\rangle$, then the approximate $\bar Z=-1$ eigenstate $|1_C\rangle$ is prepared. The narrowly peaked functions have width $\Delta^2= (J_0C)^{-1/2}$ in $\varphi$ space (though, because the potential is a cosine rather than harmonic, the tail of the peaked function decays more slowly than the tail of a Gaussian), and width $\kappa^2 = (L/C)^{-1/2}$ in $Q$ space. Hence the intrinsic logical $\bar X$ error of the approximate codewords, which is central to our estimate of the error in the phase gate, is suppressed by the factor $\exp\left(-\frac{1}{4}\sqrt{L/C}\right)$.

After the approximate codeword is prepared, the Gaussian unitary operator $e^{-i t\varphi^2/2L} = e^{-i\tilde t\varphi^2/2\pi}$ is applied (where $\tilde t= \pi t /L$ is a rescaled time variable). This unitary operator rotates the code space, transforming the stabilizer generator $M_X=e^{-2\pi iQ}$ according to
\be
M_X &\rightarrow& M_X'= e^{-i\tilde t\varphi^2/2\pi}e^{-2\pi i Q}e^{i\tilde t\varphi^2/2\pi}\nonumber\\
&=&e^{-2\pi i (Q+\varphi \tilde t/\pi)} = M_X e^{-2i \varphi \tilde t}e^{-2\pi i \tilde t}.
\ee
Recalling that $M_Z^{-1}=e^{-2i\varphi}$ is also a stabilizer generator, we see that the state returns to the code space at (rescaled) time $\tilde t=1$, but during its excursion the codeword acquires a Berry phase, and thus a nontrivial logical operation is applied. Specifically, the logical operator $\bar X$ is transformed according to
\be
\bar X &\rightarrow& \bar X'= e^{-i\varphi^2/2\pi}e^{-i \pi Q}e^{i\varphi^2/2\pi}\nonumber\\
&=&e^{- i\pi  (Q+\varphi /\pi)} = \bar X e^{-i \varphi }e^{ -i\pi /2}\nonumber\\
&=& -i \bar X\bar Z = \exp\left (i\tfrac{\pi}{4} \bar Z \right)\bar X \exp\left (-i\tfrac{\pi}{4} \bar Z \right)
\ee
while $\bar Z$ remains invariant; hence the logical operation acting on the code space is $\exp\left (i\tfrac{\pi}{4} \bar Z \right)$. An error in the logical gate arises if the coupling remains on for too long or too short a time ({\em i.e.} if $\tilde t$ is not precisely 1). However, this error is correctable with high probability if the timing error is small. We will study the consequences of overrotation/underrotation in Sec.~\ref{sec:overrotation}. 

\section{Imperfect grid states}
\label{sec:overrotation}

Now we will analyze the intrinsic phase errors in approximate codewords of the continuous variable code, and in particular how the phase error is affected by errors in the timing of the pulse that executes the phase gate. 

In Sec.~\ref{sec:code} we considered approximate codewords that can be described as  ``Gaussian grid states'' --- the codeword is a superposition of narrow Gaussian peaks governed by a broad Gaussian envelope. The Fourier transform of such a wave function is also a Gaussian grid state, so that both logical bit flip errors and logical phase errors are suppressed.

But it actually suffices for the approximate codeword to be a superposition of narrow functions with a broad envelope; neither the peak nor the envelope needs to be Gaussian. Even a non-Gaussian grid is mapped to a conjugate non-Gaussian grid by the Fourier transform, so there is good protection against both $\bar X$ and $\bar Z$ errors. A Gaussian grid state could result from coherently applying Gaussian-distributed $\varphi$ and $Q$ shifts to an ideal codeword, assuming large shifts are suppressed. But we can also get a reasonable approximate codeword by applying more general small errors to the ideal codeword, with a distribution that is not necessarily Gaussian. What is important is that large shifts in both $\varphi$ and $Q$ are improbable, not the detailed form of the distribution.

This observation will be useful when we consider in Sec.~\ref{sec:temperature} the execution of the protected phase gate in the case where the initial state of the harmonic oscillator is an excited state rather than the ground state. In that case, the envelope of the approximate codeword in $\varphi$ space is not strictly Gaussian, but rather Gaussian modulated by a Hermite polynomial, and its Fourier transform is also Gaussian modulated by a Hermite polynomial. Thus, in $Q$ space, the narrow functions peaked at integer values of $Q$ are also oscillator excited states. These functions have highly suppressed tails, ensuring that encoded phase errors are rare. In any event, considering more general kinds of grid states helps to clarify conceptually why the phase gate is robust.

\subsection{Bit-flip and phase errors}
\label{subsec:intrinsic}

Let $f$ denote a narrow function in $\varphi$ space, and $\tilde F$ denote a broad envelope function in $\varphi$ space. We express the approximate codewords as 
\begin{widetext}
\begin{eqnarray}\label{eq:approx-codewords-zero-one}
|0_C\rangle &=& \sqrt{2\pi}~\sum_{n~{\rm even}} \tilde F(\pi n) \int d\varphi~f(\varphi-\pi n)|\varphi\rangle,\nonumber\\
|1_C\rangle &=& \sqrt{2\pi}~\sum_{n~{\rm odd}}\tilde F(\pi n)\int d\varphi~f(\varphi-\pi n)|\varphi\rangle.
\end{eqnarray}
\end{widetext}
The function $f$ is normalized so that
\begin{equation}
\int d\varphi~|f(\varphi)|^2 = 1,
\end{equation}
and if the overlap between peaks centered at distinct integer multiples of $\pi$ can be neglected, then $|0_C\rangle$ and $|1_C\rangle$ are normalized provided
\begin{equation}
2\pi\sum_{n~{\rm even}} |\tilde F(\pi n)|^2 \approx 1,\quad 2\pi \sum_{n~{\rm odd}}|\tilde F(\pi n)|^2\approx 1.
\end{equation}
The intrinsic bit-flip error of the approximate codeword $|0_C\rangle$ arises from the probability that $\varphi$ lies closer to an odd multiple of $\pi$ than to an even multiple, which can be estimated as
\begin{eqnarray}\label{eq:bit-flip-error}
P_{\rm error}^{|0_C\rangle} &\approx& \left(2\pi \sum_{n~{\rm even}} |\tilde F(\pi n)|^2\right)\nonumber\\
&& \times \left(\int_{-\infty}^{-\pi/2}d\varphi~|f(\varphi)|^2 +\int_{\pi/2}^{\infty}d\varphi~|f(\varphi)|^2 \right)\nonumber\\
&\approx& \int_{-\infty}^{-\pi/2}d\varphi~|f(\varphi)|^2 +\int_{\pi/2}^{\infty}d\varphi~|f(\varphi)|^2;
\end{eqnarray}
the intrinsic error in $|1_C\rangle$ can be estimated similarly. Thus logical bit-flip errors are highly suppressed if $f(\varphi)$ is a narrow, rapidly decaying function supported near zero.

The approximate codewords in the conjugate basis are
\begin{eqnarray}
|+_C\rangle &=& \frac{1}{\sqrt{2}}\left(|0_C\rangle + |1_C\rangle\right)\nonumber\\
&=&\sqrt{\pi}~\sum_n \tilde F(\pi n) \int d\varphi~f(\varphi - \pi n)|\varphi\rangle,\nonumber\\
|-_C\rangle &=& \frac{1}{\sqrt{2}}\left(|0_C\rangle - |1_C\rangle\right)\nonumber\\
&=&\sqrt{\pi}~\sum_n\tilde F(\pi n) (-1)^n \int d\varphi~f(\varphi-\pi n)|\varphi\rangle,\nonumber\\
\end{eqnarray} 
where
\begin{equation}\label{eq:F-n-norm}
\pi\sum_n |\tilde F(\pi n)|^2\approx 1.
\end{equation}
We show in Appendix~\ref{app:overrotation} that these codewords can be expressed as
\begin{eqnarray}\label{eq:code-in-Q-space}
|+_C\rangle &=&\sqrt{2}\int dQ~\tilde f(Q)\sum_{m~{\rm even}}F(Q -m) |Q\rangle,\nonumber\\
&  \approx&\sqrt{2}~\sum_{m~{\rm even}} \tilde f(m) \int dQ~ F(Q -m) |Q\rangle,\nonumber\\
|-_C\rangle &=&\sqrt{2}\int dQ~\tilde f(Q)\sum_{m~{\rm odd}}F(Q -m) |Q\rangle,\nonumber\\
&\approx& \sqrt{2}~\sum_{m~{\rm odd}} \tilde f(m) \int dQ~ F(Q-m) |Q\rangle.\nonumber\\
\end{eqnarray} 
where
\begin{eqnarray}
2 \sum_{m~{\rm even}} |\tilde f(m)|^2 &\approx& \int dQ~|\tilde f(Q)|^2 \approx 1 ,\nonumber\\
2\sum_{m~{\rm odd}} |\tilde f(m)|^2 &\approx& \int dQ~|\tilde f(Q)|^2 \approx 1 .
\end{eqnarray}
The intrinsic phase error of the approximate codeword $|+_C\rangle$ arises from the probability that $Q$ lies closer to an odd integer than an even integer, which can be estimated as
\begin{eqnarray}\label{eq:unprimed-phase-error}
P_{\rm error}^{|+_C\rangle} &&\approx \left(2\sum_{m~{\rm even}}|\tilde f(m)|^2\right)\nonumber\\
&& \times \left(\int_{-\infty}^{-1/2}dQ~|F(Q)|^2 +\int_{1/2}^{\infty}dQ~|F(Q)|^2 \right)\nonumber\\
&&\approx\int_{-\infty}^{-1/2}dQ~|F(Q)|^2 +\int_{1/2}^{\infty}dQ~|F(Q)|^2;
\end{eqnarray}
the intrinsic error in $|-_C\rangle$ is estimated similarly. Thus logical phase errors are highly suppressed if $F(Q)$ is a narrow, rapidly decaying function supported near zero.

We see that for a good approximate codeword, the particular form of the narrow function $f(\varphi)$ and the broad function $\tilde F(\varphi)$ is not so important. Instead, what matters is that the {\em same} narrow function $f$ appears at each of the periodically spaced peaks (apart from the slow modulation in the normalization, governed by $F$). Then when we Fourier transform, constructive interference occurs for values of $Q$ that are reciprocally related to the spacing ({\em e.g.}, even integer values of $Q$ if the spacing in $\varphi$ space is $\pi$), since for such $Q$ values the various peaks in $\varphi$ space add together with a common phase. 

\subsection{Gate error estimate}
\label{subsec:gate-error}

As explained in Sec.~\ref{sec:sketch}, the error in the protected phase gate can be expressed as 
\be
|1-\langle \psi_1^{\rm fin}|\psi_0^{\rm fin}\rangle|
\ee
where $|\psi_{0,1}^{\rm fin}\rangle$ denotes the final state of the oscillator (modulo the state dependent phase $-i$ applied by the gate) when the state of the 0-$\pi$ qubit is $|0\rangle,|1\rangle$. Under conditions enumerated in Sec.~\ref{sec:sketch}, this quantity can be well approximated by the modulus of 
\be\label{eta-define-again}
\eta \equiv 1-\langle \psi_1^{\rm end}|\bar X|\psi_0^{\rm end}\rangle,
\ee
where $|\psi_{0,1}^{\rm end}\rangle$ denotes the state of the oscillator as the coupling $J(t)$ between oscillator and qubit starts to turn off, and $\bar X = \Pi^Q_{\rm even} - \Pi^Q_{\rm odd}$ denotes the error-corrected logical operator. 

Let us suppose that the states $|\psi_{0,1}^{\rm begin}\rangle$ prepared when the coupling $J(t)$ turns on are the approximate codewords $|0_C\rangle, |1_C\rangle$ depicted in Eq.~(\ref{eq:approx-codewords-zero-one}), where $f(\varphi)$ is a narrow rapidly decreasing function and $\tilde F(\varphi)$ is a broad envelope function. The coupling remains on for time $t= (1+\varepsilon)\tfrac{L}{\pi}$, where  $\varepsilon$ is the fractional error in the timing of the gate. Then as explained in Appendix \ref{app:overrotation}, the states $|\psi_{0,1}^{\rm end}\rangle$ have the same form as $|0_C\rangle,|1_C\rangle$, but with $\tilde F(\varphi)$ replaced by the function
\be
\tilde F_\varepsilon(\varphi) = e^{-i\varepsilon\varphi^2/2\pi}\tilde F(\varphi).
\ee

We define states
\be
|\psi_\pm^{\rm end}\rangle = \frac{1}{\sqrt{2}}\left(|\psi_0^{\rm end}\rangle \pm |\psi_1^{\rm end}\rangle\right);
\ee
note that $|\psi_{0,1}^{\rm end}\rangle$ are normalized, since each is obtained by applying a unitary time evolution operator to the normalized state $|\psi^{\rm in}\rangle$ of the oscillator, but they are not necessarily orthogonal and hence the states $|\psi_\pm^{\rm end}\rangle$ are not necessarily normalized. We may write
\be
\langle\psi_1^{\rm end}|\bar X|\psi_0^{\rm end}\rangle
= &&\frac{1}{2}\Big(\langle\psi_+^{\rm end}|\bar X|\psi_+^{\rm end}\rangle
- \langle\psi_-^{\rm end}|\bar X|\psi_-^{\rm end}\rangle\nonumber\\
&&+\langle\psi_+^{\rm end}|\bar X|\psi_-^{\rm end}\rangle
-\langle\psi_-^{\rm end}|\bar X|\psi_+^{\rm end}\rangle\Big),\nonumber\\
\ee
which, using Eq.~(\ref{barX-define}), has real part
\be
&&{\rm Re}~ \langle\psi_1^{\rm end}|\bar X|\psi_0^{\rm end}\rangle \nonumber\\
&&=\frac{1}{2}\Big(\langle\psi_+^{\rm end}|I-2\Pi^Q_{\rm odd}|\psi_+^{\rm end}\rangle
- \langle\psi_-^{\rm end}|2\Pi^Q_{\rm even}-I|\psi_-^{\rm end}\rangle\Big)\nonumber\\
&&= 1- \langle\psi_+^{\rm end}|\Pi^Q_{\rm odd}|\psi_+^{\rm end}\rangle
- \langle\psi_-^{\rm end}|\Pi^Q_{\rm even}|\psi_-^{\rm end}\rangle.
\ee

Therefore, using Eq.~(\ref{eta-define-again}) we may estimate the real part of the gate error as in Eq.~(\ref{eq:unprimed-phase-error}):
\be\label{eq:real-eta}
{\rm Re}~ \eta_\varepsilon\approx 2\left(\int_{-\infty}^{-1/2}dQ~|F_\varepsilon(Q)|^2 +\int_{1/2}^{\infty}dQ~|F_\varepsilon(Q)|^2\right) .\nonumber\\
\ee
To estimate the imaginary part we note that 
\be
{\rm Im} ~\eta_\varepsilon 
=\frac{1}{2i}\Big(\langle\psi_-^{\rm end}|\bar X|\psi_+^{\rm end}\rangle-\langle\psi_+^{\rm end}|\bar X|\psi_-^{\rm end}\rangle
\Big).
\ee
As in Eq.~(\ref{eq:code-in-Q-space}), we express
\begin{widetext}
\be
|\psi_-^{\rm end}\rangle &   \approx &\sqrt{2}\Big(\int_{[Q]~{\rm even}}dQ +\int_{[Q]~{\rm odd}}dQ\Big)\tilde f(Q)\sum_{m~{\rm odd}} F_\varepsilon(Q -m) |Q\rangle,\nonumber\\
\bar X|\psi_+^{\rm end}\rangle &   \approx &\sqrt{2}\Big(\int_{[Q]~{\rm even}}dQ -\int_{[Q]~{\rm odd}}dQ\Big)\tilde f(Q)\sum_{m~{\rm even}} F_\varepsilon(Q -m) |Q\rangle,\nonumber\\
\ee
\end{widetext}
and therefore obtain
\be
&&\langle\psi_-^{\rm end}|\bar X |\psi_+^{\rm end}\rangle\approx\nonumber\\
&&2 \int_{[Q]~{\rm even}}dQ |\tilde f(Q)|^2\sum_{{m~{\rm even}}\atop{n~{\rm odd}}}F_\varepsilon(Q-n)^*F_\varepsilon(Q-m)\nonumber\\
&& - 2\int_{[Q]~{\rm odd}}dQ |\tilde f(Q)|^2\sum_{{m~{\rm even}}\atop{n~{\rm odd}}}F_\varepsilon(Q-n)^*F_\varepsilon(Q-m).\nonumber\\
\ee
Thus ${\rm Im} ~\eta_\varepsilon$
is dominated by overlaps between sharply peaked functions $\{F_\varepsilon(Q-m)\}$ centered at neighboring values of $m$; we show in Appendix \ref{app:overrotation} that 
\be\label{eq:imag-eta}
{\rm Im} ~\eta_\varepsilon \approx 4~{\rm Im}\int_{0}^\infty dQ ~{\rm Odd}\left[F_\varepsilon(Q+\tfrac{1}{2})^*F_\varepsilon(Q-\tfrac{1}{2})\right]
\ee
where ${\rm Odd}[G(Q)]\equiv\tfrac{1}{2}\left(G(Q) - G(-Q)\right)$ denotes the odd part of the function $G(Q)$.

\subsection{Gaussian case}
\label{subsec:overrotation}

The gate error estimate in Eq.~(\ref{eq:real-eta}) and Eq.~(\ref{eq:imag-eta}) is expressed in terms of the narrow function $F_\varepsilon(Q)$, whose Fourier transform $\tilde F_\varepsilon(\varphi)$ is the broad envelope function that governs the grid state of the oscillator. To be concrete, let us now suppose that this function is Gaussian, the relevant case where the oscillator's initial state $|\psi^{\rm in}\rangle$ is the ground state.

The normalized ground state wave function is
\begin{eqnarray}\label{eq:F-and-F-tilde}
&&\tilde F(\varphi)=\left(\frac{\kappa^2}{\pi}\right)^{1/4}e^{-\kappa^2\varphi^2/2}, \nonumber\\
&&F(Q)= \left(\frac{1}{\pi\kappa^2}\right)^{1/4}e^{-Q^2/2\kappa^2},
\end{eqnarray}
where 
\be
\kappa^{-2} = \sqrt{\frac{L}{C}}.
\ee
Therefore, for $\varepsilon=0$ the probability of an intrinsic phase error in the grid state is 
\begin{eqnarray}\label{eq:gaussian-phase-flip-error}
 P_{\rm error}^{|+_C\rangle} \approx 2\int_{1/2}^{\infty}dQ~|F(Q)|^2\approx 2\sqrt{\frac{\kappa^2}{\pi}}e^{-1/4\kappa^2},
\end{eqnarray}
using the leading asymptotic approximation to the error function.

To find the probability of a phase error for $\varepsilon \ne 0$, we evaluate
\begin{eqnarray}
F_\varepsilon(Q) &=& \frac{1}{\sqrt{2\pi}}\int d\varphi~e^{-iQ\varphi}\tilde F_\varepsilon(\varphi)\nonumber\\
&=& \frac{1}{\sqrt{2\pi}}\int d\varphi~e^{-iQ\varphi}\left(\frac{\kappa^2}{\pi}\right)^{1/4}e^{-\kappa^2\varphi^2/2}e^{-i\varepsilon\varphi^2/2\pi}\nonumber\\
&=& \left(\frac{\kappa^2}{\pi\kappa'^4}  \right)^{1/4}\exp\left(-Q^2/2\kappa'^2\right),
\end{eqnarray}
where 
\begin{eqnarray}
\kappa'^2 = \kappa^2 +  \frac{i\varepsilon}{\pi}.
\end{eqnarray}
Thus
\begin{eqnarray}
\kappa'^{-2} &=& \kappa^{-2}\left( 1 +\frac{i\varepsilon}{\pi\kappa^2}\right)^{-1}\nonumber\\
&=&\bar\kappa^{-2}\left( 1 -\frac{i\varepsilon}{\pi\kappa^2}\right),\nonumber\\
\end{eqnarray}
where 
\begin{eqnarray}
\bar\kappa^2=\kappa^{2}\left( 1 +\frac{\varepsilon^2}{\pi^2\kappa^4}\right);
\end{eqnarray}
therefore
\begin{eqnarray}
|F_\varepsilon(Q)|^2&=&\left(\frac{\kappa^2}{\pi|\kappa'|^4}  \right)^{1/2}\exp\left(-Q^2~{\rm Re}(\kappa'^{-2})\right)\nonumber\\
&=& \frac{1}{\sqrt{\pi\bar\kappa^2}}e^{-Q^2/\bar\kappa^2},
\end{eqnarray}
From Eq.~(\ref{eq:real-eta}), our estimate of the real part of the gate error becomes
\begin{eqnarray}\label{eq:phase-error-gaussian}
{\rm Re}~\eta_\varepsilon\approx \frac{4}{\sqrt{\pi\bar\kappa^2}}\int_{1/2}^\infty dQ ~ e^{-Q^2/\bar\kappa^2}\approx 4\sqrt{\frac{\bar\kappa^2}{\pi}}e^{-1/4\bar\kappa^2}.
\end{eqnarray}

For $\varepsilon$ small compared to $\pi\kappa^2$, we may expand
\begin{equation}
\bar\kappa^{-2} = \kappa^{-2} -\frac{\varepsilon^2}{\pi^2\kappa^6} + \cdots,
\end{equation}
so that 
\begin{equation}\label{eq:gaussian-eta-real}
{\rm Re}~\eta_\varepsilon \approx \exp\left(\frac{\varepsilon^2}{4\pi^2\kappa^6}\right) {\rm Re}~\eta_{\varepsilon=0};
\end{equation}
the overrotation of the gate has little effect on the real part of the gate error for $\varepsilon \ll 2\pi\kappa^3$. On the other hand, when $\varepsilon$ is large compared to $\pi\kappa^2$, we have
\begin{eqnarray}
\bar \kappa^{-2}\approx\frac{\pi^2\kappa^2}{\varepsilon^2};
\end{eqnarray}
thus ${\rm Re}~\eta_\varepsilon=O(1)$ for $\varepsilon\approx \pi\kappa$.

To see that these results are reasonable, note that
\begin{eqnarray}
e^{-i\varepsilon  \varphi^2/2\pi}  Q e^{i\varepsilon  \varphi^2/2\pi} =  Q +\varphi\varepsilon/\pi.
\end{eqnarray}
Thus, crudely speaking, overrotation by $\varepsilon$ shifts $ Q$ by 
\be\label{eq:delta-Q-epsilon}
\delta Q\approx (\varepsilon/\pi)\langle \varphi^2\rangle^{1/2}=\varepsilon/(\pi\sqrt{2\kappa^2}).
\ee
We expect this shift to have a small effect if the amount of the shift is small compared to the width $\langle  Q^2\rangle^{1/2}= \sqrt{\kappa^2/2}$ of the narrow peaks in $Q$-space, {\em i.e.} for $\varepsilon \ll \pi\kappa^2$. On the other hand, for $\varepsilon \approx \pi\kappa$, the shift in $ Q$ is O(1), and we expect the error probability to be large.

To estimate the imaginary part of the gate error, we note that
\be
&&F_\varepsilon(Q+\tfrac{1}{2})^*F_\varepsilon(Q-\tfrac{1}{2})\nonumber\\
&=&\frac{1}{\sqrt{\pi\bar\kappa^2}}
\exp\left[-\frac{(Q-\tfrac{1}{2})^2}{2\kappa'^2}-\frac{(Q+\tfrac{1}{2})^2}{2\kappa'^{*2}}\right]\nonumber\\
&=&\frac{1}{\sqrt{\pi\bar\kappa^2}}
\exp\left[-(Q^2+\tfrac{1}{4})~{\rm Re}(\kappa'^{-2})
+iQ ~{\rm Im}(\kappa'^{-2})\right]\nonumber\\
&=&\frac{1}{\sqrt{\pi\bar\kappa^2}}e^{-1/4\bar\kappa^2}e^{-Q^2/\bar\kappa^2}
\exp\left( \frac{-i\varepsilon Q}{\pi\bar\kappa^2\kappa^2}\right),\nonumber\\
\ee
and from Eq.~(\ref{eq:imag-eta}) we obtain
\be\label{eq:phase-error-imag}
&&{\rm Im}~\eta_\varepsilon \nonumber\\
&&= -\frac{4}{\sqrt{\pi\bar\kappa^2}}~e^{-1/4\bar\kappa^2}\int_0^\infty dQ~e^{-Q^2/\bar\kappa^2}\sin\left(\varepsilon Q/\pi\bar\kappa^2\kappa^2\right)\nonumber\\
&&= -\frac{4}{\sqrt{\pi}}~e^{-1/4\bar\kappa^2}I\left(\frac{\varepsilon}{\pi\bar\kappa\kappa^2}\right),
\ee
where
\be
I(\alpha) \equiv \int_0^\infty dx ~e^{-x^2}\sin(\alpha x).
\ee
The integral $I(\alpha)$ can be expressed in terms of Gamma functions with imaginary arguments, but for our purposes it will suffice to observe some of its properties. For small $\alpha$ it has the power series expansion
\be
I(\alpha) = \frac{\alpha}{2} -\frac{\alpha^3}{12} + \cdots,
\ee
for large $\alpha$ it has the asymptotic expansion
\be
I(\alpha) = \frac{1}{\alpha} + \frac{2}{\alpha^3} + \cdots,
\ee
and it attains its maximum value $I=.5410\dots$ at $\alpha = 1.8483\dots$. 

Combining the real part in Eq.~(\ref{eq:phase-error-gaussian}) with the imaginary part in Eq.~(\ref{eq:phase-error-imag}), our estimate of the gate error becomes
\be \label{eq:overrotation-error-complete}
|\eta_\varepsilon| = \frac{4}{\sqrt{\pi}}e^{-1/4\bar \kappa^2}\sqrt{\bar\kappa^2+\left[I\left(\frac{\varepsilon}{\pi\bar\kappa\kappa^2}\right)\right]^2}
\ee
The ratio of the imaginary and real parts is 
\be
\frac{{\rm Im}~\eta_\varepsilon}{{\rm Re}~\eta_\varepsilon}=-\bar\kappa^{-1}I\left(\frac{\varepsilon}{\pi\bar\kappa\kappa^2}\right)\approx -\frac{\varepsilon}{2\pi\bar\kappa^2\kappa^2},
\ee
expanding to linear order in $\alpha$. Thus the imaginary part of the error is smaller than the real part when $\varepsilon$ is sufficiently small, but dominates by a factor of order $\kappa^{-1}$ for $\varepsilon \sim \pi\kappa^3$. The error $|\eta_\varepsilon|$ is bounded above by $e^{-1/4\bar\kappa^2}\times O(1)$ for all $\varepsilon$, and hence by $e^{-1/4\kappa^2}\times O(1)$ for $\varepsilon < 2\pi \kappa^3$.

\begin{figure}[tb]
\begin{center}
\begin{tabular}{c}
\includegraphics[width=0.45\textwidth]{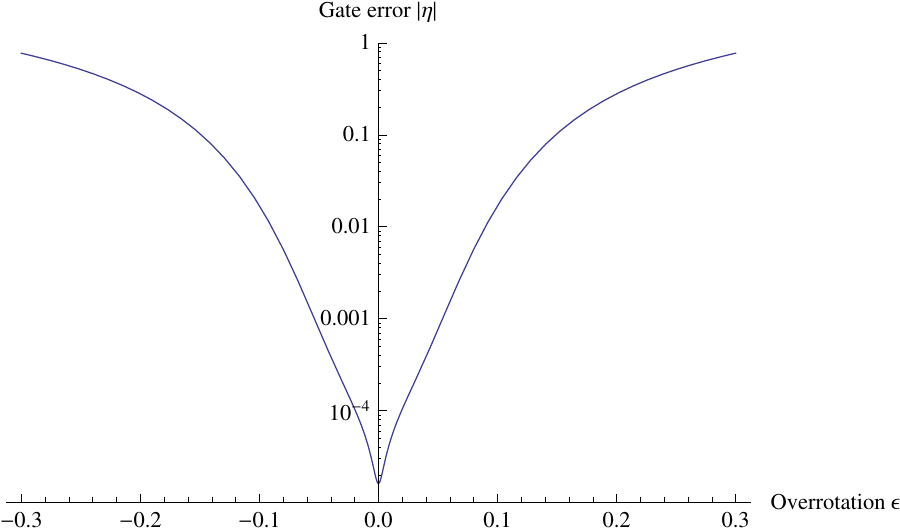}
\end{tabular}
\end{center}\caption{\label{fig:kappa30} (Color online) The estimated gate error $|\eta_\varepsilon|$ (on a log scale) as a function of the rotation error $\varepsilon$, for $\kappa^{-2}= 40$.}
\label{fig:30}
\end{figure}

\begin{figure}
\begin{center}
\begin{tabular}{c}
\includegraphics[width=0.45\textwidth]{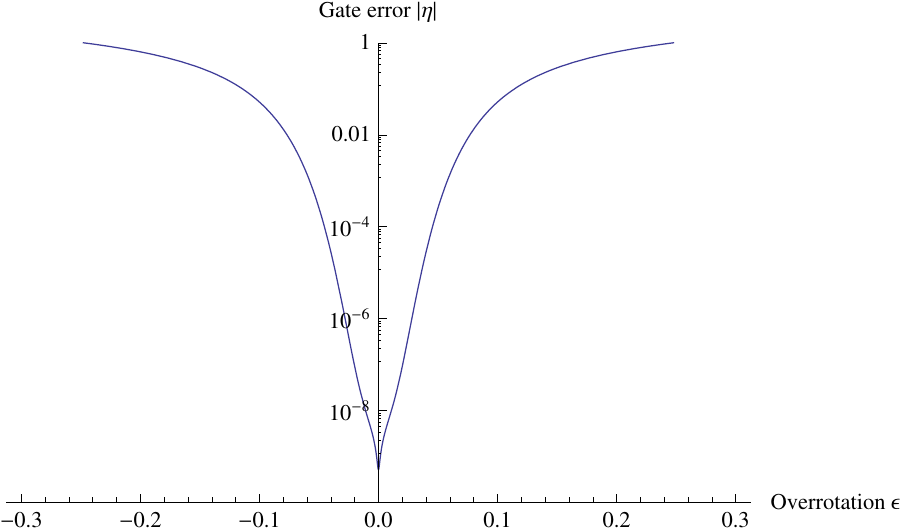}
\end{tabular}
\end{center}\caption{\label{fig:kappa80} (Color online) The estimated gate error $|\eta_\varepsilon|$  (on a log scale) as a function of the rotation error $\varepsilon$, for $\kappa^{-2}= 80$.}
\label{fig:100}
\end{figure}

In Fig.~\ref{fig:kappa30} and \ref{fig:kappa80}, we plot the gate error estimate $|\eta_{\varepsilon}|$ as a function of $\varepsilon$ for $\kappa^{-2}=40$ and $\kappa^{-2}=80$. Recall that, in the case where the $LC$ circuit is initially in its ground state, we can identify $\kappa^{-2}$ with $\sqrt{L/C}$.

\section{Diabatic error}
\label{sec:diabatic}

As explained in Sec.~\ref{sec:sketch}, the protected phase gate is very accurate if the final state vector of the oscillator depends only very weakly on the state of the 0-$\pi$ qubit: $|\psi_1^{\rm fin}\rangle\approx |\psi_0^{\rm fin}\rangle$. Our argument establishing high gate accuracy has two elements --- we show that $|\psi_1(t)\rangle \approx \bar X |\psi_0(t)\rangle$ at each stage of the oscillator's evolution, and also that $\bar X|\psi_0^{\rm fin}\rangle \approx |\psi_0^{\rm fin}\rangle$. In Sec.~\ref{sec:overrotation} we have seen that the condition $|\psi_1(t)\rangle \approx \bar X |\psi_0(t)\rangle$ is stable with respect to imperfections in the timing of the pulse that executes the gate. Now we will consider rare diabatic transitions, occurring as the coupling $J(t)$ ramps on and off, that contribute to the deviation of $\bar X|\psi_0^{\rm fin}\rangle$ from $|\psi_0^{\rm fin}\rangle$

While the coupling $J(t)$ turns on or off, the harmonic $\varphi^2/2L$ term in the potential can be treated perturbatively, where in first approximation the Hamiltonian is given by Eq.~(\ref{eq:truncated-H}); we will consider the consequences of the harmonic term in Sec.~\ref{sec:squeezing}. This Hamiltonian commutes with the operator $e^{-2\pi i Q}$, which translates $\varphi$ by $2\pi$; therefore $e^{-2\pi i Q}$ and the Hamiltonian can be simultaneously diagonalized. We express the eigenvalue of this translation operator as $e^{-2\pi i q}$, where $q = Q-[Q] \in [-\frac{1}{2},\frac{1}{2}]$ is the conserved Bloch momentum, and the integer $[Q]$ labels the distinct bands in the Hamiltonian's spectrum. 

A diabatic transition between bands may be excited while $J(t)$ varies, changing the value of $[Q]$ by an integer, typically $\pm 1$. If such transitions occur with nonnegligible probability, the final state of the oscillator will contain, in addition to a primary peak supported near $Q=0$ (where $\bar X=1$), also secondary peaks supported near $Q=\pm 1$ (where $\bar X = -1)$. We will discuss the probability of a transition between bands while $J(t)$ ramps on; a similar analysis applies to transitions occurring as $J(t)$ ramps down. 

The probability of a diabatic transition can be computed most reliably for $q$ close to $\pm\tfrac{1}{2}$, since in that case the splitting between the lowest band and the first excited band is small when $J(t)$ is small, and the continuous variable system can be well approximated by a two-level system. For example when $J=0$, the state in the lowest band with Bloch momentum $q$ slightly less than $\tfrac{1}{2}$ has charge $Q=q$, while the state in the first excited band has $Q= q-1$. Hence the splitting between bands is
\be
\frac{1}{2C}\left((q-1)^2 - q^2\right)= \frac{\tfrac{1}{2} -q}{C}.
\ee
Since $e^{\mp i\varphi}$ translates $Q$ by $\pm 1$, the perturbation $J(t)\cos\varphi$ has matrix elements
\be
\langle q-1|J(t)\cos\varphi|q\rangle = \frac{J(t)}{2} = \langle q|J(t)\cos\varphi|q-1\rangle,
\ee
and the effective two-level Hamiltonian is
\be
H_{0,1}^{\rm eff} = -\frac{\tfrac{1}{2} -q}{2C}~\sigma^Z \mp \frac{J(t)}{2}~\sigma^X,
\ee
where $\sigma^{Z,X}$ are Pauli matrices. The energy eigenstates are $\sigma^Z$ eigenstates for $J(t)\rightarrow 0$ and $\sigma^X$ eigenstates for $J(t)\to \infty$.

The time-dependent Schr\"odinger equation for this effective Hamiltonian can be solved exactly if $J(t)$ increases exponentially with time, as 
\be
J(t) = J_0~\exp(t/\tau_J^{\rm eff});
\ee
we show in Appendix \ref{app:landau} that if the initial state as $t\rightarrow -\infty$ is the ground state, then the probability that the final state is excited as $t\rightarrow\infty$ is
\be\label{eq:landau-tanh}
P_{\rm diab}(q,\tau_J^{\rm eff})&=& \frac{1}{2} - \frac{1}{2}\tanh\left(\pi\left(\tfrac{1}{2} - q\right)\frac{\tau_J^{\rm eff}}{2C}\right)\nonumber\\
&\approx& \exp\left(-\pi\left(\tfrac{1}{2} - q\right)\frac{\tau_J^{\rm eff}}{C}\right) ,
\ee
where the second equality holds when the argument of the $\tanh$ is large and positive.

We recall that if the initial state of the oscillator is the ground state or a low-lying excited state, then the probability distribution for $q$ decays as
\be
P(q) \sim \exp\left( -\sqrt{\frac{L}{C}}~ q^2\right);
\ee
hence expanding in $\delta = \tfrac{1}{2} -q$ we find
\be
P(q)~P_{\rm diab}(q,\tau_J^{\rm eff})&\sim & \exp\left(\left(\delta-\tfrac{1}{4}\right)\sqrt{\frac{L}{C}}\right)\nonumber\\
&\times &\exp\left(-\pi\delta\frac{\tau_J^{\rm eff}}{C}\right).\nonumber\\
\ee
Therefore, if
\be\label{eq:tau-J-inequality}
\tau_J^{\rm eff} > \frac{1}{\pi}\sqrt{LC},
\ee
the most likely diabatic transitions occur for $q\approx \tfrac{1}{2}$, where the two-level approximation is reasonable; we conclude in that case that the probability of a diabatic transition is suppressed by the factor $\exp(-\tfrac{1}{4}\sqrt{L/C})$. If on the other hand $\tau_J^{\rm eff} < \frac{1}{\pi}\sqrt{LC}$, then the most likely diabatic transitions occur for $q$ far from $\pm\tfrac{1}{2}$ and the two-level approximation cannot be justified. 

If $J(t)$ does not ramp on exponentially, then the exact solution in Appendix \ref{app:landau} does not apply directly. To estimate roughly the probability of a diabatic transition for more general pulse shapes, we note that the transition typically occurs when $\sigma^Z$ and $\sigma^X$ in the effective Hamiltonian have comparable coefficients, so that 
\be\label{eq:tauJ-effective-define}
\tau_J^{\rm eff} \approx \left(\frac{J}{\dot J} \right)_{JC\approx 1- 2q}
\ee
If $J(t)$ turns on like an error function with width $\tau_J$, then
\be\label{eq:J(t)-asymptote}
J(t) = \frac{J_0}{\sqrt{\pi}}\int_{-\infty}^{t/\tau_J}dx~e^{-x^2}\approx \frac{\tau_J J_0}{2|t|\sqrt{\pi}}e^{-t^2/\tau_J^2}
\ee
asymptotically for $t/\tau_J\rightarrow -\infty$, and we have
\be\label{eq:JdotJ}
J/\dot J \approx \tau_J^2/ 2|t|;
\ee
Combining Eq.~(\ref{eq:tauJ-effective-define}),(\ref{eq:J(t)-asymptote}),(\ref{eq:JdotJ}) we obtain
\be\label{eq:tauJ-effective-error-func}
\frac{2\tau_J^{\rm eff}}{\tau_J} = \left(\ln\left(\tfrac{J_0 C}{1-2q}\right)-O\left(\ln\ln\left(\tfrac{J_0 C}{1-2q}\right)\right)\right)^{-1/2};\nonumber\\
\ee
though $\tau_J^{\rm eff}$ given by Eq.~(\ref{eq:tauJ-effective-error-func}) does not satisfy Eq.~(\ref{eq:tau-J-inequality}) when $q$ is very close to $\frac{1}{2}$, the dominant diabatic transitions may still occur for $q\approx \tfrac{1}{2}$, where the two-level approximation is applicable, provided $\tau_J - \frac{1}{\pi}\sqrt{LC}$ is positive and sufficiently large. Otherwise, if the dominant value of $q$ is far from $\tfrac{1}{2}$, we can anticipate that typical diabatic transitions occur for $J(t) C=O(1)$, where the band gap is $O(1/C)$ and the transition probability is
\be
P_{\rm diab}(\tau_J) = \exp\left(-O\left(\frac{\tau_J}{C}\right)\right);
\ee

In the two-level approximation, which applies for $|Q|\approx \tfrac{1}{2}$, the probability of a jump from the lowest band ($|Q| < \tfrac{1}{2}$) to the first excited band ($|Q| > \tfrac{1}{2} $) matches the probability of a jump from the first excited band to the lowest band. Therefore, neglecting transitions to other bands and the higher-order probability of multiple transitions, and also ignoring other sources of error aside from diabatic jumps, we infer from Eq.~(\ref{eq:landau-tanh}) that the probability of $\bar X = 1$ ({\em i.e.}, $|Q| <\tfrac{1}{2}$) in the final state of the oscillator can be expressed as 
\be
&&P(|Q^{\rm fin}| < \tfrac{1}{2})\nonumber\\
&&\approx \int dQ^{\rm in}~P(Q^{\rm in})\tanh\left(\pi\left(\tfrac{1}{2} - |Q^{\rm in}|\right)\frac{\tau_J^{\rm eff}}{2C}\right);\nonumber\\
\ee
a factor of two has been included to take into account that the transition could occur during either the ramping-up phase or the ramping-down phase. Because of the enhanced probability of a transition for $|Q|\approx \tfrac{1}{2}$, the $Q^{\rm in}$ integral has support extending beyond the range $[-\tfrac{1}{2},\tfrac{1}{2}]$; the $\tanh$ function smooths out the sharp edges at $Q= \pm \tfrac{1}{2}$, replacing them by rounded steps with width of order $C/\tau_J^{\rm eff}$.

\section{Squeezing error}
\label{sec:squeezing}

In Sec.~\ref{sec:sketch} we discussed how the state of the oscillator evolves as the coupling $J(t)$ ramps on and off. There we used the idea that, because the oscillator's period is long compared to the time scale $\tau_J$ for the coupling to turn on and off, we may as a first approximation ignore the $\varphi^2/2L$ term in the potential as in Eq.~(\ref{eq:truncated-H}). Under that assumption we concluded that
\be
\langle \psi_1^{\rm begin}|\bar X|\psi_0^{\rm begin}\rangle = \langle \psi^{\rm in}|\bar X|\psi^{\rm in}\rangle\approx 1,
\ee
where $|\psi^{\rm in}\rangle$ is the oscillator's initial state, and $|\psi_{0,1}^{\rm begin}\rangle$ denotes the state just after the coupling turns on, where $|0\rangle,|1\rangle$ is the state of the 0-$\pi$ qubit. The second equality follows if the initial state of the oscillator has negligible support outside the interval $Q\in [-\tfrac{1}{2},\tfrac{1}{2}]$.

How is this conclusion affected when the quadratic term $\varphi^2/2L$ is included? If the coupling turns on slowly enough, this term can cause some squeezing of the wave function in $\varphi$ space and correspondingly some spreading in $Q$ space. To model crudely the effect of the spreading, consider first turning on $J(t)$ using $H_0$ or $H_1$ in Eq.~(\ref{eq:truncated-H}), then applying the operator $e^{-i\alpha \varphi}$ (which shifts $Q$ by an amount $\alpha$ that does not depend on the state of the 0-$\pi$ qubit). Denoting the time evolution operator as $J(t)$ turns on by $U_0$ or $U_1=\bar X U_0 \bar X$ as in Sec.~\ref{sec:sketch}, we then have
\begin{eqnarray}
|\psi_0^{\rm begin}\rangle &=& e^{-i\alpha\varphi} U_0 |\psi^{\rm in}\rangle, \nonumber\\
| \psi_1^{\rm begin} \rangle &=& e^{-i\alpha\varphi} \bar X U_0\bar X |\psi^{\rm in}\rangle, \nonumber\\
\end{eqnarray}
and therefore
\begin{eqnarray}
\langle \psi_1 ^{\rm begin}|\bar X | \psi_0^{\rm begin}\rangle = \langle \psi^{\rm in}|\bar X U_0^{-1}\bar X e^{i\alpha \varphi} \bar X e^{-i\alpha \varphi} U_0 |\psi^{\rm in}\rangle.\nonumber\\
\end{eqnarray}

Now we note that
\begin{eqnarray}
\bar X e^{i\alpha \varphi} \bar X e^{-i\alpha \varphi} &=& (-1)^{-[Q]}e^{i\alpha \varphi} (-1)^{[Q]} e^{-i\alpha \varphi} \nonumber\\
&=& (-1)^{-[Q]}(-1)^{[Q-\alpha]} = (-1)^{[Q-\alpha] - [Q]};\nonumber\\
\end{eqnarray}
furthermore, $[Q-\alpha] - [Q]$ commutes with $\cos\varphi$ and hence with $U_0$, because $e^{-i\varphi}$ acting by conjugation increases both $[Q-\alpha$] and $[Q]$ by 1 (while $e^{i\varphi}$ decreases both by 1). Hence we find
\begin{eqnarray}
\langle \psi_1 ^{\rm begin}|\bar X | \psi_0^{\rm begin}\rangle &=& \langle \psi^{\rm in}|e^{i\alpha\varphi}\bar X e^{-i\alpha\varphi}|\psi^{\rm in}\rangle\nonumber\\
&=& \langle \psi^{\rm in}|(-1)^{[Q-\alpha]}|\psi^{\rm in}\rangle;
\end{eqnarray}
in particular if $|\psi^{\rm in}\rangle$ is almost fully supported in the interval $Q\in [-1/2+|\alpha|, 1/2 - |\alpha|~]$, then $|\psi_1^{\rm begin}\rangle$ is very close to $\bar X|\psi_0^{\rm begin}\rangle$. If $|\psi^{\rm in}\rangle$ is the Gaussian ground state with $\langle Q^2\rangle = \tfrac{1}{2}\sqrt{C/L}$, the deviation of $\langle \psi_1 ^{\rm begin}|\bar X | \psi_0^{\rm begin}\rangle$ from 1 is suppressed by the exponential factor
\be
&&|1-\langle \psi_1 ^{\rm begin}|\bar X | \psi_0^{\rm begin}\rangle| \sim \exp\left(\left(\tfrac{1}{2} - |\alpha|\right)^2\sqrt{\tfrac{L}{C}}\right)\nonumber\\
&&\approx \exp\left(|\alpha|\sqrt{\tfrac{L}{C}}\right)\exp\left(-\tfrac{1}{4}\sqrt{\tfrac{L}{C}}\right).\nonumber\\
\ee

How much spreading in $Q$ space should be expected? To make a crude estimate of how the harmonic term affects the distribution in $Q$ space, we note that 
\begin{equation}
e^{-i\beta\varphi^2}Q e^{i\beta\varphi^2} = Q +2\beta \varphi,
\end{equation}
and choose $\beta \approx \tau_J/2L$ where $\tau_J$ is the time scale for the coupling to turn on; using $\langle \varphi^2\rangle = \tfrac{1}{2}\sqrt{L/C}$ in the Gaussian ground state
we infer $Q$ is shifted by an amount of order 
\be
\alpha\sim \frac{\tau_J}{L}\left(\frac{L}{C}\right)^{1/4}.
\ee
Assuming $\tau_J\sim \sqrt{LC}$ in order to suppress the diabatic error, we find that squeezing enhances the gate error by a factor 
\be
\exp\left(({\rm constant})\left(\frac{L}{C}\right)^{1/4}\right);
\ee
that is, it contributes a subleading correction to the logarithm of the gate error.

More realistically, treating the harmonic term as a perturbative correction to the zeroth-order Hamiltonian which has $\varphi\rightarrow\varphi + 2\pi$ periodicity, the dynamics is governed by the effective Hamiltonian 
\be
H_{\rm eff} = \epsilon_{J,C}(q) + \frac{\varphi^2}{2L}
\ee
where $q\in [-\tfrac{1}{2},\tfrac{1}{2}]$ is the Bloch momentum and $\epsilon_{J,C}(q)$ is the energy of the lowest band. Expanding this band energy to quadratic order, we have
\be
\epsilon_{J,C}(q) \approx \frac{q^2}{2C_{\rm eff}}.
\ee
The effective capacitance $C_{\rm eff}$ is approximately $C$ for $J$ small, but for $JC\approx 1$, the band curvature begins to flatten rapidly; correspondingly $C_{\rm eff}$ increases sharply, as does the oscillator's period $2\pi\sqrt{LC_{\rm eff}}$. The oscillator evolves adiabatically for $J$ small, but its evolution freezes when its period becomes comparable to $\tau_J$, the characteristic time scale for the variation of the Hamiltonian. Therefore, the squeezing error is determined by the wave function's width in $q$-space at the time when the oscillator freezes; hence,

\be
&&P_{\rm sq}\sim \exp\left( -({\rm constant})\sqrt{\frac{L}{C_{\rm eff}}}\right)\nonumber\\
&&\sim \exp\left( -({\rm constant})\frac{L}{\tau_J}\right)
\sim  \exp\left( -({\rm constant})\sqrt{\frac{L}{C}}\right), \nonumber\\
\ee
where we obtain the last equality by choosing $\tau_J \sim \sqrt{LC}$ to suppress the diabatic error. Thus the contribution to the gate error due to squeezing is comparable to the other sources of error. We can use a similar argument to conclude that the squeezing error arising as the coupling $J(t)$ turns off is also of the same order.

\section{Simulations}
\label{sec:numerics}

We have compared the predictions from Sec.~\ref{sec:overrotation} and \ref{sec:sketch} to numerical simulations of the single-qubit phase gate $\exp(i\frac{\pi}{4}Z)$. We solved the time-dependent Schr\"odinger equation for the Hamiltonians $H_{0,1}$ in Eq.~(\ref{eq:hamiltonian}), assuming the oscillator starts in the ground state (excited states will be considered in Sec.~\ref{sec:temperature}). These simulations were done in MATLAB using the fourth-order split-operator method, which is based on the identity 
\begin{align}
	&e^{i t(A + B)} \nonumber\\
&= e^{i \frac{\gamma}{2} t A} e^{i \gamma t B} e^{i \frac{1-\gamma}{2} t A} e^{i (1-2\gamma) t B} e^{i \frac{1-\gamma}{2} t A} e^{i \gamma t B} e^{i \frac{\gamma}{2} t A} \nonumber\\
&\quad + \mathcal{O}(t^4)
\nonumber\\
\end{align}
where $\gamma = \frac{1}{2 - \sqrt[3]{2}}$, and $A$, $B$ are the portions of the Hamiltonian that are diagonal in the position, momentum eigenbases, respectively. The full time evolution is broken up into many small steps with the Hamiltonian alternating between $A$ and $B$ and the Fourier transform or its inverse applied between successive steps.

We assume that the coupling $\mp J(t) \cos \varphi$ between the oscillator and the 0-$\pi$ qubit turns on with an error-function profile,
\begin{equation}
	J(t) = J_0 \left(\frac{1}{2} + \frac{1}{2} \erf(t/\tau_J)\right),
\end{equation}
and that the coupling turns off after the time delay $\tau\approx L/\pi$ according to the time-reversed function $J(\tau-t)$;
the time scale $\tau_J$ for turning the coupling on and off and the time delay $\tau$ were chosen to optimize the gate accuracy. Then we varied the time delay $\tau = \tau_0\left( 1+\varepsilon\right)$, where $\tau_0$ is the optimal value, to study the effect of  overrotation/underrotation on the gate accuracy. 

\begin{figure}[!ht]
\begin{center}
\includegraphics[width=0.5\textwidth]{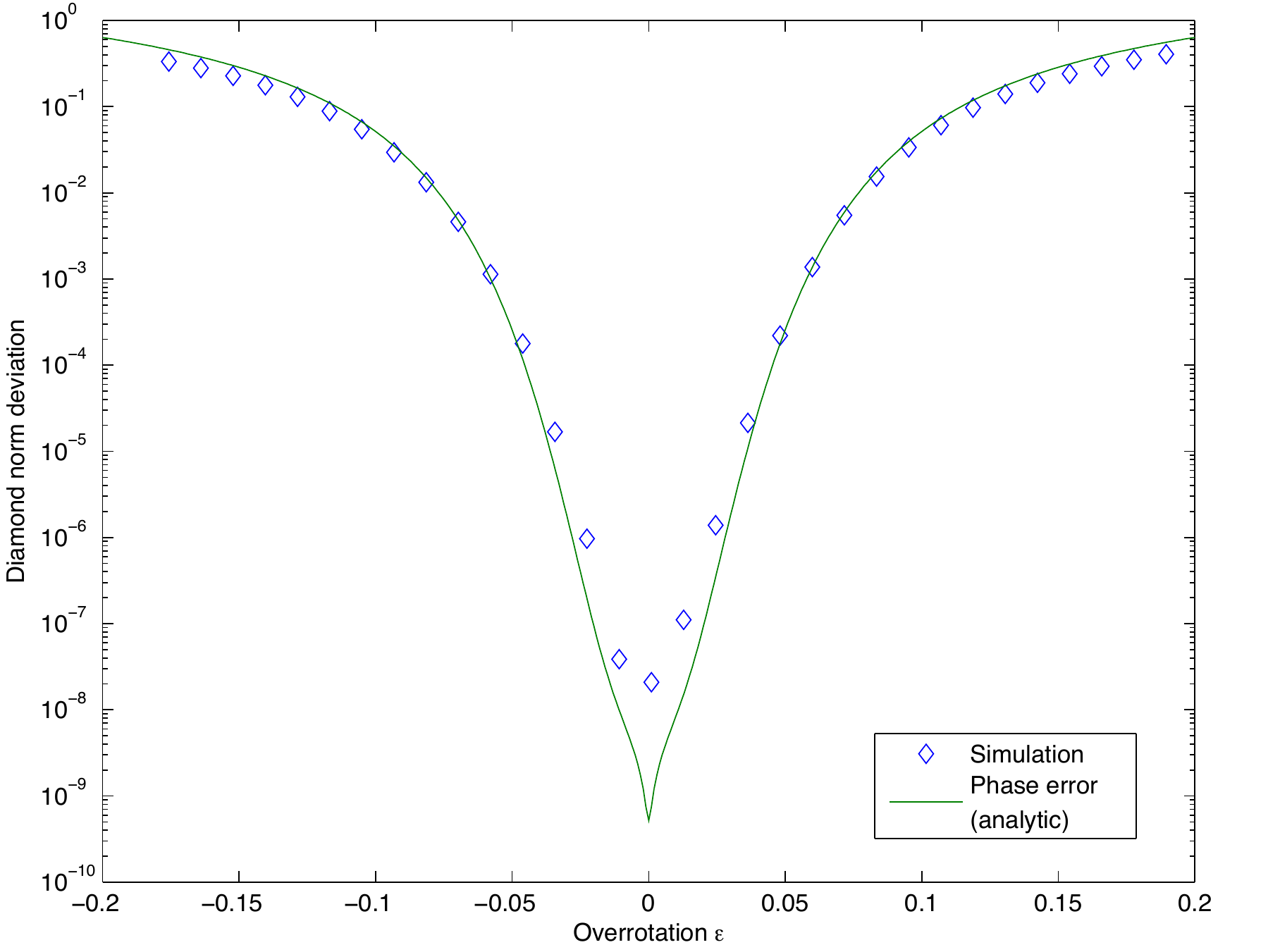}
\end{center}
\caption{\label{fig:diamond-norm} (Color online) Blue diamonds show on a logarithmic scale the numerically computed diamond-norm deviation from the ideal phase gate, for $\kappa^{-2}=\sqrt{L/C}= 80$, $\Delta^{-2} = \sqrt{J_0 C}=8$, and $\tau_J/C = 80$, as a function of the overrotation parameter $\varepsilon$. The solid green line is the analytic prediction from Eq.~(\ref{eq:overrotation-error-complete}) for the overrotation error alone. The discrepancy for small $\varepsilon$ arises from corrections due to diabatic transitions and squeezing.}
\end{figure} 

Fig.~\ref{fig:diamond-norm} shows the results for a series of simulations with $\sqrt{L/C} = 80$, $\sqrt{J_0 C }= 8$ and $\tau_J = \sqrt{LC}$, plotted together with the analytic prediction Eq.~(\ref{eq:overrotation-error-complete}) arising from the overrotation error alone. For small $\varepsilon$, the gate error is
\be
|\eta_\varepsilon| \approx 2 \times 10^{-8}\approx 10 ~ \exp\left(-\frac{1}{4}\sqrt{\frac{L}{C}}\right),
\ee
substantially larger than the analytic prediction $|\eta_\varepsilon|\approx 5\times 10^{-10}$, but roughly compatible with our expectations for the scale of the error due to diabatic transitions and squeezing, the dominant errors in this regime, when we choose $\tau_J\approx \sqrt{LC}$.
For larger values of $\varepsilon$ the overrotation error dominates and the numerical results agree well with the analytic prediction.


The simulations verify that the protected phase gate is much more robust than an unprotected gate. For an unprotected gate, executed by coupling the 0-$\pi$ qubit to a Josephson junction as described in Sec.~\ref{sec:gate}, achieving a gate error less $10^{-4}$ in the diamond norm requires pulse timing accuracy of order $10^{-4}$. In contrast, for a protected gate with the parameters specified above,  4\% accuracy in the pulse timing suffices for achieving a gate error less than $10^{-4}$. When the pulse timing accuracy is better than 1\%, the gate error is below $4\times 10^{-8}$.



\section{Nonzero temperature}
\label{sec:temperature}
Our estimate of the intrinsic gate error in Eq.~(\ref{eq:real-eta}) and Eq.~(\ref{eq:imag-eta}) applies to any grid-like state of the form Eq.~(\ref{eq:code-in-Q-space}). In Eq.~(\ref{eq:phase-error-gaussian}) and Eq.~(\ref{eq:phase-error-imag}) we treated the specific case of a Gaussian grid state, which is what arises during the execution of the protected phase gate in the case where the initial state $|\psi_{\rm in}\rangle$ of the harmonic oscillator is the ground state. Let us now consider the case where the initial state is excited, as occurs with nonzero probability at any nonzero temperature. To keep things simple, we will ignore the effects of overrotation, setting $\varepsilon = 0$. In that case, ${\rm Im}~\eta_\varepsilon = 0$ because $F_\varepsilon(Q)$ is real, so we only need to worry about the real part of the gate error. In this section we also neglect the corrections due to diabatic transitions and squeezing.

When the coupling to the 0-$\pi$ qubit is turned off, the harmonic oscillator Hamiltonian is 
\begin{equation}
H = \frac{Q^2}{2C} + \frac{\varphi^2}{2L} ,
\end{equation}
and its ground state $|\psi_0\rangle$ obeys
\begin{eqnarray}
\langle \varphi |\psi_0\rangle &=& \left(\frac{1}{\pi}\sqrt{\frac{C}{L}}  \right)^{1/4}e^{-\frac{1}{2}\sqrt{\frac{C}{L}}~\varphi^2}, \nonumber\\
\langle Q|\psi_0\rangle &=& \left(\frac{1}{\pi}\sqrt{\frac{L}{C}}  \right)^{1/4}e^{-\frac{1}{2}\sqrt{\frac{L}{C}}~Q^2}.
\end{eqnarray}
Our estimate diamond norm deviation from the ideal gate becomes
\begin{eqnarray}
|\eta(0)| &\approx& 4\int_{1/2}^{\infty}dQ ~|\langle Q|\psi_0\rangle|^2\nonumber\\
&=& 4 \left(\frac{1}{\pi}\sqrt{\frac{L}{C}}  \right)^{1/2}\int_{1/2}^\infty dQ ~e^{-\sqrt{\frac{L}{C}}~Q^2}\nonumber\\
&\approx& \frac{4}{\sqrt{\pi}}\left(\frac{L}{C}\right)^{-1/4} e^{-\frac{1}{4}\sqrt{\frac{L}{C}}}.
\end{eqnarray}
Numerically, we have,  {\em e.g.}, $|\eta(0)| \approx 1.6 \times 10^{-5}$ for $\sqrt{L/C} = 40$ and $|\eta(0)| \approx 5.2 \times 10^{-10}$ for $\sqrt{L/C} = 80$.

The oscillator's $n$th excited state is 
\begin{equation}
|\psi_n\rangle = \frac{\left(a^\dagger\right)^n}{\sqrt{n!}}|\psi_0\rangle, 
\end{equation}
where
\begin{equation}
a^\dagger = \left(\frac{C}{4L}\right)^{1/4}\varphi - i \left(\frac{L}{4C}\right)^{1/4} Q,
\end{equation}
which becomes
\begin{equation}
a^\dagger = i\left(\frac{C}{4L}\right)^{1/4}\frac{d}{dQ} - i \left(\frac{L}{4C}\right)^{1/4} Q
\end{equation}
in the $Q$ representation. Therefore, the $n$th harmonic oscillator excited state can be expressed in $Q$ space as
\begin{eqnarray}
&&\langle Q|\psi_n\rangle = \tfrac{1}{\sqrt{n!}}\left(-\left(\tfrac{C}{4L}\right)^{1/4}\tfrac{d}{dQ} + \left(\tfrac{L}{4C}\right)^{1/4} Q\right)^n \langle Q|\psi_0\rangle\nonumber\\
&&= \tfrac{2^{n/2}\pi^{-1/4}}{\sqrt{n!}} \left(  \sqrt{\tfrac{L}{C}}~\right)^{\frac{n}{2} + \frac{1}{4}}\left(Q^n+ \cdots\right)e^{-\frac{1}{2}\sqrt{\frac{L}{C}}~Q^2},
\end{eqnarray}
where in the second line we have retained only the leading power of $Q$ in the prefactor of the exponential. To estimate the probability of a logical phase error, we assume that this leading power dominates, and we also use the leading term in the asymptotic expansion
\begin{equation}
\int_x^\infty dt~ t^{2n} e^{-\alpha t^2} = \left(\frac{x^{2n-1}}{2\alpha }\right)e^{-\alpha x^2}\left(1-O(1/x^2)\right)
\end{equation}
to calculate 
\begin{eqnarray}\label{eq:phase-error-excited}
|\eta(n)| &\approx& 4\int_{1/2}^{\infty}dQ ~|\langle Q|\psi_n\rangle|^2\nonumber\\
&\approx& 
\tfrac{2^{n+2}}{n!\sqrt{\pi}} \left(  \frac{L}{C}\right)^{\frac{n}{2} + \frac{1}{4}}\int_{1/2}^\infty dQ ~Q^{2n}e^{-\sqrt{\frac{L}{C}}~Q^2}\nonumber\\
&\approx &
\tfrac{1}{2^{n-2}n!\sqrt{\pi}} \left(  \frac{L}{C}\right)^{\frac{n}{2} - \frac{1}{4}}e^{-\frac{1}{4}\sqrt{\frac{L}{C}}}\nonumber\\
&=& \frac{1}{2^{n}n!} \left(\frac{L}{C}\right)^{n/2}|\eta(0)|.
\end{eqnarray}
Thus, for example, the intrinsic gate error for the first excited ($n=1$) state is enhanced relative to the ground state by the factor $\frac{1}{2}\sqrt{L/C}$.
This approximation is applicable when $n$ is not too large, so that the leading power of $Q$ dominates the prefactor of the exponential in the tail of the wave function at $|Q| > 1/2$; in particular we require that
\begin{equation}
\langle Q^2\rangle_n = n \langle Q^2\rangle_0 = \frac{n}{2}\left(\frac{L}{C}\right)^{-1/2} \ll \frac{1}{2},
\end{equation}
or
\begin{equation}
n \ll \sqrt{L/C}.
\end{equation}

The energy of the $n$th oscillator state is $E_n = n/\sqrt{LC}$; therefore, in the thermal ensemble with inverse temperature $\beta$, the probability that the oscillator is in the $n$th state is 
\begin{equation}
P_n = \left(1-e^{-\beta/\sqrt{LC}}\right)e^{-n\beta/\sqrt{LC}}.
\end{equation}
Thus, if the oscillator is in a thermal state, while the intrinsic phase error probability for the $n$th  state is enhanced by the factor $(L/C)^{n/2}/2^{n}n!$, it is also suppressed by the Boltzmann factor $e^{-n\beta/\sqrt{LC}}$. Summing up the error probabilities for all oscillator states, with the appropriate Boltzmann weights, we find 
\begin{widetext}
\begin{eqnarray}
|\eta(\beta)|=\left(1-e^{-\beta/\sqrt{LC}}\right)\sum_{n=0}^\infty e^{-n\beta/\sqrt{LC}}~|\eta(n)|
&\approx& 
\left(1-e^{-\beta/\sqrt{LC}}\right)|\eta(0)|\sum_{n=0}^\infty e^{-n\beta/\sqrt{LC}}~\frac{1}{2^{n}n!} \left(\frac{L}{C}\right)^{n/2}\nonumber\\
&=& \left(1-e^{-\beta/\sqrt{LC}}\right)\exp\left( \frac{1}{2}\sqrt{\frac{L}{C}}~e^{-\beta/\sqrt{LC}}\right)|\eta(0)|;\nonumber\\
\end{eqnarray}
\end{widetext}
%
%
the real part of $\eta$ (and in the case we are considering there is no imaginary part) is essentially the probability of a logical phase error in the grid state, and hence to compute the gate error we need only compute this logical error probability for the thermal ensemble.
Thus, the error at finite temperature is comparable to the zero-temperature error, provided that 
\begin{eqnarray}
\sqrt{\frac{L}{C}}~e^{-\beta/\sqrt{LC}} \ll 1.
\end{eqnarray}
If, for example, $\beta/\sqrt{LC}\approx 3$, then compared to the zero-temperature case, thermal effects enhance the phase error probability by the factor  $2.6$ for $\sqrt{L/C}=40$ and $7.0$ for $\sqrt{L/C}=80$. We expect, then, that the protected phase gate remains reasonably robust provided the temperature is smaller than or comparable to the frequency of the superinductive $LC$ circuit.

\begin{figure}[ht]
\begin{center}
\includegraphics[width=0.45\textwidth]{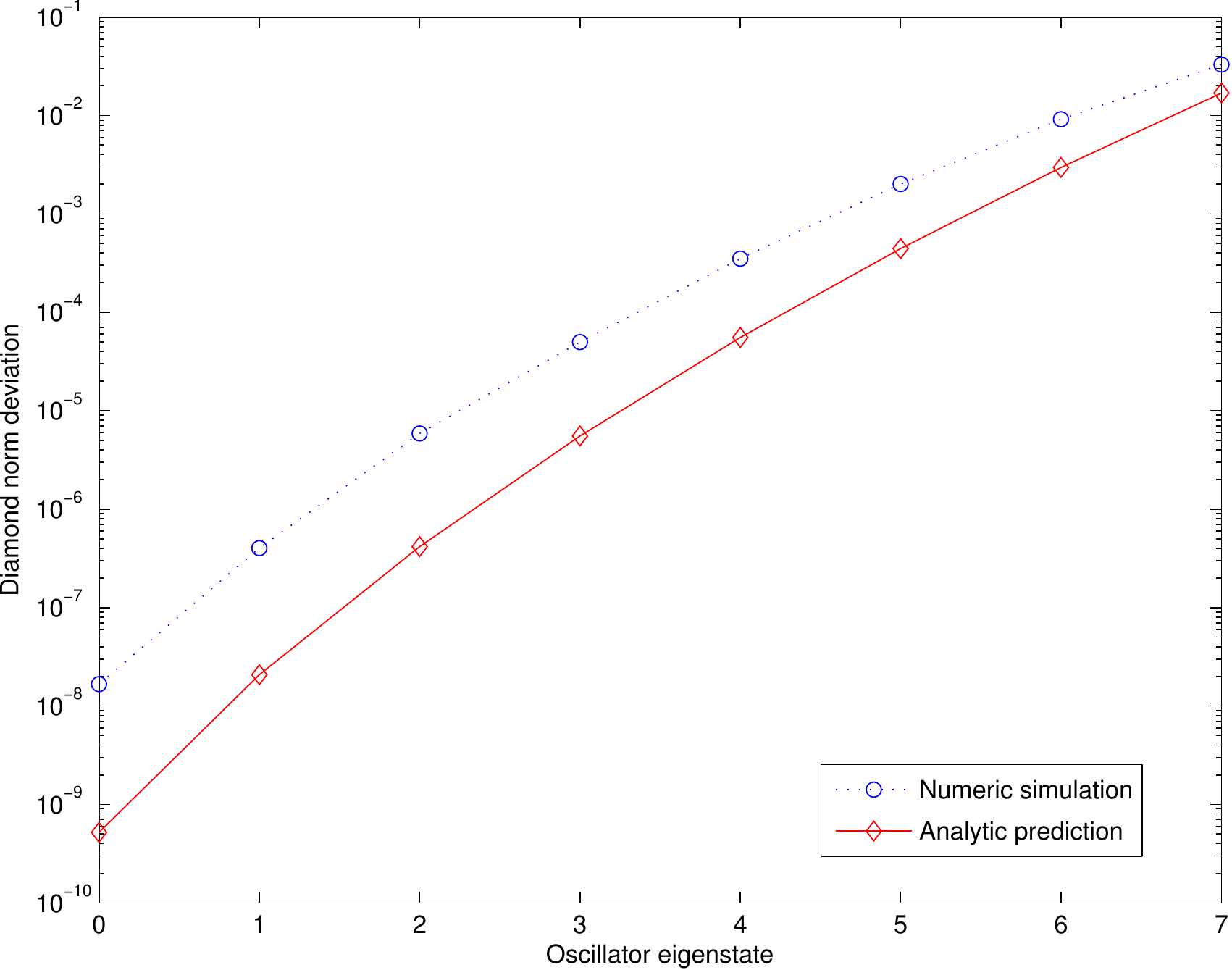}
\end{center}
\caption{\label{fig:diamond-norm-vs-n} (Color online) The minimum diamond-norm deviation of the protected phase gate from the ideal gate, as a function of the initial oscillator eigenstate $n$, for $\sqrt{L/C} = 80$ and $\sqrt{J_0C}= 8$. Results from numerical simulations are shown in blue, and the analytic prediction Eq.~(\ref{eq:phase-error-excited}) is shown in red. The discrepancy arises from corrections due to diabatic effects and squeezing, which are neglected in the derivation of Eq.~(\ref{eq:phase-error-excited}).}
\end{figure}

Numerical results, plotted in Fig.~\ref{fig:diamond-norm-vs-n}, show that the gate performance remains robust for excited eigenstates. In addition to the enhancement of the intrinsic phase error predicted by Eq.~(\ref{eq:phase-error-excited}), there is also a contribution to the gate error arising from diabatic transitions and squeezing, which becomes less important for more highly excited eigenstates.   

Up until now we have addressed how the accuracy of the phase gate is affected if, due to thermal fluctuations, the {\em initial} state of the $LC$ oscillator is not the ground state, but we have not considered thermally excited transitions that might occur while the qubit and oscillator are coupled {\em during} the execution of the gate. A thermally activated transition between bands, like a diabatic transition, could flip the value of $\bar X = (-1)^{[Q]}$ and cause the gate to fail. But as in Sec.~\ref{sec:diabatic}, the relevant band gap is of order $1/C$, so such transitions are suppressed by a Boltzmann factor
\be
P_{\rm thermal}(\beta) &=& \exp\left(-O\left(\frac{\beta}{C}\right)\right)\nonumber\\
&=& \exp\left(-O\left(\frac{\beta}{\sqrt{LC}}\cdot \sqrt{\frac{L}{C}}\right)\right),
\ee
exponentially small in $\sqrt{L/C}$ if $\beta/\sqrt{LC} = O(1)$. On the other hand, spontaneous decay of the oscillator during the execution of the gate is not likely to flip the value of $\bar X$, and hence has little impact on the gate accuracy.

\section{Perturbative stability}
\label{sec:anharmonic}
\begin{figure}[!ht]
\begin{center}
\includegraphics[width=0.45\textwidth]{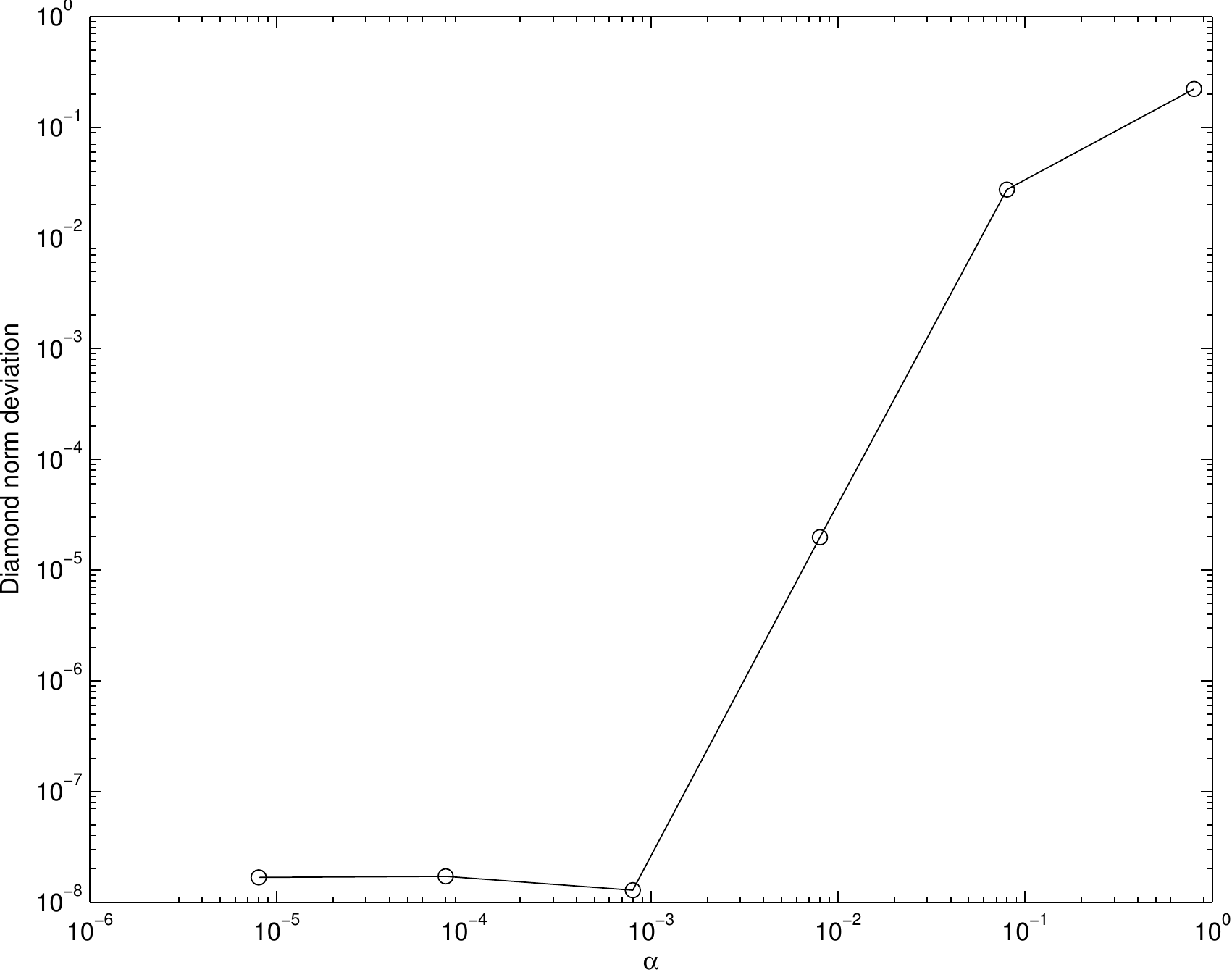}
\end{center}
\caption{\label{fig:diamond-norm-vs-alpha} The minimum diamond-norm deviation of the protected phase gate from the ideal gate, as a function of the oscillator's anharmonicity parameter $\alpha = \lambda L\sqrt{L/C}$. Here, as in Fig.~\ref{fig:diamond-norm}, $\sqrt{L/C} = 80$ and $\sqrt{J_0C}= 8$.}
\end{figure}

Aside from its stability with respect to pulse timing errors and thermal effects, we also expect the protected phase gate to be robust against small deformations in the Hamiltonian of the $LC$ oscillator and of the switch coupling the oscillator to the qubit. Suppose for example that the oscillator's potential energy $V$ includes a small anharmonic term so that 
\be
V = \frac{\varphi^2}{2L} + \lambda \varphi^4.
\ee
Over time $t$, the effect of the anharmonic term on the charge $Q$ is
\be
e^{-i\lambda\varphi^4 t} Q e^{i\lambda \varphi^4 t} = Q + 4\lambda t \varphi^3; 
\ee
thus the charge spreads by an amount
\be
\delta Q \approx 4\lambda t \langle \varphi^6\rangle^{1/2} = 4\lambda t \sqrt{15}~\langle \varphi^2\rangle^{3/2}.
\ee
Comparing to the contribution in Eq.~(\ref{eq:delta-Q-epsilon}) to $\delta Q$ arising from overrotating the gate, and choosing $t \approx L/\pi$, we see that the effect of the anharmonic term is roughly comparable to the effect of an overrotation error 
\be
\varepsilon \approx 4\sqrt{15} ~\lambda L \langle\varphi^2\rangle = 2\sqrt{15}~ \lambda L \sqrt{L/C}\approx 7.75\alpha,
\ee
where $\alpha \equiv \lambda L\sqrt{L/C}$ is a dimensionless parameter characterizing the strength of the anharmonic correction. For $\sqrt{L/C} = 80$ and $\sqrt{J_0C}=8$ as in Sec.~\ref{sec:numerics}, numerical results plotted in Fig.~\ref{fig:diamond-norm-vs-alpha} confirm that the gate accuracy is not much affected by the anharmonic term for $\alpha \lesssim 10^{-3}$, as expected. In these simulations, we assumed that the initial state of the oscillator is the ground state of the unperturbed oscillator Hamiltonian, which for $\alpha$ small has a large overlap with the ground state of the perturbed Hamiltonian.

The Josephson coupling between the 0-$\pi$ qubit and the oscillator is a periodic function of $\varphi$ with period $2\pi$, but need not be a pure cosine potential. If we include a next-to-leading harmonic correction, the time-dependent Hamiltonian has the form
\begin{equation}\label{eq:beta-pert}
	H_{0,1}(t) = \frac{Q^2}{2C} + \frac{\varphi^2}{2L} \mp f(t) J_0\left[ \cos\varphi \pm \beta \cos 2\varphi \right],
\end{equation}
where $f(t)$ varies between $0$ and $1$ as the coupling turns on or off, and $\beta$ is a dimensionless parameter characterizing the strength of the perturbation. (Note that shifting $\varphi$ by $\pi$ changes the sign of $\cos\varphi$ but not the sign of $\cos 2\varphi $.) For $\sqrt{J_0C} \gg 1$ and $\beta \ll 1$, we expect the wave function to be well localized near even or odd integer multiples of $\pi$ in $\varphi$ space (depending on whether the Hamiltonian is $H_0$ or $H_1$) while the coupling is turned on; hence it should be a good approximation to expand the potential in a power series about these local minima, and therefore the perturbation is roughly equivalent to rescaling $J_0$ according to
\be
J_0\rightarrow J_0\left(1+ 4\beta \right).
\ee
The precise value of $J_0$ does not strongly influence the phase error probability, as long as it is large enough to allow discrete peaks to form in $\varphi$ space ({\em i.e.}, to strongly suppress phase slips); it instead determines the probability of an intrinsic bit flip error, as in Eq.~(\ref{eq:bit-flip-error}).

Numerical simulations confirm that the phase gate accuracy is insensitive to the perturbation in Eq.~(\ref{eq:beta-pert}) when $\beta$ is small. For $\sqrt{L/C} = 80$ and $\sqrt{J_0 C}=8$, we find that the effect on the gate error is negligible for $| \beta | \lesssim 0.05$. 

\section{Universality}
\label{sec:universal}

We have focused so far on performing the single-qubit gate $\exp(i\frac{\pi}{4}Z)$ and the two-qubit gate $\exp(i\frac{\pi}{4}Z\otimes Z)$; in principle these can be executed with very high fidelity by coupling a 0-$\pi$ qubit or a pair of such qubits to a superinductive $LC$ circuit. But unfortunately these gates are not adequate by themselves for universal quantum computing. 

One way to obtain a universal gate set is to augment these gates by the following operations \cite{Kitaev2006}: (1) Preparation of the single-qubit states $|0\rangle$ and $|+\rangle$ (the $Z=1$ and $X=1$ eigenstates).
(2) Measurement of the single-qubit Pauli operators $Z$ and $X$. (3) The single-qubit gate $\exp(i\frac{\pi}{8}Z)$. (In fact (1) need not be regarded as independent of (2), since repeated noncommuting measurements can be used to achieve the state preparation, but we list these operations separately for clarity and completeness.)


The operation $\exp(i\frac{\pi}{8}Z)$ could be executed by coupling the 0-$\pi$ qubit to a Josephson junction for a specified time, as in Fig.~\ref{fig:unprotected}. This unprotected gate might be fairly noisy. However, if all the other operations in the universal set were perfect, then scalable quantum computing would be possible provided the noisy $\exp(i\frac{\pi}{8}Z)$ gate meets the loose fidelity criterion $F > .93$ \cite{BravyiKitaev2005}. Therefore if the gates $\exp(i\frac{\pi}{4}Z)$ and $\exp(i\frac{\pi}{4}Z\otimes Z)$ are well protected, it follows that highly reliable universal quantum computation can be achieved provided that the measurements, like these gates, have a very low error rate.

In practice, the measurements are likely to be noisy. However, if they can be performed nondestructively (with only a very low probability of changing the eigenvalue of measured operator), then they can be repeated multiple times to improve reliability. 

We note that the {\sc cphase} gate, the two-qubit gate, diagonal in the computational basis, with eigenvalues $\{1,1,1,-1\}$, can be constructed from protected gates using the decomposition
\be
&&\mathrm{CPHASE} = \exp\left(i\tfrac{\pi}{4}\left(Z-I\right)\otimes\left(Z-I\right)\right)\nonumber\\
&& = \exp\left(i\tfrac{\pi}{4}Z\otimes Z\right)\exp\left(-i\tfrac{\pi}{4}Z\otimes I\right)\exp\left(-i\tfrac{\pi}{4}I\otimes Z\right)\nonumber\\
\ee
(up to an overall phase). One way to perform a nondestructive measurement of $Z$ is to use the property
\be
\mathrm{CPHASE}: &|0\rangle\otimes |+\rangle \rightarrow |0\rangle\otimes |+\rangle,\nonumber\\
&|1\rangle\otimes |+\rangle \rightarrow |1\rangle\otimes |-\rangle;
\ee
we may apply {\sc cphase} to the target qubit (the one to be measured) and an ancilla qubit prepared in the state $|+\rangle$, then perform $X$ measurement on the ancilla qubit. If the {\sc cphase} gate is not likely to induce a bit flip on the target qubit, this procedure can be repeated many times, then the measurement result determined by a majority vote of the outcomes. 

If we are limited to using our protected gates, we cannot use the same trick to amplify an $X$ measurement. Perhaps the charge measurement described in Sec.~\ref{sec:qubit}, though the outcome is noisy, can be done fault tolerantly, meaning that measurement procedure is not likely to flip the eigenvalue of $X$. In that case, amplification by repetition and majority voting will work. Otherwise, there are alternative ways to boost the measurement accuracy, using repetition coding.

For example, Ref. \cite{AliferisPreskill2008,Aliferis2009,Brooks2012} describes a scheme for universal fault-tolerant quantum computing built from the {\sc cphase} gate, $|+\rangle$ preparation, $X$ measurement, and, in addition, preparation of the single-qubit states
\be
|{-}i\rangle &=& \frac{1}{\sqrt{2}}\left(|0\rangle -i |1\rangle\right),\nonumber\\
|e^{-i\pi/4}\rangle &=& \frac{1}{\sqrt{2}}\left(|0\rangle +e^{-i\pi/4} |1\rangle\right),\nonumber
\ee
which can be achieved by applying $\exp(i\frac{\pi}{4}Z)$ or $\exp(i\frac{\pi}{8}Z)$ to $|+\rangle$. The main point of \cite{AliferisPreskill2008,Aliferis2009,Brooks2012} is that this scheme works effectively when the noise in the {\sc cphase} gate is highly biased, {\em i.e.}, when $Z$ errors are much more common than $X$ errors.
The point we wish to emphasize here is that the  scheme remains effective when the $X$ measurement error rate is much higher than the {\sc cphase} gate error rate.

\begin{figure}[tb]
\begin{center}
\mbox{\Qcircuit @C=1.0em @R=0.2em {
	\lstick{\ket{+}} & \ctrl{2} & \ctrl{3} & \ctrl{4} & \ctrl{6} & \ctrl{7} & \ctrl{8} & \qw & \qw & \qw & \gate{\mathcal{M}_{X}} \\
	*{\vphantom{\rule{0em}{1em}}}\\
	\lstick{\ket{+}} & \control \qw & \qw & \qw & \qw & \qw & \qw & \qw & \qw & \qw & \qw & *+<.6em>{\vphantom{{\mathcal{M}_{X}}}}\qw  \\
	\lstick{\ket{+}} & \qw & \control \qw & \qw & \qw & \qw & \qw & \qw & \qw & \qw & \qw & *+<.6em>{\vphantom{{\mathcal{M}_{X}}}}\qw  \\
	\lstick{\ket{+}} & \qw & \qw & \control \qw & \qw & \qw & \qw & \qw & \qw & \qw & \qw & *+<.6em>{\vphantom{{\mathcal{M}_{X}}}}\qw  \\
	*{\vphantom{\rule{0em}{1em}}}\\
	& \control \qw & \qw & \qw & \control \qw & \qw & \qw & \qw & \qw & \qw & \gate{\mathcal{M}_{X}}  \\
	& \qw & \control \qw & \qw & \qw & \control \qw & \qw & \qw & \qw & \qw & \gate{\mathcal{M}_{X}} \\
	& \qw & \qw & \control \qw & \qw & \qw & \control \qw & \qw & \qw & \qw & \gate{\mathcal{M}_{X}} \\
	*{\vphantom{\rule{0em}{1em}}}\\
	& \qw & \qw & \qw & \control \qw & \qw & \qw & \qw & \qw & \qw & \gate{\mathcal{M}_{X}} \\
	& \qw & \qw & \qw & \qw & \control \qw & \qw & \qw & \qw & \qw & \gate{\mathcal{M}_{X}} \\
	& \qw & \qw & \qw & \qw & \qw & \control \qw & \qw & \qw & \qw & \gate{\mathcal{M}_{X}} \\
	*{\vphantom{\rule{0em}{1em}}}\\ 
	\lstick{\ket{+}} & \qw & \qw & \qw & \qw & \qw & \qw & \control \qw & \qw & \qw & \qw & *+<.6em>{\vphantom{{\mathcal{M}_{X}}}}\qw  \\
	\lstick{\ket{+}} & \qw & \qw & \qw & \qw & \qw & \qw & \qw & \control \qw & \qw & \qw & *+<.6em>{\vphantom{{\mathcal{M}_{X}}}}\qw  \\
	\lstick{\ket{+}} & \qw & \qw & \qw & \qw & \qw & \qw & \qw & \qw & \control \qw & \qw & *+<.6em>{\vphantom{{\mathcal{M}_{X}}}}\qw  \\
	*{\vphantom{\rule{0em}{1em}}}\\
	\lstick{\ket{+}} & \ctrl{-12} & \ctrl{-11} & \ctrl{-10} & \ctrl{-8} & \ctrl{-7} & \ctrl{-6} & \ctrl{-4} & \ctrl{-3} & \ctrl{-2} & \gate{\mathcal{M}_{X}} \\
}}
\end{center}
\caption{\label{fig:cnot}
Logical {\sc cnot} gate acting on two blocks of the repetition code, shown here for code length $n=3$. Ancilla qubits are prepared in the $X=1$ eigenstate $|+\rangle$, interact via {\sc cphase} gates with the data qubits, then are measured in the $X$ basis. (These preparations, gates, and measurements are repeated several times, and the result is determined by a majority vote; the measurement repetition is not shown.) Finally, each qubit in the two input blocks is measured in the $X$ basis, the results are decoded by a majority vote in each block, and logical Pauli errors in the output blocks are inferred from the results.}
\end{figure}
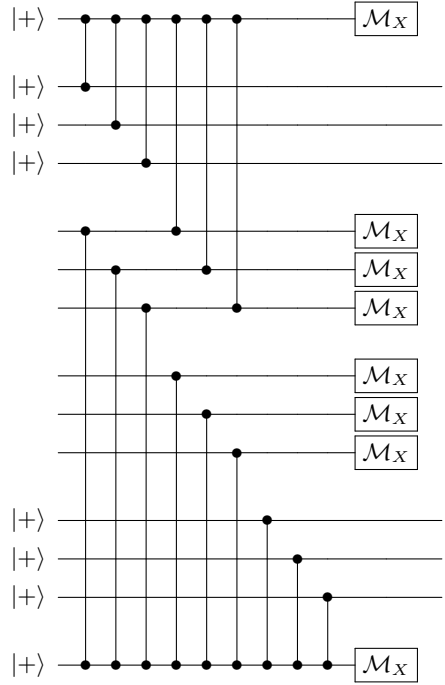

The crucial element of the construction in \cite{AliferisPreskill2008,Aliferis2009} is a ``teleported'' encoded {\sc cnot} gate acting on blocks of a repetition code, shown in Fig.~\ref{fig:cnot}. (In \cite{Brooks2012} this construction is extended to Bacon-Shor codes.) The probability of a logical error in the {\sc cnot} gate can be bounded above as \cite{AliferisPreskill2008}
\be
&&\varepsilon_{\mathrm{CNOT}} \le 4{n \choose  \frac{n+1}{2}}\left(3n\varepsilon_g+2\varepsilon_m\right)^{(n+1)/2} +7n^2\varepsilon_g'\nonumber\\
&& \quad \le \sqrt{\frac{8}{\pi n}}\left( 12 n\varepsilon_g + 8 \varepsilon_m\right)^{(n+1)/2}+7n^2\varepsilon_g',
\ee
where $\varepsilon_g$ is the probability of a dephasing error and $\varepsilon_g'$ is the probability of a bit flip error in the {\sc cphase} gate, and $\varepsilon_m$ is the error probability in a $|+\rangle$ preparation or an $X$ measurement. Here $n$ is the length of the repetition code, and we assume that each measurement is repeated $n$ times; the prefactor of 4 arises because the {\sc cnot} gadget contains four measurements decoded by majority vote, any of which might fail, and the factor $3n\varepsilon_g+2\varepsilon_m$ is an upper bound on the probability of error for each bit in each decoded block, neglecting the bit flip errors. (The second inequality is obtained from the Stirling approximation.) The term $7n^2\varepsilon_g'$ is an upper bound on the probability that one or more {\sc cphase} gates in the {\sc cnot} gadget have bit flip errors.

Just to illustrate the robustness of the {\sc cnot} gadget with respect to measurement and preparation errors (assuming bit flips are highly suppressed), note that for $\varepsilon_m = .01$, $\varepsilon_g = 10^{-5}$, and $\varepsilon_g'= 10^{-9}$, by choosing $n=11$ we find $\varepsilon_{\rm CNOT} < 10^{-6}$. The analysis in \cite{AliferisPreskill2008} then shows that, by applying the state distillation ideas from \cite{BravyiKitaev2005}, scalable quantum computing can achieved by augmenting the CNOT gadget with $|-i\rangle$ and $|e^{-i\pi/4}\rangle$ preparations having the relatively high error rate $\varepsilon_m$. This example indicates one possible way in which noisy measurements can be tolerated if appropriate gates are highly reliable.

\section{Conclusions}
\label{sec:conclusions}

A ``0-$\pi$ qubit,'' is a two-lead superconducting device whose energy is minimized when the superconducting phase difference between the leads is either 0 or $\pi$. This qubit, if properly designed, can be very robust with respect to weak local noise \cite{DoucotVidal2002,IoffeFeigelman2002,Kitaev2006}. In this paper, we have taken it for granted that  near-perfect 0-$\pi$ qubits are attainable, and have asked whether quantum information encoded in such qubits can be processed fault tolerantly. Our conclusion, fleshing out a suggestion in \cite{Kitaev2006}, is that highly accurate nontrivial quantum gates can be executed by coupling one or two qubits to a superconducting $LC$ oscillator with very large inductance. In principle the gate error becomes exponentially small when $\sqrt{L/C}$ is large compared to $\hbar/4e^2 \approx 1.03 ~ k\Omega$, where $L$ is the inductance and $C$ is the capacitance of the oscillator. 

We have estimated the gate accuracy using both analytic arguments and numerical simulations. The analytic arguments use approximations, in particular for the analysis of errors due to diabatic transitions and squeezing, that are validated by the simulations. The point of the simulations is not necessarily to capture fully the behavior of realistic devices, but rather to verify that the analytic arguments are on solid ground. The analysis applies to any sufficiently robust 0-$\pi$ qubit, regardless of its internal structure. 

The protected gates are the single-qubit phase gate $\exp\left(i\frac{\pi}{4}Z \right)$ and the two-qubit phase gate $\exp\left(i\frac{\pi}{4}Z\otimes Z \right)$. In both cases, the oscillator starts out in a low-lying energy state; then turning on a tunable Josephson coupling between the qubit(s) and the oscillator prepares a state of the oscillator protected by a continuous variable quantum error-correcting code. The coupling is kept on for a specified time, during which the code state acquires a nontrivial Berry phase, inducing a nontrivial encoded operation, with error probability exponentially small in $\sqrt{L/C}$. Then the coupling turns off, returning the oscillator to a low-lying state slightly different from the initial one. 

The ramping of the coupling on and off takes place over a time scale $\tau_J = O(\sqrt{LC})$, chosen to achieve an optimal compromise between errors due to diabatic transitions (which favor large $\tau_J$) and errors due to squeezing (which favor small $\tau_J$). The gates are robust against generic small errors in the Hamiltonian and thermal effects, due to the good error-correcting properties of the quantum code. Entropy due to noise is mostly absorbed by the oscillator, inflicting little damage on the 0-$\pi$ qubit. These protected phase gates are not a universal gate set by themselves, but a universal fault-tolerant scheme can be built from the protected gates together with single-qubit measurements and noisy unprotected phase gates. 

We do not know whether a very robust 0-$\pi$ qubit, and/or the superinductive $LC$ circuit needed for protected phase gates, will prove to be feasible in superconducting devices. We hope that other settings will be found in which highly reliable quantum gates can be realized by using tunable couplings between qubits and oscillators.

\acknowledgments
We thank David DiVincenzo for helpful discussions. 
This work was supported in part by the Intelligence Advanced Research Projects Activity (IARPA) via Department of Interior National Business Center contract number D11PC20165. The U.S. Government is authorized to reproduce and distribute reprints for Governmental purposes notwithstanding any copyright annotation thereon. The views and conclusions contained herein are those of the author and should not be interpreted as necessarily representing the official policies or endorsements, either expressed or implied, of IARPA, DoI/NBC
or the U.S. Government.
We also acknowledge support from NSF grant PHY-0803371, DOE grant DE-FG03-92-ER40701, and NSA/ARO grant W911NF-09-1-0442. The Institute for Quantum Information and Matter (IQIM) is an NSF Physics Frontiers Center with support from the Gordon and Betty Moore Foundation. 

\appendix

\section{Adiabatic switch}
\label{app:switch}
Here we describe the static properties of a single rung of the two-rung circuit depicted in Fig.~\ref{fig:four-leads}. This single-rung circuit is of independent interest; because it's effective Josephson energy depends sensitively on circuit parameters, it can serve as an ``adiabatic switch'' which turns on and off the coupling between superconducting qubits.

\begin{figure}[!ht]
\begin{center}
\includegraphics[width=0.4\textwidth]{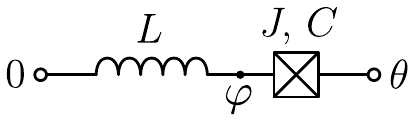}
\end{center}
\caption{\label{fig:LJC} Adiabatic switch. For $\sqrt{L/C} \gg 1$, the effective Josephson energy $J_{\rm eff}$ of the switch is depends sensitively on circuit parameters.}
\end{figure} 

The circuit shown in Fig.~\ref{fig:LJC} is described by the Hamiltonian
\begin{equation}\label{eq:CLJ}
H=\frac{Q^{2}}{2C}+\frac{\varphi^{2}}{2L}-J\cos(\varphi-\theta),\quad 
\text{where}\quad Q=-i\frac{\partial}{\partial\varphi}.
\end{equation}
The dynamical variable is the superconducting phase $\varphi$ at the indicated point, and the phase difference $\theta$ between the leads is fixed; $Q$ is the electric charge operator in units of $2e$. While $\theta$ is defined modulo $2\pi$, the variable $\varphi$ is a real number because the states $|\varphi\rangle$ and $|\varphi+2\pi\rangle$ differ by the phase winding in the inductor. 
We choose units such that $\hbar=1$ and express the capacitance and inductance in rationalized units such that $C^{-1}$ and $L^{-1}$ have dimensions of energy (or inverse time). Specifically,
\begin{equation}
C=\frac{C_{\conv}}{(2e)^2},\qquad\quad
L=\frac{L_{\conv}}{(\hbar/2e)^2},
\end{equation}
where the subscript ``conv'' refers to conventional units. We can also relate the parameters $J$, $C$ to the commonly used charging energy $E_C$ and Josephson energy $E_J$:
\begin{equation}
E_C=\frac{e^2}{2C_{\conv}}=\frac{1}{8C},\qquad\quad
E_J=J_{\conv}=J.
\end{equation}
In the main text of the paper we have studied this same Hamiltonian, where $\theta=0$ or $\pi$ and $J$ pulses on and off during the execution of a quantum gate. But here we focus on the case where $J$ is time independent.

In the d.c.\ regime, the whole device behaves like an effective Josephson element --- its ground state energy $E(\theta)$ depends on the phase difference $\theta$ between the leads. The effective Josephson energy can be characterized by
\begin{equation}
J_{\eff}=E''(0)=\left.\frac{\partial^{2}E}{\partial\theta^2}\right|_{\theta=0}.
\end{equation}
We will see that this number varies by orders of magnitude when the circuit parameters change in a much narrower range. 

Suppose that $C$ and $L$ are fixed, while $J$ may vary from very small to very large values. In the limit $J\to 0$, we may treat the Josephson energy as a small perturbation of the $LC$ oscillator, obtaining 
\begin{equation}
E(\theta)=-J_{\eff}\cos\theta+\const,
\end{equation}
where
\begin{equation}\label{eq:cosPhi}
J_{\eff}=J\langle\cos\varphi\rangle,
=Je^{-\langle\varphi^2\rangle/2}=J\exp\left(-\tfrac{1}{4}\sqrt{L/C}\right).
\end{equation}
In the opposite limit $J\to\infty$, the dynamical phase is locked: $\varphi\equiv\theta\pmod{2\pi}$, and therefore
\begin{equation}
E(\theta)=\min_{n}\frac{(\theta+2\pi n)^2}{2L}+\const,\quad
J_{\eff}=L^{-1}.
\end{equation}
If $L$ is large, the effective Josephson coupling is suppressed in both limits, but the suppression is exponential only in the limit $J\to 0$.

We say that the circuit is superinducting if the dimensionless characteristic impedance is large, or equivalently if the impedance in conventional units is large compared to the superconducting impedance quantum  $R_Q = \hbar/(2e)^2\approx 1.03\,\mathrm{k\Omega}$. Reaching this superinducting regime is a significant engineering challenge, quite hard to achieve using geometric inductance (except perhaps by constructing a coil with a very large number of turns). Indeed, the inductance of a loop of wire is accompanied by a parasitic capacitance such that $\sqrt{L_{\conv}/C_{\conv}}\sim 4\pi/c\approx 377\,\Omega$ (where $c$ is the speed of light). This is the impedance of the free space, smaller than $R_Q$ by the factor $16\pi\alpha$ where $\alpha = e^2/\hbar c \approx 1/137$ is the fine structure constant. One possible way to realize a superinductor is to build a long chain of Josephson junctions \cite{Devoret2009,Masluk2012,Gershenson2012}. Another is to use a long wire, thick enough to suppress phase slips, built from a material with large kinetic inductance. 

In the case where $\sqrt{L/C}\gg 1$, we can compute the ground state energy of the Hamiltonian Eq.~(\ref{eq:CLJ}) semi-analytically. First we note that in the limit $L\to \infty$  the problem reduces to particle moving in a periodic potential:
\begin{equation}\label{eq:CJ} 
H'=\frac{Q^{2}}{2C}-J\cos\widetilde{\varphi},\qquad
\text{where}\quad \widetilde{\varphi}=\varphi-\theta.
\end{equation}
This approximate Hamiltonian $H'$ preserves the quasimomentum $q=Q\bmod 1$, and therefore can be solved using Bloch wave functions. In the lowest Bloch band, the energy can be expressed as a function of the quasimomentum $q\in \left[-\tfrac{1}{2},\tfrac{1}{2}\right]$ in two different limits:
\begin{align}
&JC\ll 1: &&
\eps(q)=\frac{q^2}{2C},\\
&JC\gg 1: &&
\eps(q)=-J+\frac{\omega}{2}-2\lambda\cos(2\pi q),
\label{eq:tunamp}
\end{align}
where
\begin{align}\label{eq:tunnel-amp}
\omega=\sqrt{J/C},\quad
\lambda=\frac{4}{\sqrt{\pi}}\,J^{3/4}C^{-1/4}\,e^{-8\sqrt{JC}}.
\end{align}
In the case $JC\gg 1$, the system stays near a minimum of the cosine potential at $\widetilde{\varphi}=2\pi n$ and  occasionally tunnels to an adjacent minimum through the potential barrier. In Eq.~(\ref{eq:tunnel-amp}), $\omega$ is the angular frequency for small oscillations about the potential minimum, and $\lambda$ is the amplitude tunneling amplitude. 

To compute $\lambda$ we recall the semiclassical analysis \cite{Coleman} for a particle of mass $m$ tunneling through a symmetric double-well potential $V(x)$ with minima at $x=a,b$, which yields
\begin{equation}\label{eq:tungen}
\lambda=\frac{\omega}{\sqrt{\pi}}e^{-S+\omega\tau/2},
\end{equation}
where
\begin{equation}
S=\int_{a}^{b}dx\sqrt{2mV(x)},\quad
\tau=\int_{a+\Delta x}^{b-\Delta x}dx\sqrt{\frac{m}{2V(x)}};
\end{equation}
here $\omega=\sqrt{V''(a)/m}$ and $\Delta x=(V''(a)\,m)^{-1/4}$ is the width of the ground state localized around the potential minimum. We obtain Eq.~(\ref{eq:tunamp}) using $m=C$ and $V(\widetilde{\varphi})=J(1-\cos\widetilde{\varphi})$.

\begin{figure}[!ht]
\begin{center}
\includegraphics[width=0.5\textwidth]{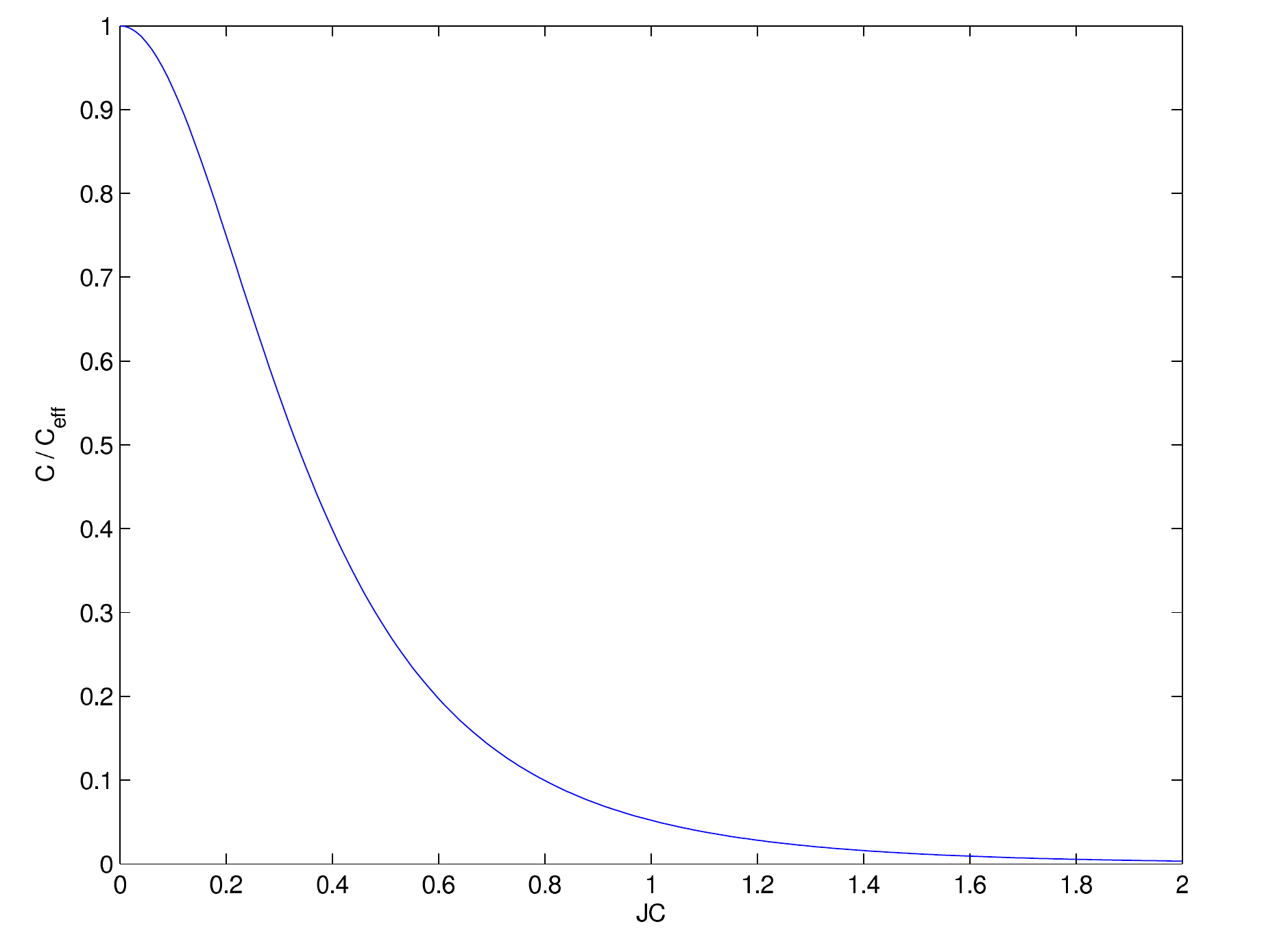}
\end{center}
\caption{\label{fig:Ceff} (Color online) The inverse effective capacitance of a Josephson junction as a function of $JC$.}
\end{figure}

Defining the effective capacitance $C_{\rm eff}$ by
\begin{equation}
\frac{1}{C_{\eff}}=\eps''(0),
\end{equation}
we obtain from Eq.~(\ref{eq:tunamp},\ref{eq:tunnel-amp}) the asymptotic values
\begin{equation}
\frac{C}{C_{\eff}}=\left\{\begin{array}{l@{\quad}l}
1, &\text{if } JC\ll 1,\\
32\pi^{3/2}(JC)^{3/4}e^{-8\sqrt{JC}}, &\text{if } JC\gg 1.
\end{array}\right.
\end{equation}
The numerically computed value of $C_{\rm eff}$ as a function of $JC$ is plotted in Fig.~\ref{fig:Ceff}.

Now we return to the original Hamiltonian $H$ in Eq.~(\ref{eq:CLJ}). For small but nonzero $L^{-1}$. the quasimomentum $q$, though not exactly conserved, is a slow variable which can be treated using the adiabatic approximation. We obtain the effective Hamiltonian
\begin{equation}
H_{\eff}=\frac{\varphi^2}{2L}+\eps(q),\qquad
\text{where}\quad \varphi=i\frac{\partial}{\partial q}+\theta.
\end{equation}
This problem is similar to the one we have just solved, Eq.~(\ref{eq:CJ}); now the variable $q$ is periodic (defined modulo $1$), and the parameter $\theta$ plays the role of quasimomentum. 

If $\sqrt{L/C_{\eff}}\gg 1$, {\em i.e.}\ $JC\ll\ln(L/C)$, the ground state energy is given by the formula $E(\theta)=-J_{\eff}\cos\theta+\const$, where $J_{\eff}$ is twice the tunneling amplitude. This expression is analogous to Eq.~(\ref{eq:tunamp}), except that now we consider tunneling through the periodic ``potential'' $\eps(q)$ from an integer value of $q$ to an adjacent integer value. From Eq.~(\ref{eq:tungen}) we find
\begin{equation}\label{eq:Jeff-tunnel}
J_{\eff}=\nu\,C_{\eff}^{-3/4}L^{-1/4} \exp\left(-\mu\sqrt{L/C_{\eff}}\right),
\end{equation}
where 
\begin{align}
\mu&=2\int_{0}^{1/2}\sqrt{2C_{\eff}\bigl(\eps(q)-\eps(0)\bigr)}\,dq,\\
\nu&=\frac{1}{\sqrt{\pi}}\,\exp\left(\int_{0}^{1/2}
\Bigl(\bigl(2C_{\eff}\bigl(\eps(q)-\eps(0)\bigr)\bigr)^{-1/2}
-q^{-1}\Bigr)dq\right).
\end{align}
The parameters $\mu$ and $\nu$ are numbers of order $1$: As $JC$ increases from zero to infinity,
$\mu$ changes from $1/4=0.25$ to $2/\pi^2\approx 0.2026$, and $\nu$ changes from $1/\sqrt{\pi}\approx 0.564$ to $4/\pi^{3/2}\approx 0.718$.

For Eq.~(\ref{eq:Jeff-tunnel}) to apply we also need that $J$ not be too small: $JC\gg(L/C)^{-1/4}$. For smaller $J$, the adiabatic approximation breaks down near the point $q=\pm 1/2$ where two branches of the parabola $\eps(q)=q^2/(2C)$ meet each other. However, the exponential factor $e^{-S}$ is still correct and coincides with $\langle\cos\varphi\rangle$ from Eq.~(\ref{eq:cosPhi}). 

\begin{figure}[!ht]
\begin{center}
\includegraphics[width=0.5\textwidth]{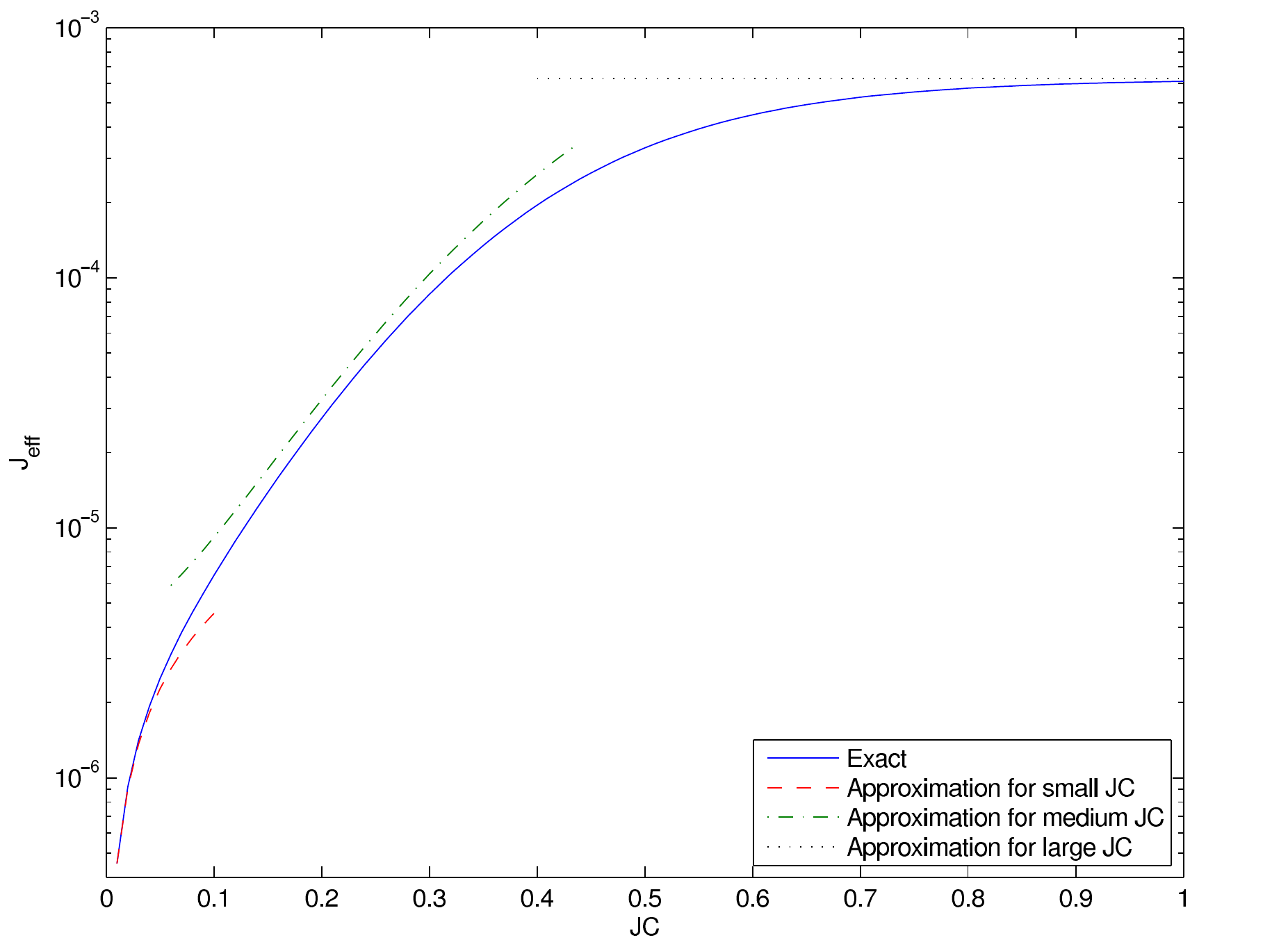}
\end{center}
\caption{\label{fig:Jeff} (Color online) The effective Josephson parameter of the adiabatic switch as a function of $JC$ for $\sqrt{L/C}=40$.}
\end{figure}

To summarize, we have calculated the effective Josephson coupling in three different regimes, finding
\begin{align}
\label{eq:Jeff_smallJ}
JC&\ll(L/C)^{-1/4}:\nonumber\\
&
J_{\eff}=J\exp\left(-\tfrac{1}{4}\sqrt{L/C}\right),&&\\[3pt]
\label{eq:Jeff_mediumJ}
(L/C)^{-1/4}&\ll JC\ll\ln(L/C):\nonumber\\ 
&
J_{\eff}=\nu\,C_{\eff}^{-3/4}L^{-1/4}
\exp\left(-\mu\sqrt{L/C_{\eff}}\right),&&\\[3pt]
\label{eq:Jeff_largeJ}
JC&\gg\ln(L/C):\nonumber\\
&
J_{\eff}=L^{-1}.&&
\end{align}
The numerical results for $\sqrt{L/C}=40$, together with plots of Eqs.~(\ref{eq:Jeff_smallJ}--\ref{eq:Jeff_largeJ}), are shown in Figure~\ref{fig:Jeff}.

\section{Quantifying the gate error}
\label{app:error}

The protected phase gate is executed by coupling the qubit to an oscillator for a prescribed time interval. We assume that the initial state of qubit and oscillator is a product state $|\psi\rangle\otimes |\psi_{\rm in}\rangle$, where $|\psi\rangle = a|0\rangle + i b|1\rangle$ is the initial (normalized) state of the qubit and $|\psi_{\rm init}\rangle$ is the initial state of the oscillator. After the coupling between qubit and oscillator is switched off, the joint state of qubit and oscillator becomes
\begin{equation}
|\psi'\rangle =a|0\rangle\otimes|\psi_0\rangle +b|1\rangle\otimes |\psi_1\rangle,
\end{equation}
where $\langle\psi_0|\psi_0\rangle = \langle\psi_1|\psi_1\rangle=1$.
We assume that there are no bit flip errors, but there may be a phase error. If the gate is ideal, then $\langle \psi_0|\psi_1\rangle=1$, and the gate rotates the phase of $|1\rangle$ by the angle $-\pi/2$ relative to the phase of $|0\rangle$. We wish to quantify the error, using some appropriate measure of the deviation of $\langle \psi_0|\psi_1\rangle$ from 1.

\subsection{Fidelity}
One way to quantify the error is to use the fidelity of the actual state with the ideal state. Tracing out the oscillator, we obtain the final density operator for the qubit
\begin{eqnarray}
\rho &=& {\rm tr}_{\rm osc}|\psi'\rangle\langle\psi'|
= \left( \begin{array}{cc}
\langle\psi_0|\psi_0\rangle |a|^2 & \langle\psi_1|\psi_0\rangle ab^* \\
\langle\psi_0|\psi_1\rangle a^*b & \langle\psi_1|\psi_1\rangle |b|^2 \end{array} \right)\nonumber\\
&=& \left( \begin{array}{cc}
|a|^2 & Ce^{-i\phi}ab^* \\
Ce^{i\phi}a^*b & |b|^2 \end{array} \right),
\end{eqnarray}
where
\begin{equation}
\langle\psi_0|\psi_1\rangle = Ce^{i\phi},
\end{equation}
and $C=|\langle\psi_0|\psi_1\rangle|$ is real and nonnegative.
The fidelity $F$ with the ideal state $|\psi_{\rm ideal}\rangle = a|0\rangle + b|1\rangle$ is
\begin{eqnarray}
F &=& \langle \psi_{\rm ideal}|\rho|\psi_{\rm ideal}\rangle \nonumber\\
&=& |a|^4 +|a|^2|b|^2\left(Ce^{i\phi} + Ce^{-i\phi}\right)+|b|^4\nonumber\\
&=&\left(|a|^2+|b|^2\right)^2 -2\left(1-C\cos\phi\right)|a|^2|b|^2
\end{eqnarray}
Thus the ``infidelity'' (the deviation of $F$ from 1) is maximal when $|a|^2=|b|^2=1/2$, and we conclude that
\begin{equation}
1-F \le \frac{1}{2}\left(1-C\cos\phi\right).
\end{equation}
Denoting $\delta = 1-C$, we have
\begin{equation}
1-F\approx \frac{1}{2}\delta + \frac{1}{4}\phi^2.
\end{equation}
assuming $\delta,\phi \ll 1$.

\subsection{Trace norm}

Another useful measure is the deviation of $\rho$ from the ideal density operator $\rho_{\rm ideal}=|\psi_{\rm ideal}\rangle\langle \psi_{\rm ideal}|$ in the trace norm. We see that
\begin{eqnarray}\label{eq:rho-rho-ideal}
\rho_{\rm ideal} -\rho 
= \left( \begin{array}{cc}
0 & \left(1-Ce^{-i\phi}\right)ab^* \\
\left(1-Ce^{i\phi}\right) a^*b & 0 \end{array} \right),
\end{eqnarray}
whose eigenvalues
\begin{equation}
\pm ~|a|~|b|~\sqrt{1+C^2-2C\cos\phi}
\end{equation}
have maximal absolute value for $|a|=|b| = 1/\sqrt{2}$. Hence the trace norm satisfies
\begin{eqnarray}
\|\rho_{\rm ideal} -\rho \|_1 &\le& \sqrt{1+C^2-2C\cos\phi}\nonumber\\
&=&\left| 1-\langle\psi_0|\psi_1\rangle\right|.
\end{eqnarray}
Denoting $\delta=1-C$, we have
\begin{eqnarray}
\|\rho_{\rm ideal} -\rho \|_1 &\le& \sqrt{(1-C)^2+2C(1-\cos\phi)}\nonumber\\
&\approx& \sqrt{\delta^2 +\phi^2},
\end{eqnarray}
assuming $\delta,\phi \ll 1$.

\subsection{Kraus operators}

It is also useful to have a Kraus operator decomposition of the noise process. We may define states $|\psi_0'\rangle$ and $|\psi_1'\rangle$ such that
\begin{equation}
|\psi_0\rangle = e^{-i\phi/2}|\psi_0'\rangle,\quad |\psi_1\rangle = e^{i\phi/2}|\psi_1'\rangle
\end{equation}
and therefore
\begin{equation}
\langle\psi_0'|\psi_1'\rangle =C \ge 0.
\end{equation}
In terms of these states,
\begin{eqnarray}
|\psi'\rangle &=&a|0\rangle\otimes|\psi_0\rangle +b|1\rangle\otimes |\psi_1\rangle \nonumber\\
&=& \left(e^{-i\phi/2}a|0\rangle +be^{i\phi/2}|1\rangle\right)\otimes \frac{1}{2}\left(|\psi_0'\rangle + |\psi_1'\rangle\right) \nonumber\\
&+& \left(e^{-i\phi/2}a|0\rangle -be^{i\phi/2}|1\rangle\right)\otimes \frac{1}{2}\left(|\psi_0'\rangle -|\psi_1'\rangle\right),\nonumber\\
\end{eqnarray}
where
\begin{eqnarray}
&&\left\|\frac{1}{2}\left(|\psi_0'\rangle \pm |\psi_1'\rangle\right)\right\|=\sqrt{\frac{1}{2}(1\pm C)},\nonumber\\
&&\left(\langle \psi_0'| + \langle\psi_1'|\right)\left(|\psi_0'\rangle - |\psi_1'\rangle\right)=0.
\end{eqnarray}
Therefore,
\begin{eqnarray}\label{eq:cal-N-define}
\rho = {\cal N}(\rho_{\rm ideal})=M_0\rho_{\rm ideal}M_0^\dagger + M_1\rho_{\rm ideal}M_1^\dagger,
\end{eqnarray}
where
\begin{eqnarray}
M_0 &=& \sqrt{\frac{1}{2}(1+C)}
\left( \begin{array}{cc}
e^{-i\phi/2} & 0 \\
0 & e^{i\phi/2} \end{array} \right), \nonumber\\
M_1 &=& \sqrt{\frac{1}{2}(1-C)}
\left( \begin{array}{cc}
e^{-i\phi/2} & 0 \\
0 & -e^{i\phi/2} \end{array} \right).
\end{eqnarray}
Note that we have rotated away the ideal phase gate in our definition of the noise operation ${\cal N}$, so that ${\cal N}=I$ corresponds to the ideal gate. 

\subsection{Diamond norm}

In some versions of the quantum accuracy threshold theorem, the strength of Markovian noise is characterized by the deviation 
\begin{equation}
\varepsilon = \| {\cal N}- U\|_\diamond .
\end{equation}
of a noisy gate ${\cal N}$ from the corresponding ideal gate $U$ in the ``diamond norm'' \cite{KitaevDiamond}. The advantage of the diamond norm is that it quantifies the damage inflicted by an operation that acts on a subsystem that might be entangled with a complementary subsystem, {\em e.g.}, a noisy gate acting on a qubit or pair of qubits that is entangled with the rest of the qubits in a quantum computer.
The diamond norm $\|{\cal E}\|_\diamond$ is defined as the $L^1$ norm of the extended operator ${\cal E} \otimes I$; that is
\begin{equation}
\|{\cal E}\|_\diamond = \max_\sigma \|{\cal E}\otimes I(\sigma)\|_1.
\end{equation}
If ${\cal E}$ acts on a Hilbert space ${\cal H}$ of dimension $d$, then $I$ denotes the identity operator acting on another Hilbert space ${\cal H}'$ of dimension $d$, and $\sigma$ is a state on ${\cal H}\otimes{\cal H}'$. 


For the operation defined by Eq.~(\ref{eq:cal-N-define}), the two-qubit state $\sigma$ that maximizes the $L^1$ distance between $({\cal N}\otimes I)(\sigma)$ and $\sigma$ is a maximally entangled pure state, which we may choose to be $|\phi^+\rangle=\frac{1}{\sqrt{2}}\left(|00\rangle + |11\rangle\right)$. Letting ${\cal N}$ act on the first qubit, we obtain an ensemble of two pure states, 
\begin{eqnarray}
&\frac{1}{\sqrt{2}}\left(e^{-i\phi/2}|00\rangle + e^{i\phi/2} |11\rangle\right), \quad &{\rm prob}=\frac{1}{2}(1+C),\nonumber\\
&\frac{1}{\sqrt{2}}\left(e^{-i\phi/2}|00\rangle - e^{i\phi/2} |11\rangle\right), \quad &{\rm prob}=\frac{1}{2}(1-C),\nonumber\\
\end{eqnarray}
and the density operator can be expressed as a $2\times 2$ matrix acting on the span of $|00\rangle$ and $|11\rangle$:
\begin{widetext}
\begin{eqnarray}
{\cal N}\otimes I(|\phi^+\rangle\langle \phi^+|)
= \left( \begin{array}{cc}
\frac{1}{2} & \frac{1}{2}Ce^{-i\phi} \\
\frac{1}{2}Ce^{i\phi} & \frac{1}{2} \end{array} \right),
\quad
({\cal N}\otimes I-I\otimes I)(|\phi^+\rangle\langle \phi^+|)
= \left( \begin{array}{cc}
0 & \frac{1}{2}(Ce^{-i\phi} -1)\\
\frac{1}{2}(Ce^{i\phi}-1) & 0 \end{array} \right).
\end{eqnarray}
\end{widetext}
Comparing with Eq.~(\ref{eq:rho-rho-ideal}) in the case $a=b=\frac{1}{\sqrt{2}}$, we find
\begin{eqnarray}
&&\|{\cal N}-I\|_\diamond = \max \|\rho_{\rm ideal} -\rho \|_1 \nonumber\\
&&= \sqrt{1+C^2-2C\cos\phi}
= \left| 1-\langle\psi_0|\psi_1\rangle\right|.
\end{eqnarray}
Evidently, extending ${\cal E}$ to ${\cal E}\otimes I$ does not increase its maximal $L^1$ norm; hence the noisy gate's deviation from the ideal gate in the diamond norm coincides with the maximal trace distance deviation of the density operator $\rho$ from the ideal density operator $\rho_{\rm ideal}$.

The accuracy of the two-qubit phase gate can be analyzed in the same way. Now the final state of the oscillator depends on the total phase difference across a pair of 0-$\pi$ qubits; it is $|\psi_0\rangle$ for the two-qubit states $|00\rangle$ and $|11\rangle$, and $|\psi_1\rangle$ for the two-qubit states $|01\rangle$ and $|10\rangle$. The eigenvalues of $\rho_{\rm ideal} - \rho$ become doubly degenerate, and hence the diamond norm deviation from the ideal gate is twice as large as for the single-qubit phase gate.

Both the fidelity and the diamond norm are useful measures of the gate error. The significant difference is that the diamond norm deviation ($\approx \sqrt{(1-C)^2+ \phi^2}$) is linear in $\phi$ (for $\phi$ small and $C\approx 1$), while the infidelity is quadratic in $\phi$. The threshold theorem establishes a sufficient condition for scalable quantum computing, expressed as an upper bound on the diamond norm, and it applies under the pessimistic assumption that phase errors accumulate linearly with the circuit size. But if the phase errors are actually random, we might expect them to add in quadrature, and in that case the infidelity may be a more appropriate way to quantify the gate error. 

\medskip

\section{Grid states}
\label{app:overrotation}

Here we provide additional details concerning the properties of grid states, which were omitted from the discussion in Sec.~\ref{sec:overrotation}. 

\subsection{Approximate codewords in $\varphi$ space and $Q$ space}

Let $f$ denote a narrow function in $\varphi$ space, and $\tilde F$ denote a broad envelope function in $\varphi$ space. We express the approximate codewords of the continuous variable code as
\begin{widetext}
\begin{eqnarray}
|0_C\rangle &=& \sqrt{2\pi}~\sum_{n~{\rm even}} \tilde F(\pi n) T(\pi n)\int d\varphi~f(\varphi)|\varphi\rangle,\nonumber\\
|1_C\rangle &=& \sqrt{2\pi}~\sum_{n~{\rm odd}}\tilde F(\pi n) T(\pi n) \int d\varphi~f(\varphi)|\varphi\rangle,
\end{eqnarray}
\end{widetext}

where $T(a)$ denotes the  $\varphi$ translation operator whose action is $T(a)|\varphi\rangle = |\varphi + a\rangle$. The function $f$ is normalized so that
\begin{equation}
\int d\varphi~|f(\varphi)|^2 = 1,
\end{equation}
and if the overlap between peaks centered at distinct integer multiples of $\pi$ can be neglected, then $|0_C\rangle$ and $|1_C\rangle$ are normalized provided
\begin{equation}
2\pi\sum_{n~{\rm even}} |\tilde F(\pi n)|^2 \approx 1,\quad 2\pi \sum_{n~{\rm odd}}|\tilde F(\pi n)|^2\approx 1.
\end{equation}

The approximate codewords in the conjugate basis are
\begin{eqnarray}
|+_C\rangle &=& \frac{1}{\sqrt{2}}\left(|0_C\rangle + |1_C\rangle\right)\nonumber\\
&=&\sqrt{\pi}~\sum_n \tilde F(\pi n) T(\pi n)\int d\varphi~f(\varphi)|\varphi\rangle,\nonumber\\
|-_C\rangle &=& \frac{1}{\sqrt{2}}\left(|0_C\rangle - |1_C\rangle\right)\nonumber\\
&=&\sqrt{\pi}~\sum_n\tilde F(\pi n) (-1)^nT(\pi n) \int d\varphi~f(\varphi)|\varphi\rangle,\nonumber\\
\end{eqnarray} 
where
\begin{equation}
\pi\sum_n |\tilde F(\pi n)|^2\approx 1.
\end{equation}
To express these states in the $Q$ basis we use
\be
\int d\varphi ~f(\varphi)|\varphi\rangle &=& \int dQ\int d\varphi ~f(\varphi) |Q\rangle\langle Q|\varphi\rangle \nonumber\\
&=& \int dQ ~\tilde f(Q)|Q\rangle,
\ee
where
\begin{eqnarray}\label{eq:Fourier-define}
\tilde f(Q) &=& \frac{1}{\sqrt{2\pi}}\int d\varphi ~e^{-iQ\varphi}f(\varphi),\nonumber\\
 f(\varphi) &=& \frac{1}{\sqrt{2\pi}}\int dQ ~e^{iQ\varphi}\tilde f(Q);\quad 
\end{eqnarray}
since the function $f$ is narrow in $\varphi$ space, its Fourier transform $\tilde f$ is broad in $Q$ space.
The translation operator is represented in $Q$ space as $T(a)= \exp(-iQa)$, and therefore 
\begin{eqnarray}\label{eq:plus-minus-codewords}
|+_C\rangle &=&  \sqrt{\pi}~\sum_n \tilde F(\pi n) T(\pi n)\int d\varphi~f(\varphi)|\varphi\rangle \nonumber\\
&=& \sqrt{\pi}~\sum_n \tilde F(\pi n) \int dQ~e^{-i\pi nQ}\tilde f(Q)|Q\rangle,\nonumber\\
|-_C\rangle &=& \sqrt{\pi}~\sum_n\tilde F(\pi n) (-1)^nT(\pi n) \int d\varphi~f(\varphi)|\varphi\rangle\nonumber\\
&=& \sqrt{\pi}~\sum_n \tilde F(\pi n) \int dQ~e^{-i\pi n(Q-1)}\tilde f(Q)|Q\rangle.\nonumber\\
\end{eqnarray} 

Now we reverse the order of the summation and integration, and use the Poisson summation formula, in the form (see Appendix \ref{app:poisson}):
\begin{equation}\label{eq:poisson-summation}
\sqrt{\pi}~\sum_n e^{-i\pi n Q }\tilde F(\pi n) = \sqrt{2}~\sum_m F(Q-2m),
\end{equation}
where
\begin{eqnarray}\label{eq:2-pi-fourier}
F(Q) &=& \frac{1}{\sqrt{2\pi}}\int d\varphi~e^{-iQ\varphi}\tilde F(\varphi), \nonumber\\
 \tilde F(\varphi)&=& \frac{1}{\sqrt{2\pi}}\int dQ ~e^{iQ\varphi }F(Q).
\end{eqnarray}
The codewords $|+_C\rangle$ and $|-_C\rangle$ can now be expressed as
\begin{eqnarray}\label{eq:codeword-plus-minus-F}
|+_C\rangle &   = &\sqrt{2}\int dQ~\tilde f(Q)\sum_{m~{\rm even}} F(Q -m) |Q\rangle,\nonumber\\
|-_C\rangle &= & \sqrt{2}\int dQ~\tilde f(Q)\sum_{m~{\rm odd}} F(Q-m) |Q\rangle.\nonumber\\
\end{eqnarray} 
Here $F$ is a narrow function centered at zero in $Q$ space, so that $|+_C\rangle$ has support near even integer values of $Q$, and $|-_C\rangle$ has support near odd integer values of $Q$. Because $\tilde f(Q)$ is slowly varying, it is nearly constant within each peak, and the codewords can be well approximated as
\begin{eqnarray}
|+_C\rangle &   \approx &\sqrt{2}~\sum_{m~{\rm even}} \tilde f(m) \int dQ~ F(Q -m) |Q\rangle,\nonumber\\
|-_C\rangle &\approx& \sqrt{2}~\sum_{m~{\rm odd}}\tilde f(m) \int dQ~ F(Q-m) |Q\rangle.\nonumber\\
\end{eqnarray} 
where
\begin{eqnarray}
2 \sum_{m~{\rm even}} |\tilde f(m)|^2 &\approx& \int dQ~|\tilde f(Q)|^2 \approx 1 ,\nonumber\\
2\sum_{m~{\rm odd}} |\tilde f(m)|^2 &\approx& \int dQ~|\tilde f(Q)|^2 \approx 1 .
\end{eqnarray}

\subsection{Phase gate overrotation}

We wish to study the performance of the encoded phase gate, which rotates the relative phase of $|0_C\rangle$ and $|1_C\rangle$ by $-\pi/2$. For ideal codewords, this gate is achieved by applying the unitary operator $e^{-i\varphi^2/2\pi}$, which has the value 1 when $\varphi$ is an even multiple of $\pi$, and has the value $-i$ when $\varphi$ is an odd multiple of $\pi$. Loosely speaking, this operation arises from the harmonic potential of the superinductor, turned on for a specified time interval. An error could occur because the  timing is not precisely correct, so that $e^{-i\varphi^2(1+\varepsilon)/2\pi}$ is applied instead, where $\varepsilon \ll 1$.

In $\varphi$ space, each narrow peak in the approximate codeword is stabilized by the cosine potential, but when the phase gate is overrotated, the relative phases of the peaks are modified, with the peak localized near $\varphi = n\pi$ acquiring the phase $e^{-i\varepsilon \pi n^2 /2}$. Thus, instead of Eq.~(\ref{eq:plus-minus-codewords}), the approximate $\bar X$ eigenstates become
\begin{eqnarray}
|\pm_C\rangle_\varepsilon&=&  \sqrt{\pi}~\sum_n (\pm 1)^n e^{-i\varepsilon \pi n^2 /2}\tilde F(n\pi)\nonumber\\
&&\quad \quad \times \int dQ~e^{-i\pi nQ}\tilde f(Q)|Q\rangle.
\end{eqnarray} 
We again use the Poisson summation formula
\begin{equation}
\sqrt{\pi}~\sum_n e^{-i\pi n Q }\tilde F_\varepsilon(\pi n) = \sqrt{2}~\sum_m F_\varepsilon(Q-2m),
\end{equation}
but now applied to the modified function
\begin{equation}
\tilde F_\varepsilon(\varphi) =e^{-i\varepsilon\varphi^2/2\pi} \tilde F(\varphi) 
\end{equation}
such that 
\begin{equation}
\tilde F_\varepsilon(n\pi) = e^{-i\varepsilon \pi n^2 /2}\tilde F(n\pi).
\end{equation}
Therefore, as in Eq.~(\ref{eq:code-in-Q-space}), the approximate codewords $|\pm_C\rangle_\varepsilon$ can be expressed as
\begin{eqnarray}\label{eq:overrotated-plus_codeword}
|+_C\rangle_\varepsilon &\approx& \sqrt{2}~\sum_{m~{\rm even}} \tilde f(m) \int dQ~ F_\varepsilon(Q -m) |Q\rangle,\nonumber\\
|-_C\rangle_\varepsilon &\approx& \sqrt{2}~\sum_{m~{\rm odd}} \tilde f(m) \int dQ~ F_\varepsilon(Q -m) |Q\rangle,\nonumber\\
\end{eqnarray} 
where 
\begin{eqnarray}
F_\varepsilon(Q) = \frac{1}{\sqrt{2\pi}}\int d\varphi~e^{-iQ\varphi}\tilde F_\varepsilon(\varphi).
\end{eqnarray}

\subsection{Imaginary part of overrotation error}
As explained in Sec.~\ref{subsec:gate-error}, the imaginary part of the gate error due to overrotation is estimated as 
\be
{\rm Im}~\eta_\varepsilon = {\rm Im}~\langle \psi_-^{\rm end}|\bar X|\psi_+^{\rm end}\rangle,
\ee
where
\be
&&\langle\psi_-^{\rm end}|\bar X |\psi_+^{\rm end}\rangle\approx\nonumber\\
&&2 \int_{[Q]~{\rm even}}dQ |\tilde f(Q)|^2\sum_{{m~{\rm even}}\atop{n~{\rm odd}}}F(Q-n)^*F(Q-m)\nonumber\\
&& - 2\int_{[Q]~{\rm odd}}dQ |\tilde f(Q)|^2\sum_{{m~{\rm even}}\atop{n~{\rm odd}}}F(Q-n)^*F(Q-m).\nonumber\\
\ee
We evaluate this expression as follows:
\begin{widetext}
\be
\langle\psi_-^{\rm end}|\bar X |\psi_+^{\rm end}\rangle&\approx&2 \sum_{m~{\rm even}} |\tilde f(m)|^2 \int_{m-\tfrac{1}{2}}^{m+\tfrac{1}{2}}dQ~\left(F(Q-m-1)^* + F(Q-m+1)^*\right)F(Q-m)\nonumber\\
&-&2 \sum_{n~{\rm odd}} |\tilde f(n)|^2 \int_{n-\tfrac{1}{2}}^{n+\tfrac{1}{2}}dQ~F(Q-n)^* \left(F(Q-n-1)+F(Q-n+1)\right)\nonumber\\
&\approx&  \left(2\sum_{m~{\rm even}} |\tilde f(m)|^2\right)\left(\int_{-1}^0 dQ~F(Q-\tfrac{1}{2})^*F(Q+\tfrac{1}{2}) +\int_0^1 dQ~F(Q+\tfrac{1}{2})^*F(Q-\tfrac{1}{2})\right)\nonumber\\
&-&  \left(2\sum_{n~{\rm odd}} |\tilde f(n)|^2\right)\left(\int_{-1}^0 dQ~F(Q+\tfrac{1}{2})^*F(Q-\tfrac{1}{2}) +\int_0^1 dQ~F(Q-\tfrac{1}{2})^*F(Q+\tfrac{1}{2})\right)\nonumber\\
&\approx& 2\int_{0}^1 dQ~\left({\rm Odd}\left[F(Q+\tfrac{1}{2})^*F(Q-\tfrac{1}{2})\right] - c.c.\right)\nonumber\\
&\approx& 2\int_{0}^\infty dQ~\left({\rm Odd}\left[F(Q+\tfrac{1}{2})^*F(Q-\tfrac{1}{2})\right] - c.c.\right),
\ee
\end{widetext}
To obtain the first equality we suppose that the integral is dominated by the overlaps of peaked functions $\{F(Q-m)\}$ centered at neighboring integer values of $m$, and approximate the slowly varying function $\tilde f(Q)$ by a constant in each integral. We obtain the second equality by shifting the integration variable in each integral, and the third equality by using the normalization condition $2 \sum_{m~{\rm even}} |\tilde f(m)|^2 \approx 1\approx 2 \sum_{n~{\rm odd}} |\tilde f(n)|^2 $, while noting that the integral of the even part of $F(Q+\tfrac{1}{2})^*F(Q-\tfrac{1}{2})$ cancels between the integrals over $[-1,0]$ and $[0,1]$. (${\rm Odd}[G(Q)]\equiv\tfrac{1}{2}\left(G(Q) - G(-Q)\right)$ denotes the odd part of the function $G(Q)$, and $c.c.$ denotes the complex conjugate.) Finally, because the integrand decays rapidly we make a negligible error by extending the upper limit of integration from 1 to infinity. Thus we conclude that
\be
{\rm Im} ~\eta_\varepsilon \approx 4~{\rm Im}\int_{0}^\infty dQ ~{\rm Odd}\left[F(Q+\tfrac{1}{2})^*F(Q-\tfrac{1}{2})\right].\nonumber\\
\ee

\section{Poisson summation formula}
\label{app:poisson}

To derive Eq.~(\ref{eq:poisson-summation}), we note that 
\begin{equation}\label{eq:periodic-define}
G(Q) \equiv \sum_m F(Q-2m)
\end{equation}
is a periodic function of $Q$ with period two, and therefore has a Fourier series expansion
\begin{equation}\label{eq:fourier-series}
G(Q)= \sum_n e^{-i\pi n Q} \tilde G_n,
\end{equation}
where
\be
\tilde G_n&=&  \frac{1}{2}\int_0^2 dQ~e^{i\pi nQ} G(Q)\nonumber\\
&=& \frac{1}{2}\int_0^2 dQ \sum_m F(Q-2m) e^{i \pi n Q } e^{-i2\pi mn}
\ee
(in the last equality we have inserted $e^{-i2\pi  m n}=1$). Now we can combine the integral over $Q$ from 0 to 2 and the sum over $m$, obtaining an integral over $Q$ from $-\infty$ to $\infty$; therefore,
\begin{equation}
\tilde G_n=  \frac{1}{2}\int_{-\infty}^{\infty} dQ~ F(Q) e^{i \pi Q n}= \sqrt{\frac{\pi}{2}} ~\tilde F(\pi n),
\end{equation}
and combining Eq.~(\ref{eq:periodic-define}) with Eq.~(\ref{eq:fourier-series}) yields
\begin{equation}\label{eq:Poisson}
G(Q) = \sum_m F(Q-2m)= \sqrt{\frac{\pi}{2}}
~\sum_n e^{-i\pi nQ} \tilde F(\pi n).
\end{equation}

A more general formula is also sometimes useful:
\begin{eqnarray}\label{eq:Poisson-generalized}
&&\sum_m F(Q-2m)e^{i\pi(Q-2m)\alpha}\nonumber\\
&&= \sqrt{\frac{\pi}{2}}
~\sum_n e^{-i\pi Qn} \tilde F(\pi \left(n +\alpha\right)).
\end{eqnarray}
We may obtain Eq.~(\ref{eq:Poisson-generalized}) from Eq.~(\ref{eq:Poisson}) by observing that multiplying $F(Q)$ by $e^{i\pi Q\alpha}$ is equivalent to shifting the argument of its Fourier transform $\tilde F(\varphi)$ by $\pi\alpha$.

\section{Diabatic transitions in a two-level system}
\label{app:landau}

Here we derive a formula for the probability of a diabatic transition in a time dependent two-level system, used in Sec.~\ref{sec:diabatic} to estimate the probability of a transition in the oscillator as the coupling between the oscillator and the 0-$\pi$ qubit ramps on or off.

We consider the Schr\"odinger equation
\be
\frac{d}{dt}|\psi(t)\rangle = -i~H(t)|\psi(t)\rangle
\ee
(with $\hbar = 1$) for the two-level time-dependent Hamiltonian
\be
H(t) = -\Delta ~\sigma^Z - V_0 e^{t/\tau}\sigma^X,
\ee
where $\sigma^Z$,$\sigma^X$ are the Pauli matrices. If we express the time $t$ in units of $\tau$, and absorb $V_0$ by shifting the time variable, the Hamiltonian becomes (assuming $V_0 > 0$)
\be 
H(t) = -u~\sigma^Z - e^{t}\sigma^X,
\ee
where
\be
u = \tau \Delta.
\ee
In the limit $t\rightarrow -\infty$, the second term is negligible, and the general solution becomes
\be
|\psi(t)\rangle = c_0 ~e^{iut}\left(
\begin{array}{c}
1 \\
0
\end{array}
\right)
+ 
c_1 ~e^{-iut}\left(
\begin{array}{c}
0 \\
1
\end{array}
\right),
\ee
while in the limit $t\rightarrow \infty$ the first term is negligible and the solution is
\be
|\psi(t)\rangle = c_+ ~e^{i{e^t}}
\frac{1}{\sqrt{2}}\left(
\begin{array}{c}
1 \\
1
\end{array}
\right)
+ 
c_- ~e^{-i{e^t}}
\frac{1}{\sqrt{2}}\left(
\begin{array}{c}
1 \\
{-}1
\end{array}
\right).
\ee
Our goal is the find the S-matrix relating these two asymptotic solutions:
\be
\left(
\begin{array}{c}
c_+ \\
c_-
\end{array}
\right)
= S\left(
\begin{array}{c}
c_0 \\
c_1
\end{array}
\right).
\ee

Defining
\be
|\psi(t)\rangle \equiv \left(\begin{array}{c}
c_0(t) \\
c_1(t)
\end{array}
\right)
= \left(\begin{array}{c}
e^{iut}~\tilde c_0(t) \\
e^{-iut}~\tilde c_1(t)
\end{array}
\right),
\ee
the Schr\"odinger equation becomes
\be
\frac{d\tilde c_0}{dt} &=& i ~ e^{(1-2iu)t} ~\tilde c_1 \nonumber\\
\frac{d\tilde c_1}{dt} &=& i ~ e^{(1+2iu)t} ~\tilde c_0.
\ee
Assuming $u> 0$, the solution $|\psi^{(0)}(t)\rangle$ that starts out in the ground state obeys the initial conditions $\tilde c_0 \rightarrow 1$ and $\tilde c_1\rightarrow 0$ as $t\rightarrow -\infty$; expanded as a power series in $e^t$, this solution is
\be
\tilde c_0(t) & = & \sum_{n=0}^\infty \frac{(-1)^n}{n!}\frac{\Gamma(\tfrac{1}{2}+ i u)}{\Gamma(\tfrac{1}{2} + iu +n)}\left(\frac{e^t}{2}\right)^{2n}\nonumber\\
\tilde c_1(t) & = & \left(\frac{i}{2}\right)e^{(1+2iu)t}\sum_{n=0}^\infty \frac{(-1)^n}{n!}\frac{\Gamma(\tfrac{1}{2}+ i u)}{\Gamma(\tfrac{3}{2} + iu +n)}\left(\frac{e^t}{2}\right)^{2n}\nonumber\\
\ee
Matching this formula to the power series expansion for the Bessel function
\be
J_\nu(x)=\sum_{n=0}^\infty \frac{1}{n!~\Gamma(\nu + 1 + n)}\left(\frac{x}{2}\right)^{\nu+2n},
\ee
 we find
\be
\tilde c_0(t) & = & \Gamma(\tfrac{1}{2} + iu) \left(\frac{e^t}{2}\right)^{\frac{1}{2} -iu}J_{{-}\frac{1}{2} +iu}(e^t),\nonumber\\
\tilde c_1(t) & = & \frac{i}{2}~e^{(1+2iu)t}~\Gamma(\tfrac{1}{2} + iu) \left(\frac{e^t}{2}\right)^{{-}\frac{1}{2} -iu}J_{\frac{1}{2} +iu}(e^t),\nonumber\\
\ee
and therefore
\be
|\psi^{(0)}(t)\rangle = 2^{{-}\frac{1}{2} + iu}~e^{t/2}~\Gamma(\tfrac{1}{2} + iu)
\left(
\begin{array}{c}
 J_{{-}\frac{1}{2} +iu}(e^t) \\
i~J_{\frac{1}{2} +iu}(e^t)
\end{array}
\right).\nonumber\\
\ee
To find the solution $|\psi^{(1)}(t)\rangle$ that starts out in the excited state, it suffices to change the sign of $u$ and interchange $c_0$, $c_1$; hence
\be
|\psi^{(1)}(t)\rangle = 2^{{-}\frac{1}{2} - iu}~e^{t/2}~\Gamma(\tfrac{1}{2} - iu)
\left(
\begin{array}{c}
i~J_{\frac{1}{2} -iu}(e^t)\\
 J_{{-}\frac{1}{2} -iu}(e^t) 
\end{array}
\right).\nonumber\\
\ee

From the asymptotic behavior
\be
J_\nu(x)\approx \sqrt{\frac{2}{\pi x}}~\cos\left(x - (\nu +\tfrac{1}{2})\tfrac{\pi}{2}\right)
\ee
of the Bessel function as $x\rightarrow \infty$, we find how the solutions $|\psi^{(0,1)}(t)\rangle$ behave for $t\rightarrow \infty$:
\begin{widetext}
\be
|\psi^{(0)}(t)\rangle &\approx& \frac{2^{iu}~\Gamma(\tfrac{1}{2} + iu)}{\sqrt{2\pi}}\left(e^{\frac{\pi}{2}u} ~e^{i{e^t}}
\frac{1}{\sqrt{2}}\left(
\begin{array}{c}
1 \\
1
\end{array}
\right)
+ 
e^{-\frac{\pi}{2}u}~e^{-i{e^t}}
\frac{1}{\sqrt{2}}\left(
\begin{array}{c}
1 \\
{-}1
\end{array}
\right)\right),\nonumber\\
|\psi^{(1)}(t)\rangle &\approx& \frac{2^{-iu}~\Gamma(\tfrac{1}{2} - iu)}{\sqrt{2\pi}}\left(e^{-\frac{\pi}{2}u} ~e^{i{e^t}}
\frac{1}{\sqrt{2}}\left(
\begin{array}{c}
1 \\
1
\end{array}
\right)
+ 
e^{\frac{\pi}{2}u}~e^{-i{e^t}}
\frac{1}{\sqrt{2}}\left(
\begin{array}{c}
{-}1 \\
1
\end{array}
\right)\right).
\ee
\end{widetext}
Hence we conclude that the S-matrix is
\be
\left(
\begin{array}{cc}
f(u) & f(-u) \\
f(-u)^* & -f(u)^*
\end{array}
\right),
\ee
where
\be
f(u) = \frac{1}{\sqrt{2\pi}}~ 2^{iu}~ \Gamma(\tfrac{1}{2} + iu)~e^{\frac{\pi}{2} u}.
\ee

The probability of a transition from the ground state $|0\rangle$ to the excited state $|-\rangle$, or from the excited state to the ground state, is
\be
P(0\rightarrow -)  = |f(-u)|^2 \ = \frac{1}{2\pi}~ \Gamma(\tfrac{1}{2}+iu)\Gamma(\tfrac{1}{2} - iu) e^{-\pi u};\nonumber\\
\ee
from the identity $\Gamma(x)\Gamma(1-x) = \tfrac{\pi}{\sin\pi x}$, we obtain
\be
P(0\rightarrow -) = P(1\rightarrow +) = \frac{e^{-\pi u}}{2 \cos(i\pi u)} \nonumber\\
=\frac{e^{-\pi u}}{e^{\pi u}+e^{-\pi u}}= \frac{1}{2}\left( 1-\tanh \pi u\right).
\ee
(The probability that no transition occurs is given by the same formula, but with $u$ replaced by $-u$.)
For $u$ large, {\em i.e.}, when the initial energy splitting $2\Delta$ is large compared to the time scale $\tau$ for the perturbation to turn on, the transition probability is exponentially suppressed:
\be
P(0\rightarrow -) =P(1\rightarrow +) \approx e^{-2\pi u}= e^{-2\pi\tau \Delta}.
\ee

\end{document}